\documentclass[aps,preprint,superscriptaddress,nofootinbib,preprintnumbers,prd,eqsecnum]{revtex4-2}

\usepackage{bm,amsmath,amsfonts,amssymb,slashed,graphicx,xspace,placeins,rotating}
\usepackage{xcolor,colortbl}
\usepackage{multirow,array,makecell}
\usepackage[inline]{enumitem}
\usepackage{setspace}
\usepackage[font={small,stretch=1}, justification=centerlast, compatibility=false]{caption}
\usepackage{subcaption}
\usepackage{stackengine,scalerel}
\makeatletter
\renewcommand\centerlast{%
  \let\\\@centercr
  \leftskip\z@\@plus 1fil%
  \rightskip\z@\@plus -1fil%
  \parfillskip\z@\@plus 1fill%
}
\makeatother

\allowdisplaybreaks
\DeclareMathAlphabet{\mathpzc}{OT1}{pzc}{m}{it}

\definecolor{nicered}{rgb}{0.7,0.1,0.1}
\definecolor{nicegreen}{rgb}{0.1,0.5,0.1}
\definecolor{niceblue}{rgb}{0.1,0.1,0.5}
\usepackage[bookmarks=false]{hyperref}
\hypersetup{colorlinks,citecolor= nicegreen,linkcolor= niceblue,urlcolor= niceblue}
\usepackage{breakurl}

\newcommand{\nn}{\nonumber}
\newcommand{\GeV}{\text{GeV}}
\newcommand{\MeV}{\text{MeV}}

\newcommand{\g}{\gamma}
\newcommand{\mn}{\mu\nu}
\newcommand{\ab}{\alpha\beta}
\newcommand{\ceq}{\stackrel{\to}{=}}

\makeatletter
\newcommand{\overleftrightsmallarrow}{\mathpalette{\overarrowsmall@\leftrightarrowfill@}}
\newcommand{\overarrowsmall@}[3]{%
  \vbox{%
    \ialign{%
      ##\crcr
      #1{\smaller@style{#2}}\crcr
      \noalign{\nointerlineskip}%
      $\m@th\hfil#2#3\hfil$\crcr
    }%
  }%
}
\def\smaller@style#1{%
  \ifx#1\displaystyle\scriptstyle\else
    \ifx#1\textstyle\scriptstyle\else
      \scriptscriptstyle
    \fi
  \fi
}
\makeatother

\newcommand{\genbar}[1]{\,\overline{\!#1}{}}
\newcommand{\gentilde}[1]{\,\widetilde{\!#1}{}}

\makeatletter
	\let\vec@temp\vec
	\renewcommand{\vec}[1]{\!\vec@temp{\,#1}}
\makeatother
\stackMath
\def\dhat#1{\ThisStyle{\setbox0=\hbox{$\SavedStyle#1$}%
  \stackengine{0pt}{\SavedStyle#1}{\SavedStyle\hspace{.2\ht0}%
  \hat{\vphantom{#1}}\kern\dimexpr2.2\LMpt+.7pt\relax\hat{\vphantom{#1}}}{O}{c}{F}{T}{L}}%
}

\newcommand{\Dx}{D^{(*)}}
\newcommand{\Bx}[1][]{B^{(*)#1}}

\newcommand{\cbar}{\bar{c}}
\newcommand{\Bbar}{\genbar{B}}

\newcommand{\Bxbar}[1][]{\Bbar^{(*)#1}}
\newcommand{\Dv}[1][1]{D^{*}_{#1}}
\newcommand{\Djx}[1][]{D_J^{(*)#1}}

\newcommand{\vc}[1][]{\hat{v}^{\prime#1}}
\newcommand{\vb}[1][]{\hat{v}^{#1}}

\newcommand{\cbvp}{\cbar^{v'}_+}
\newcommand{\ctvp}{\tilde{c}^{\vc}_+}
\newcommand{\cvp}[1][\vc]{c^{#1}_+}
\newcommand{\bv}[1][\vb]{b^{#1}_+}

\newcommand{\btv}{\tilde{b}^{\vb}_+}

\newcommand{\Dslash}{\slashed{D}}
\newcommand{\vslash}[1][\vb]{\slashed{#1}}
\newcommand{\ccdot}{\!\cdot\!}
\newcommand{\vcD}{\vb \ccdot D}
\newcommand{\Qbar}{\genbar{Q}}
\newcommand{\Qtilde}{\gentilde{Q}}
\newcommand{\Jbar}{\genbar{J}}
\newcommand{\Jtilde}{\gentilde{J}}
\newcommand{\mJ}{\mathcal{J}}
\newcommand{\mL}{\mathcal{L}}

\newcommand{\mJtilde}{\gentilde{\mJ}}
\newcommand{\Hc}{H_c}

\newcommand{\Hcp}[1][\mkern-1mu]{H_{c}^{\prime\mkern1mu#1}}
\newcommand{\Hcf}[1][\!]{H_{c}^{\mkern-2mu f\,#1}}
\newcommand{\Hb}{H_b}
\newcommand{\Hbp}[1][\mkern-1mu]{H_{b}^{\prime\mkern1mu#1}}
\newcommand{\Piv}[2][\vb]{\Pi_{#2}(#1)}
\newcommand{\mLqcd}{\mL_{\text{QCD}}}

\newcommand{\Qpv}{Q_+^{\vb}}
\newcommand{\Qtpv}{\Qtilde_+^{\vb}}
\newcommand{\Qmv}{Q_-^{\vb}}
\newcommand{\Qtmv}{\Qtilde_-^{\vb}}

\newcommand{\Hbar}{\genbar{H}}

\newcommand{\Htilde}{\gentilde{H}}

\newcommand{\mf}{m_{\mkern-2mu f}}
\newcommand{\pf}{p_{\mkern-2mu f}}
\newcommand{\mfpi}{m_{f\pi}}
\newcommand{\pc}[1][]{p^{\prime#1}}
\newcommand{\mc}[1][]{m^{\prime#1}}
\newcommand{\hmc}[1][]{\hat{m}^{\prime#1}}
\newcommand{\Gc}[1][]{\Gamma^{\prime#1}}
\newcommand{\hpc}[1][]{\hat{p}^{\prime#1}}
\newcommand{\tpc}[1][]{\tilde{p}^{\prime#1}}
\newcommand{\heps}{\hat\epsilon}
\newcommand{\zbr}{z_{\text{br}}}

\newcommand{\Cv}[1]{C_{V#1}}
\newcommand{\Ca}[1]{C_{A#1}}

\newcommand{\hL}[2]{\dhat{L}_{#1}^{(#2)}}
\newcommand{\hchi}[1]{\dhat{\chi}_{#1}}
\newcommand{\heta}{\dhat{\eta}}

\newcommand{\Hha}[1]{\hat{h}_{A#1}}
\newcommand{\Hhv}{\hat{h}_V}
\newcommand{\hHa}[1]{\dhat{h}_{A#1}}
\newcommand{\hHv}{\dhat{h}_V}

\newcommand{\lqcd}{\Lambda_{\text{QCD}}}

\newcommand{\LamB}{\bar{\Lambda}}
\newcommand{\GamB}{\bar{\Gamma}}
\newcommand{\lam}[1]{\lambda_{#1}}
\newcommand{\kap}[1]{\kappa_{#1}}
\newcommand{\aS}{\alpha_s}
\newcommand{\haS}{{\hat{\alpha}_s}}
\newcommand{\ec}{\varepsilon_c}
\newcommand{\eb}{\varepsilon_b}
\newcommand{\eQ}{\varepsilon_Q}
\newcommand{\mbS}{m_b^{1S}}
\newcommand{\dmbc}{\delta m_{bc}}
\newcommand{\sellP}[1][(H)]{\sigma#1}
\newcommand{\sqLLp}{\sqrt{\mkern-2mu \LamB\LamB'}}

\newcommand{\amp}[3]{\frac{\langle #1(p') |\, #2\, | #3(p) \rangle}{\sqrt{m_{\let\overline\relax#3} m_{#1}}}}

\newcommand{\mB}{m_B}
\let\mDv\mc
\let\pDv\pc
\let\GDv\Gc
\let\hmDv\hmc
\let\hpDv\hpc
\newcommand{\yDv}[1][]{y^{\prime#1}}

\newcommand{\hrDv}[1][]{\hat{r}^{\prime#1}}
\newcommand{\mDpi}{m_{D\pi}}
\newcommand{\rDpi}{r_{D\pi}}

\newcommand{\Rg}[1]{\hat{R}_{#1}}
\DeclareMathOperator{\sech}{sech}

\newcommand{\wh}{\hat{w}}

\newcommand{\etpi}{\eta_\pi}
\newcommand{\phpi}{\phi_\pi}

\def\spnt{1}
\if\spnt1
	\newcommand{\up}{+}
	\newcommand{\dn}{-}
\else
	\newcommand{\up}{2}
	\newcommand{\dn}{1}
\fi

\newcommand{\Tr}{\text{Tr}}

\makeatletter
\g@addto@macro\bfseries{\boldmath}
\makeatother

\makeatletter
\let\tmp@footnote\footnote
\renewcommand{\footnote}[1]{\tmp@footnote{\linespread{0.9}\selectfont{}#1}}
\makeatother

\makeatletter
\let\temp@caption\caption
\renewcommand{\caption}[2][]{\temp@caption[#1]{\linespread{1.2}\selectfont{}#2}}
\makeatother

\allowdisplaybreaks
\begin{document}

\title{On-shell recursion and holomorphic HQET for heavy quark hadronic resonances}

\author{Claudio Andrea Manzari}
\author{Dean J.\ Robinson}
\affiliation{Ernest Orlando Lawrence Berkeley National Laboratory, 
University of California, Berkeley, CA 94720, USA}
\affiliation{Berkeley Center for Theoretical Physics, 
Department of Physics,
University of California, Berkeley, CA 94720, USA}

\begin{abstract}
We develop a new theoretical framework for the treatment of heavy quark (HQ) resonances within heavy quark effective theory (HQET). 
This framework uses on-shell recursion techniques to express the resonant amplitude as a product of on-shell subamplitudes, 
which allows one to employ a form-factor representation of the hadronic matrix elements and to obtain an HQ expansion, 
but at the price of introducing complex momenta. 
We construct a generalized ``holomorphic HQET'' onto which such complex-momentum matrix elements can be matched,
and we show that $PT$ symmetry ensures the Isgur-Wise functions (and the perturbative corrections) become holomorphic functions of the complex recoil parameter with real coefficients. 
They are thus an analytic continuation of the standard HQET description. 
This framework admits a HQ hadron (strong decay) width expansion. 
At second order, we show it is compatible with data for the $B_{1(2)}^{(*)}$ and $D_{1(2)}^{(*)}$ HQ doublets.
Taking the $\bar{B} \to (D_1^*(1^-) \to D\pi)l\nu$ system as an example, 
we compute the holomorphic HQET expansion to first order,
as well as the complex-momentum on-shell subamplitudes.
A toy numerical study of the resulting differential rates demonstrates that this framework generates HQ resonance lineshapes with large tails, 
resembling those seen in data.
\end{abstract}

\makeatletter
\let\temp@clearpage\clearpage
\let\clearpage\relax
\maketitle
\let\clearpage\temp@clearpage
\makeatother

\newpage
 \twocolumngrid
 {
 \fontsize{10}{8}\selectfont 
 \columnsep20pt
 	\tableofcontents
 }
 \onecolumngrid

\section{Introduction}

Anticipated future datasets for semileptonic $b \to c l\nu$ decays will be large enough 
to enable measurement of the CKM matrix element $|V_{cb}|$ and the lepton flavor universality violation (LFUV) ratios $R(\Dx)$ 
at (sub)percent statistical precision.
However, such measurements can only be realized if all large sources of systematic uncertainty are brought under control.
In this respect, the semileptonic $\Bbar \to D^{**} l \nu$ decays to orbitally-excited final states are responsible for 
a leading systematic uncertainty in precision measurements of $\Bbar \to \Dx l \nu$ exclusive decays as well as in inclusive $B \to X_c l\nu$ measurements.
In exclusive measurements they generate so-called feed-down backgrounds from $D^{**} \to \Dx \pi$ strong decays
in which the pion is not reconstructed with adequate efficiency,
resulting in systematic uncertainties at the few percent level in measurements of $|V_{cb}|$~\cite{BaBar:2007cke,Belle:2015pkj,Belle:2017rcc,Belle:2018ezy}
or $R(\Dx)$~\cite{BaBar:2013mob,Belle:2016dyj,Belle:2017ilt,Belle:2019rba,LHCb:2015gmp,LHCb:2017rln,LHCb:2023uiv,LHCb:2023zxo}.
In inclusive measurements~\cite{CLEO:2004bqt, BaBar:2009zpz,Belle:2006jtu,Belle:2021idw,Belle-II:2022evt} they contribute to the signal mode itself.

The $D^{**}$ states furnish two heavy quark (HQ) symmetry doublets
with typical widths of a few$\,\times\,100$\,MeV or few$\,\times 10\,$\,MeV, respectively.
While these states are still somewhat narrow in the sense that the width-mass ratio $\Gamma/m$ is small,
the importance of off-shell and finite-width effects---the modelling of the $D^{**}$ lineshape---is 
better characterized by the ratio of the decay width versus the \emph{splitting} 
between the resonance mass and the nearest pair production threshold
(see the ``Resonances'' chapter of Ref.~\cite{Workman:2022ynf} for a review).
In the case of $\Bbar \to (D^{**} \to DX)l\nu$, $X = \pi$, $\pi\pi$, decays
the ratios $\Gamma_{D^{**}}/|m_{D^{**}} - m_{D + X}| \sim 10$--$50$\%, 
suggesting that the modelling of the $D^{**}$ lineshapes and associated off-shell and finite-width effects 
may play an important role in reducing the net systematic uncertainty in $|V_{cb}|$, $R(\Dx)$ or $R(X_c)$ 
measurements from several percent to below the percent level.
As an example, in the recent Belle~II analysis~\cite{Belle-II:2022evt}, 
uncertainties associated with $\Bbar \to D^{**} l \nu$ decay branching fraction and lineshape modelling
account for more than $30\%$ of the total systematic uncertainty 
in regions with the greatest sensitivity to $|V_{cb}|$.

State-of-the-art theoretical descriptions for these decays, however,
rely on the assumption that the intermediate $D^{**}$ can be treated as on-shell,
based on several \emph{ad hoc} justifications and/or motivations:
First, this assumption permits the full amplitude to be factored into a $B \to D^{**} l\nu$ piece, 
a $D^{**} \to DX$ piece, and a model for the $D^{**}$ lineshape.
This is the approach used in e.g. \texttt{EvtGen}~\cite{Lange:2001uf,Ryd:2005zz}, 
which uses a Breit-Wigner parametrization with Blatt-Weisskopf centrifugal barrier factors~\cite{Blatt:1952ije,VonHippel:1972fg} for the resonance lineshape
(see again Ref.~\cite{Workman:2022ynf} for a review).
Second, treatment of the $D^{**}$ as an on-shell intermediate state permits the $\Bbar \to D^{**}$ hadronic matrix elements 
to be represented by standard bases of form factors,
i.e. those that span the set of allowed amplitudes per angular momentum and (charge-)parity selection rules for on-shell particle states.
This omits, however, longitudinal mode contributions that arise in the exchange of resonances with non-trivial spin. 
Third, calculation or parametrization of these form factors may then proceed by use of various theoretical frameworks, 
including Heavy Quark Effective Theory (HQET) or dispersive bound approaches among others,
that also require the hadronic states to be on-shell.
For instance, the HQET traces that represent the HQET matrix elements are constructed explicitly using spacetime representations for on-shell particle states. 
Thus, even if one incorporates the longitudinal mode with additional form factors, current theoretical frameworks cannot describe them.
As a result, state-of-the-art theoretical approaches not only do not, but also seemingly cannot, incorporate
the potentially large corrections from off-shell effects in $\Bbar \to (D^{**} \to DX)l\nu$ decays
in a self-consistent manner.
(The recent Ref.~\cite{Gustafson:2023lrz} attempts to incorporate longitudinal contributions via an \emph{ad hoc} partial wave expansion of the matrix elements;
this approach, however, relies on approximate factorization arguments
and cannot systematically incorporate relations between matrix elements generated by 
e.g the heavy quark expansion or dispersive bounds without formally extending the latter to describe off-shell states.)

In the general framework of HQET, the heavy quark within the HQ hadron is generically not on-shell.
Typically it has a dispersion relation of the form $p_Q = m_Q v + k$, where $v$ is the HQ velocity 
and $k$ is the residual momentum in the light degrees of freedom that dresses the heavy quark: the brown muck.
Thus on general grounds, one expects that an effective theory for HQ hadrons that incorporates off-shell hadron effects 
should still conform to the expectations from HQ symmetry and should still exhibit an HQ expansion.
Further, one should expect the Isgur-Wise functions of the HQET for off-shell hadrons to be related, somehow, to those in the standard HQET for on-shell hadrons,
because the brown muck---the light degrees of freedom whose dynamics HQET describes---cannot know whether the hadron itself is on-shell or off-shell.
 
In this paper we develop a new framework for the treatment of heavy quark resonances that is compatible with these HQ symmetry expectations,
and admits an HQET-like description of matrix elements containing an off-shell resonant exchange.
This framework uses on-shell recursion techniques to express the full QCD amplitude 
in terms of a product of ``shifted'' on-shell subamplitudes times simple pole residues.
Because the external hadron states are massive, 
the momentum two-line shift required to construct the recursion relation is itself complex.
This leads to the introduction of a complex null reference momentum for the resonance polarization states,
that in essence corresponds to choosing a polarization axis for the resonance in complex $3+1$ dimensional space.
In turn the shifted on-shell subamplitudes for the hadronic matrix elements must involve complex momenta---we say they are ``complex-shifted''---and 
thus must be computed within a complexified generalization of the usual spinor-helicity formalism.
A generalized sense of complex conjugation, that preserves the holomorphy of the matrix elements 
with respect to the resonance simple pole, 
controls the analytic properties (and discrete-symmetry transformation properties) of the matrix elements
as well as of any effective theory onto which they are matched.

The complex generalization of HQET onto which the complex-shifted QCD matrix elements may be matched is not unique.
We show there is a particular generalization of HQET that incorporates complex HQ velocities 
such that the resulting HQ power expansion has the same structure as in standard HQET for on-shell hadrons:
a ``holomorphic HQET'', so called because we show such a theory must be holomorphic with respect to the complex HQ velocity.
A central and powerful result of this work is to show that $PT$ symmetry of QCD ensures the Isgur-Wise functions (and the perturbative corrections) of this effective theory 
are holomorphic functions of the complex recoil parameter, $w$, with real coefficients.
They are thus simply an analytic continuation under $w$ of the real Isgur-Wise functions (and the real perturbative corrections) in the standard HQET description for real-momentum on-shell states.
We further show that holomorphic HQET admits a generalization of the hadron mass expansion to include hadron (strong decay) widths, 
and demonstrate this expansion is compatible with data for the widths of the $B_{1(2)}^{(*)}$ and $D_{1(2)}^{(*)}$ HQ doublets.

While we employ an on-shell recursion relation within QCD, and then match onto a complexified version of HQET,
a parallel approach to this problem has contemporaneously been discussed in the literature. In Ref.~\cite{Papucci:2024qbt} the authors instead generate an on-shell recursion relation 
within Heavy Hadron Chiral Perturbation Theory (HHChPT),
after first matching the $\Bbar \to (D^{**} \to DX)$ on-shell amplitudes (with real momenta) onto HQET.
Expressions for the $\Bbar \to DX$ amplitude are then obtained order-by-order in HHChPT 
in terms of standard real Isgur-Wise functions for $\Bbar \to D^{**}$ and their derivatives.
The underlying correspondence between the results in this work and Ref.~\cite{Papucci:2024qbt} is yet to be understood.

This paper is structured as follows: In Sec.~\ref{sec:onshellrec} 
we construct the on-shell recursion relation, 
specify a generalization of the spinor-helicity formalism for complex momenta,
and examine the implied analytic properties of the complex-shifted on-shell subamplitudes.
In Sec.~\ref{sec:holohqet} we proceed to develop holomorphic HQET from first principles,
including the HQ power expansion and radiative corrections---the 
derivation of their analytic properties under $PT$ symmetry is derived in Appendices---the 
hadron mass and width expansion, and Schwinger-Dyson relations.
As a demonstrative first application of this framework, we consider in Sec.~\ref{sec:BD1*}
the $\Bbar \to (\Dv(1^-) \to D\pi)l\nu$ radially-excited decay,
which features the same quantum numbers as the familiar $\Bbar \to \Dx$ transitions.
We show how to implement the trace formalism to represent the holomorphic HQET matrix elements, 
and explicitly compute the holomorphic HQET expansion of the Standard Model (SM) form factors to first order.
We derive and apply the on-shell recursion relation for this system, 
and examine how it constrains the form of parametrizations of the Isgur-Wise functions.
We explicitly compute the complex-shifted SM subamplitudes and 
the associated fully differential rate as a function of the $\Dv$ lineshape invariant mass.
Finally, we examine a toy numerical study of the resulting off-shell differential rates, 
and find that this framework generates HQ resonance lineshapes that are characteristic of those seen in data.
Sec. \ref{sec:summout} concludes.

\section{On-shell construction}
\label{sec:onshellrec}
\subsection{Amplitude composition}
We are interested in a decay amplitude for the resonant transition $\Hb \to \Hcp \to \Hcf X$, 
where $X$ is some light hadronic (or radiative) system and $\Hcp$ denotes an intermediate resonant state.
For the sake of concreteness, in this work we take $X = \pi$ and consider an intermediate vector resonance,
keeping in mind that the following on-shell construction and results will generalize to other light hadronic system final states and intermediate states with other spins. 
The decay amplitude of interest then has the form~\footnote{%
\label{ft:hbp}
In the $m_b/m_c \to \infty$ limit, this is the sole factorization of the $\Hb \to \Hcf X l \nu$ amplitude. 
In the finite $m_b/m_c \gg 1$ regime, there are typically subleading terms from $\Hb \to \pi(\Hbp \to \Hcf l \nu)$. See Sec.~\ref{sec:othres} below.}
\begin{align}
	\label{eqn:amplM}
	\mathcal{A}[\Hb \to (\Hcp \to \Hcf \pi) l \nu] 
	& = \begin{gathered}\includegraphics[width = 6cm]{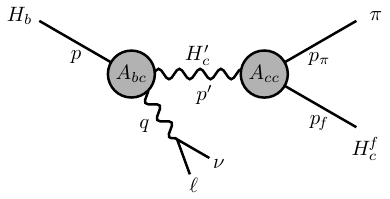}\end{gathered} \nn\\
	& = A_{bc\ell\nu}^{\mu}(p, \pc) \, \frac{g_{\mu \nu} -\pc_\mu \, \pc_\nu \big/M^2(\pc[2])}{\pc[2] - M^2(\pc[2])}  \, A_{cc}^{\nu}(\pf, p_\pi)\,,
\end{align}
in which we have also defined the kinematics associated with each leg or resonance,
following the standard convention that the incoming (outgoing) momentum of the beauty (charm) hadron in the $b \to c$ vertex is denoted $p$ ($\pc$).
$A^{\mu}_{bc\ell\nu}$ and $A^\nu_{cc}$ are amplitudes for the $b \to c\ell\nu$ electroweak and charm strong decays, respectively.
The $A^{\mu}_{bc\ell\nu}$ amplitude should be understood to involve a $b \to c$ hadronic amplitude contracted with a recoiling charged-current coupled to the leptons
(in the SM, an off-shell $W \to \ell \nu$ amplitude)
with total momentum $q$.
The external states $\Hb$, $\pi$, and $\Hcf$ are on-shell, such that $p^2 = m^2$, $p_\pi^2 = m_\pi^2$, and $\pf^2 = \mf^2$, respectively.

The momentum of the intermediate $\Hcp$ resonance is fixed by the configuration of the external states $\pc = \pf + p_\pi$,
with an off-shell invariant mass $\pc[2] = \mfpi^2 \equiv (\pf + p_\pi)^2$.
The $\Hcp$ propagator is the Green's function of the two-point 1PI function $\pc_\mu \pc_\nu - [\pc[2] - M^2(\pc[2])]g_{\mu\nu}$.
The spectral function $M^2(\pc[2])$ in the propagator is the all-orders $\Hcp$ mass term.
By virtue of the propagator, the amplitude has a simple pole at  $\pc = \hpc$ at which
\begin{equation}
	\label{eqn:simplepole}
	\hpc[2] = M^2(\hpc[2])\,,\qquad \text{with} \qquad \hpc[2] = \mc[2] - i\mc \Gc \equiv \hmc[2]\,,
\end{equation}
where $\mc$ is the pole mass and $\Gc$ is the total width of the vector $\Hcp$ state.
(As usual, the gradient $dM^{2}/d\pc[2]\big|_{\pc = \hpc} = 0$, so that the residue of the simple pole is unity.)
In addition, as the resonance is an unstable particle, $M^2(\pc[2])$ must feature a threshold singularity at 
\begin{equation}
	\label{eqn:thbrpt}
	\pc[2] = (m_f + m_{\pi})^2\,,
\end{equation}
resulting in a branch cut in $M^2(\pc[2])$ on the real axis above the branch point~\eqref{eqn:thbrpt}.

Of particular importance is the completeness relation identity
\begin{equation}
	\label{eqn:complrel}
	- g_{\mu \nu} + \frac{\pc_\mu \pc_\nu}{\pc[2]} \equiv \sum_{\lambda = \pm,0} \epsilon^\lambda_\mu(\pc) \epsilon^{\bar{\lambda}}_\nu(\pc)\,,
\end{equation}
for any $\pc[2]$, in which the $\bar{\lambda}$ superscript denotes the conjugate spin (defined in detail below).
The polarizations are transverse and orthonormal, i.e.
\begin{equation}
	 \epsilon^\lambda(\pc) \cdot \pc = 0\,, \qquad \epsilon^\lambda(\pc) \cdot \epsilon^{\bar{\kappa}}(\pc) = -\delta^{\lambda\kappa},
\end{equation}
and thus they form a particle representation of a spin-$1$ state.
Applying the completeness relation~\eqref{eqn:complrel} to the amplitude in Eq.~\eqref{eqn:amplM}, 
\begin{equation}
	\label{eqn:decampl}
	\mathcal{A} = A_{b c \ell \nu}^{\mu}(p, \pc) \, \frac{1}{\pc[2] - M^2(\pc[2])}\Big[ 
	-\sum_{\lambda = \pm,0} \epsilon^{\bar{\lambda}}_\mu(\pc) \epsilon^{\lambda}_\nu(\pc) + \pc_\mu \pc_\nu\Big(\frac{1}{\pc[2]} - \frac{1}{M^2(\pc[2])}\Big)\Big]  \, A_{cc}^{\nu}(\pf, p_\pi)\,.
\end{equation}
The standard physical interpretation is that the propagator decomposes into the exchange of an on-shell vector state of mass $\sqrt{\pc[2]} (= \mfpi)$, 
plus an off-shell longitudinal mode $\sim \pc_\mu \pc_\nu$.
The latter vanishes only at the simple pole, 
i.e. when $\pc[2] \to \hpc[2]$ per Eq.~\eqref{eqn:simplepole},
and thus is never zero if $\pc$ is real and $\Gc > 0$.

Several limits and/or approximations are typically applied to Eq.~\eqref{eqn:decampl}.
In the narrow-width \emph{limit},
$\Gc \to 0$ such that $\hmc[2] \to \mc[2]$ 
while $\Hcp$ branching ratios remain finite---i.e. $|A_{cc}|^2/\Gc$ is finite---the 
squared propagator prefactor $\big|1/(\pc[2] - M^2(\pc[2]))\big|^2 \to \pi \delta(\pc[2] - \mc[2])/(\mc \Gc)$.
This limit thus corresponds to $\Hcp$ being a real on-shell state,
and we shall therefore sometimes refer to it as the ``narrow-width real on-shell limit'' for clarity.
The narrow-width \emph{approximation} extends this to the $\Gc \ll \mc$ regime: 
the longitudinal mode is neglected
such that the intermediate resonance is treated as an on-shell particle state of varying mass $\sqrt{\pc[2]} (= \mfpi)$ 
and the lineshape is encoded in the propagator prefactor via a model for $M^2(\pc[2])$.
The narrow-width approximation formally fails as one extends it to $\pc[2]$ far from the $\mc[2]$, 
and thus it is not expected to properly capture the tails of the lineshape.
In the stricter regime where the splitting between the resonance mass and the $f\pi$ production threshold 
is much smaller than the width of the resonance, i.e. $\Gc \ll |\mc - (m_f + m_\pi)|$,
the prefactor may, however, be well approximated by a (dynamic-width) Breit-Wigner parametrization~\cite{Workman:2022ynf}.
Thus conversely, one should expect the ratio $\Gc/|\mc - (m_f + m_\pi)|$ to roughly parametrize the importance of corrections 
beyond the narrow-width and Breit-Wigner approximation.

The polarizations may be represented in spinor-helicity form with respect to a null reference momentum $k$, with $k^2 = 0$, 
that defines a lightcone projection
\begin{equation}
	\label{eqn:lcproj}
	\pc_\mu = \tpc_\mu + \frac{\pc[2]}{2 k \cdot \pc} k_\mu\,,
\end{equation}
such that $\tpc[2] =0$ and
\begin{equation}
	\epsilon^{\pm}_\mu(k; \tpc) = \pm \frac{\big\langle k^{\mp} \big| \sigma_\mu \big| \tpc[\mp] \big\rangle}{\sqrt{2} \big\langle k^{\mp} \big| \tpc[\pm] \big\rangle}\,, 
	\qquad \epsilon^{0}_\mu(k; \tpc) = i\Big( \frac{\tpc_\mu}{\sqrt{\pc[2]}} - \frac{\sqrt{\pc[2]}}{2 k \cdot \pc} k_\mu \Big)\,.
\end{equation}
By construction $k \cdot \pc = k \cdot \tpc$.
The conjugate spin in Eq.~\eqref{eqn:complrel}, indicated by the $\bar{\lambda}$ superscript,
is defined such that $\epsilon^{\bar{\pm}} = \epsilon^{\mp}$, and $\epsilon^{\bar{0}} = -\epsilon^{0}$.\footnote{%
The imaginary prefactor in the definition of $\epsilon^{0}$, and the distinction between $\epsilon^{0}$ and $\epsilon^{\bar{0}}$, is relevant when one considers spin-$2$ and higher exchanges.
For the vector exchange considered here, one could equally have defined $\epsilon^{0}_\mu(k; \tpc) = \tpc_\mu/\sqrt{\pc[2]}- \sqrt{\pc[2]}k_\mu/ 2 k \cdot \pc$ and simply $\epsilon^{\bar{0}} = \epsilon^{0}$.}
In anticipation of $k$ becoming a complex momentum below, we have defined the spin conjugation without reference to complex conjugation
(it can, however, be defined with respect to a different sense of conjugation we define below).

\subsection{On-shell recursion relation}
\label{sec:recrel}
\begin{figure}[tb]
    \centering
    \includegraphics[width = 5cm]{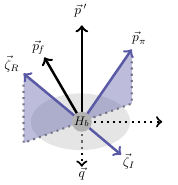}
    \caption{Configuration of a solution $\zeta = \zeta_R + i \zeta_I$ to Eq.~\eqref{eqn:zetanull} in the $\Hb$ reference frame, 
    in which the purely spacelike $\zeta_R=(0,\vec{\zeta}_R)$ lies in the plane spanned by $\vec{p}_f$ and $\vec{p}_\pi$, while purely spacelike 
    $\zeta_I=(0,\vec{\zeta}_I)$ lies perpendicular to it, such that both $\vec{\zeta}_{R,I} \cdot \vec{p}_\pi =0$. 
    See also Fig.~\ref{fig:refframes} and the explicit parametrization in Sec.~\ref{sec:BD1*}.}
    \label{fig:BRefFrame}
\end{figure}

To generate the on-shell recursion relation, we introduce a null complex momentum shift on the $b$-hadron and pion external legs through the $\Hcp$ internal line, 
\begin{align}
	p^\mu 	& \to p^\mu + z \zeta^\mu \equiv p^\mu(z) \,,\nn \\
	\pc[\mu]  	& \to  \pc[\mu] + z \zeta^\mu  \equiv \pc[\mu](z)\,, \nn \\ 
	p_\pi^\mu & \to  p_\pi^\mu + z \zeta^\mu  \equiv p_\pi^\mu(z)\,, \label{eqn:zshift}
\end{align}
with $z \in \mathbb{C}$, such that 
\begin{equation}
	\label{eqn:zetanull}
	\zeta^2 = 0 \quad \text{and} \quad  p\cdot\zeta = p_\pi \cdot \zeta = 0\,.
\end{equation}
These conditions ensure $p^2(z) = m^2$ and $p_\pi^2(z) = m_\pi^2$, i.e the $b$-hadron and pion remain on-shell.
Because both the $b$-hadron and pion are massive, the solution to Eq.~\eqref{eqn:zetanull} itself must be a complex momentum:
In the $\Hb$ rest frame, $\zeta$ must be purely space-like and lie on the plane defined by 
(i.e. orthogonal to) the pion $3$-momentum $\vec{p}_\pi$.
One may therefore write the solution in the general form $\zeta = \zeta_R + i \zeta_I$, in which $\zeta_{R,I}$ are real orthogonal space-like momenta 
satisfying $\zeta_R \cdot p_\pi = \zeta_I \cdot p_\pi = \zeta_R \cdot \zeta_I = 0$ and $\zeta_R^2 = \zeta_I^2$. 
An example solution for $\zeta_{R,I}$ is shown in Fig.~\ref{fig:BRefFrame}.
(We will return to an explicit parametrization of this solution for the example $\Bbar \to \Dv$ system considered in Sec~\ref{sec:BD1*}.)

Under this shift, the analytic continuation $\mathcal{A}(z)$ of the amplitude~\eqref{eqn:amplM} has a simple pole at $z =\hat{z}$, 
corresponding to $\pc[2](\hat{z}) = \pc[2] + 2 \hat{z} \, \zeta \cdot \pc = \hmc[2]$ from Eq.~\eqref{eqn:simplepole},
with solution
\begin{equation}
	\label{eqn:zhat}
	\hat{z} = -\big(\pc[2] - \hmc[2]\big)/(2 \zeta \cdot \pc)\,.
\end{equation}
Note $\hat{z}$ depends on the simple pole at $\hmc$ and the physical kinematics of the external states via $\pc = \pf + p_\pi$ and $\zeta$.
We use a hatted notation for all shifted momenta evaluated at $z = \hat{z}$, 
i.e. $\hpc = \pc(\hat{z})$ (the simple pole by definition) as well as
\begin{equation}
	\hat{p} = p(\hat{z}), \qquad \text{and} \quad \hat{p}_\pi = p_\pi(\hat{z})\,.
\end{equation}

At the simple pole, as noted above, the longitudinal mode contribution to the amplitude~\eqref{eqn:decampl} vanishes.
The $\Hcp$ resonance thus becomes an on-shell particle state with complex invariant mass $\hmc$, 
because it has a particle representation in terms of the three polarizations, $\epsilon^\lambda_\mu(\hpc)$.
(We construct a representation of these complex-shifted polarizations in Sec.~\ref{sec:sphlgc}.)
We thus say $\Hcp$ is on-shell at the simple pole $\pc[2](z) = \hmc[2]$, equivalent to $z = \hat{z}$.
In the narrow-width limit, this sense of on-shell reduces to the usual real on-shell condition, $\pc[2] = \mc[2]$.\footnote{%
Though the $b$-hadron and $\pi$ decay weakly and thus formally also have a non-zero width, 
we implicitly apply the narrow-width limit to these states, so that their invariant masses remain real on-shell.
We will see below in Sec.~\ref{sec:hadmassexp} that assigning a zero width to 
weakly-decaying ground-state HQ supermultiplets is well-defined order by order in the HQ expansion.}

To construct the recursion relation itself, we consider the singly-subtracted integrand $\mathcal{A}(z)/z$, 
which has simple pole residues at $z=0$ and $z=\hat{z}$.
Further, the threshold singularity per Eq.~\eqref{eqn:thbrpt} implies this integrand has a branch cut on the real axis of the complex $z$ plane for $z \ge \zbr$,
with branch point
\begin{equation}
	\label{eqn:zbr}
	\zbr = -\big(\pc[2] - (m_f + m_{\pi})^2\big)/(2 \zeta \cdot \pc)\,.
\end{equation}
A standard keyhole contour around this branch cut captures the two simple poles, such that one obtains a relation
with respect to the boundary and branch-cut discontinuity integrals
\begin{equation}
	\label{eqn:fullrecrel}
	\mathcal{A}(0) = -\text{Res}\big[\mathcal{A}(z)/z)\big]\big|_{\hat{z}} + \frac{1}{2\pi i}\bigg[\int_{|z| \to\infty} \mathcal{A}(z)/z \, dz + \int_{\zbr}^\infty \text{Disc}\big[\mathcal{A}(z)/z\big] dz\,\bigg].
\end{equation}
In generating Eq.~\eqref{eqn:fullrecrel}, it is important that the recoil momentum $q = p - \pc$ is unaffected by the shift~\eqref{eqn:zshift}, 
so that the contour integral in $z$ is agnostic to $\cbar b$-type subthreshold resonances in the $q^2$ spectrum of the $b\to c$ exclusive amplitude
(see further Sec.~\ref{sec:paramIW}).

Whether the amplitude 
$\mathcal{A} = \mathcal{A}(0)$ is ``constructible'' in terms of a recursion relation with respect to just the residue of $A(z)/z$ at $\hat{z}$
depends on whether the boundary term vanishes  in Eq.~\eqref{eqn:fullrecrel}
and whether the discontinuity integral can be neglected.
The behavior of the boundary term depends on the large-$z$ behavior of the hadronic functions within the $b \to c$ amplitude 
(see e.g. Refs.~\cite{Cheung:2015cba,Cohen:2010mi,Feng:2009ei} regarding the general topic of the constructibility of on-shell recursion relations).
To the extent that the hadronic functions characterize a wavefunction overlap of the initial and final state hadrons, 
one should expect them to vanish as the recoil $\sim p(z) \cdot \pc(z) \sim z$ grows large.
That is, one expects the $b \to c$ amplitude to vanish as $|z| \to \infty$.
Noting that one expects also $M^2(\pc[2]) \sim z$ in the large-$z$ limit, 
so that the propagator term in Eq.~\eqref{eqn:decampl} is expected to scale as $\big[z + z^2/z\big]/z \sim 1$,
then here we shall assume the hadronic functions vanish fast enough that also the full amplitude $\mathcal{A}(z)/z \to 0$ as $|z| \to \infty$,
and thus the boundary term vanishes: we leave the question of the general conditions under which this assumption is true for future investigation.

\begin{figure}[t]
   \begin{subfigure}[b]{0.4\textwidth}
		\includegraphics[width=5cm]{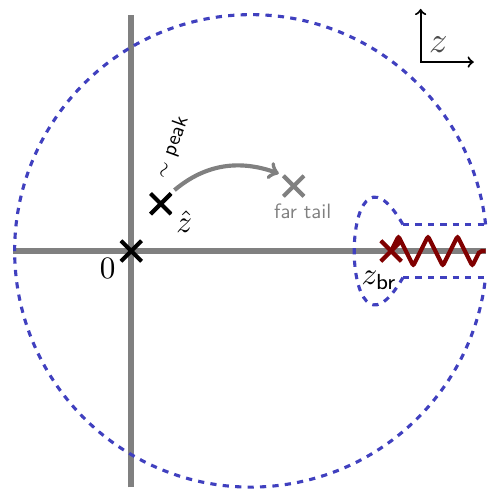}
		\caption{Near peak and far tail}
		\label{fig:cntrpeak}
   \end{subfigure}
\hfil
   \begin{subfigure}[b]{0.4\textwidth}
		\includegraphics[width=5cm]{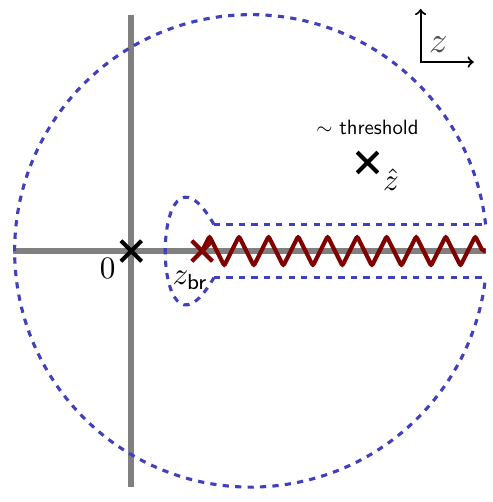}
		\caption{Near threshold}
		\label{fig:cntrthres}
   \end{subfigure}
\caption{Analytic structure of $\mathcal{A}(z)/z$. 
(\subref{fig:cntrpeak}): Near the peak $\pc[2] \sim |\hmc[2]|$, and transitioning to the far tail $\pc[2] \gg |\hmc[2]|$. 
In both cases $\pc[2] \gg(m_f+ m_\pi)^2$ such that the branch point $\zbr$~\eqref{eqn:zbr} is relatively unchanged. 
(\subref{fig:cntrthres}): Near the pair production threshold $\pc[2] \to (m_f + m_\pi^2)$. In this regime $\zbr \to 0$ while $\hat{z} \to \infty$.
}
\label{fig:cntr}
\end{figure}

The branch cut discontinuity integral must be proportional to the imaginary part of $M^2(\pc[2])$.
In light of the (assumed) large-$z$ behavior of the hadronic matrix elements scaling as $1/z^n$ for some positive $n >1$,
one then expects this term to roughly scale as $\sim \Gc/\mc  \times 1/\zbr^{n-1}$.
To understand the conditions under which the discontinuity integral can be neglected compared to the $\hat{z}$ residue,
one must consider several characteristic regimes for the (real) resonance invariant mass $\pc[2]$.
First, for $\pc[2]$ nearby the resonance peak such that $\pc[2] \sim |\hmc[2]|$, then $|\hat{z}| \sim \Gc/\mc \ll 1$ 
while from Eq.~\eqref{eqn:zbr} $\zbr \sim \mathcal{O}(1)$ (or larger),
as shown in Fig.~\ref{fig:cntrpeak}. 
In this regime, the large-$z$ behavior implies that the keyhole contour integral is small
and thus the branch cut can be neglected.

Second, for $\pc[2]$ in the far tail well above the $\hmc[2]$ pole, as also shown in Fig.~\ref{fig:cntrpeak},
then $\pc[2] \gg  |\hmc[2]| > (m_f + m_{\pi})^2$ such that $|\hat{z}|$ grows to be $\mathcal{O}(1)$ (or larger) while bounded above by $\zbr$.
The location of $\zbr$ stays relatively unchanged versus nearby the peak,
because the numerator and denominator of Eqs~\eqref{eqn:zbr} (and~\eqref{eqn:zhat}) scale similarly for $\pc[2] \gg (m_f + m_{\pi})^2$.
As a result the keyhole contour is effectively the same as that near the peak. 
Thus assuming the far tail contribution from the $\hat{z}$ residue drops relatively slowly 
(which we shall see is the case in Sec.~\ref{sec:BD1*}) the branch cut term can similarly be neglected.

Finally, by contrast, as $\pc[2]$ approaches threshold the branch point scales as $\zbr \to 0$
while $|\hat{z}|$ diverges, as shown in Fig.~\ref{fig:cntrthres}.
In this regime one does not expect the branch cut term to be neglectable.
As a (very) conservative heuristic for this behavior, 
we assume the branch cut term is non-negligible for
\begin{equation}
	\label{eqn:heurthres}
	\zbr \lesssim \mathcal{O}(1) \quad \text{and} \quad \zbr \lesssim \text{few}\times |\hat{z}|\,,
\end{equation}
while outside of this regime one has the recursion relation
\begin{equation}
	\label{eqn:recrel}
	\mathcal{A}(0) \simeq -\frac{1}{\pc[2] - \hmc[2]} A^{\mu}_{bc\ell\nu}(\hat{z}) \Big[\sum_{\lambda = \pm,0} \epsilon^\lambda_\mu(\hpc) \epsilon^{\bar{\lambda}}_\nu(\hpc)  \Big] A^\nu_{cc}(\hat{z})\,. 
\end{equation}
(The $1/[\pc[2] - \hmc[2]] = 1/[2\hat{z} \zeta \cdot \pc]$ prefactor arises via the jacobian of the contour integral.)
This recursion relation relates the full unshifted decay amplitude to the product of on-shell amplitudes at the simple pole,
times a Breit-Wigner-like factor $1/[\pc[2] - \hmc[2]] =  1/[ \pc[2] - \mc[2] + i \mc \Gc]$.

\subsection{Other resonances}
\label{sec:othres}
The preceding discussion has considered only a single charmed resonance in $\Hb \to (\Hcp \to \Hcf \pi) l \nu$; see footnote~\ref{ft:hbp}.
The description of the lineshape for the full $\Hb \to \Hcf \pi l \nu$ process, however, 
may also include terms arising from the beauty resonance, $\Hbp$,  related to the $\Hcp$ resonance by heavy quark flavor symmetry, 
via $\Hb \to \pi (\Hbp \to \Hcf l \nu)$.
If $\Hcf$ and $\Hb$ are in same HQ multiplet, then the same set of form factors and Isgur-Wise functions arise in $\Hb \to \Hcp l \nu$ and $\Hbp \to \Hcf l \nu$.
It may also feature multiple charmed resonances, as is the case for $\Bbar \to (D^{**} \to D \pi)l\nu$,
or other charmed or beauty (excited) resonances entirely.
The latter may be incorporated by including a pole for each resonance of interest.

Our chosen `$\Hb\pi$' complex shift~\eqref{eqn:zshift} on the pion and $\Hb$ external legs does not shift the $\Hbp$ resonance in $\Hb \to \pi (\Hbp \to \Hcf l \nu)$, 
such that the contribution from such a resonance under the shift in Eq.~\eqref{eqn:zshift} would amount to a finite contribution on the $|z| \to \infty$ boundary,
sometimes referred to as a pole at infinity.
Similarly, if one had imposed a `$\Hcf\pi$' shift, the $\Hcp$ resonance would instead have generated a finite contribution on the corresponding $|z| \to \infty$ boundary.
Thus to include both $\Hcp$ and $\Hbp$ contributions to the full $\Hb \to \Hcf \pi l \nu$ amplitude, one may apply the recursion relation~\eqref{eqn:fullrecrel} twice---once
under the $\Hb\pi$ shift to the $\Hb \to (\Hcp \to \Hcf \pi) l \nu$ term and once under the $\Hcf\pi$ shift to the $\Hb \to \pi (\Hbp \to \Hcf l \nu)$ term---and 
sum the results. 
No poles at infinity arise in this construction from resonances, though one must still assume each recursion relation is constructible.
Under the $\Hcf\pi$ shift, the $\Hbp$ resonance is always in the regime that $\zbr \le  |\hat{z}|$, 
so that the $\Hbp$ branch cut term cannot be neglected.

Because in practice $\Hbp$ is heavier than $\Hb$, it must be purely virtual even in the narrow-width limit,
in which case the contribution from $\Hbp$ exchange is expected to be negligible. 
For finite widths, because $m_b \gg m_c$, 
one similarly expects the $\Hbp$ resonance contribution to be negligible nearby or below the $\Hcp$ resonance peak.
However, in the deep virtual regime $\pc[2] \gg \mc[2]$, well above the $\Hcp$ lineshape peak, 
the $\Hcp$ and $\Hbp$ may achieve comparable virtualities,
such that effects from the virtual $\Hbp$ must contemplated.~\footnote{We thank M.~Papucci and R.~Plestid for pointing this out.}
To estimate this: Up to threshold effects and chiral corrections (see Sec~\ref{sec:hadmassexp} below), 
in the HQ limit one expects the $\Hb \to \Hbp \pi$ vertex should have the same coupling strength as that in $\Hcp \to \Hcf \pi$
(because the widths are expected to be approximately the same; a much narrower state will have a correspondingly weaker effective coupling squared).
One may then approximately characterize the relative virtuality via the ratio of propagators, 
and thus approximately characterize the regime in which the $\Hbp$ resonance terms become non-negligible, via
\begin{equation}
	\label{eqn:hbpreg}
	\pc[2] - \mc[2] \sim \frac{1}{\mathcal{O}(1)}|m_{\Hb}^2- 2 m_{\Hb} E_\pi^* + m_\pi^2 - m_{\Hbp}^2|\,,
\end{equation}
in which we have included a conservative $\mathcal{O}(1)$ factor (say, $\sim3$) in relative size, $E_\pi^*$ is the energy of the pion in the $\Hb$ rest frame, 
and we have neglected finite width terms in the propagator, 
which are negligible in the deep virtual regime.
This may be conservatively further rewritten as $\pc[2] \gtrsim \mc[2] + 2m_{\Hb}(m_{\Hbp} - m_{\Hb} + m_\pi)/\mathcal{O}(1)$,
where we have replaced $E_\pi^*$ by its lower bound $m_\pi$, though in practice $E_\pi^*$ is typically much larger.
Since also in practice $m_{\Hb}/\mc \sim 2$, then Eq.~\eqref{eqn:hbpreg} amounts to a regime above the lineshape peak of at least $\mathcal{O}(m_{\Hbp} - m_{\Hb} + m_\pi)$, 
which is typically several hundred MeV, such as in the case of the $B^{**}$ states.
In this work, we are mainly interested in developing the non-trivial technology required to compute finite-width effects in the $\Hb \to (\Hcp \to \Hcf \pi) l \nu$ term. 
When we later examine the (toy) numerical behavior of these effects versus data (see Sec.~\ref{sec:toynumex}),
given these estimates for the $\Hbp$ terms, we shall therefore restrict our analysis to regimes within several hundred MeV above the lineshape peak.
Deeper analysis of the $\Hbp$ resonance terms is reserved for future work.

\subsection{Spinor-helicity construction and generalized conjugation}
\label{sec:sphlgc}
Use of the recursion relation~\eqref{eqn:recrel} requires a generalization of the spinor-helicity construction for complex momenta.
For a massless complex momentum $\zeta$, one may apply the usual $SL(2,\mathbb{C})$ soldering $\zeta^\mu \sigma_\mu = | \zeta^{\pm} \rangle \langle \zeta^{\pm}|$.
All the usual spinor identities apply, with the key exception that it is no longer the case that $\langle \zeta^{\pm}|$ is the Hermitian conjugate of $| \zeta^{\pm} \rangle$,
and hence $\langle \zeta^+ | k^-\rangle^\dagger \not= \langle \zeta^- | k^+\rangle$ for some other massless momentum $k$.
Rather, $| \zeta^{\pm} \rangle^\dagger =  \langle \zeta^{*\pm}|$, i.e. with complex-conjugated momentum.

One may rewrite this as $| \zeta^{*\pm} \rangle^\dagger = \langle \zeta^{\pm}|$ and observe that the operation on the left-hand side 
is (in algebraic language) an involution that preserves holomorphy of the spinor representation with respect to the complex momentum.
That is, for any $f$ that is a holomorphic function of $z \in \mathbb{C}$, one can define a `$\sharp$' involution $f(z)^\sharp = [f(z^*)]^*$, 
such that holomorphy in $z$ is preserved: one thus may unambiguously also write $f(z)^\sharp = f^\sharp(z)$,
in the sense that $f^\sharp$ is a holomorphic function of $z$.
A self-adjoint function 
\begin{equation}
	\label{eqn:csselfadj}
	f^\sharp(z) = f(z)\,,
\end{equation}
is then a holomorphic function in $z$ with real coefficients
(in algebraic language, such an $f: \mathbb{C} \to \mathbb{C}$ would be a $*$-homomorphism with respect to complex conjugation),
which provides a natural generalization of the realness of Hermitian functions.

In the context of spinors this becomes 
\begin{equation}
	\label{eqn:csspinor}
	|\zeta^\pm\rangle^{\ddagger} = \langle \zeta^\pm|\,,
\end{equation}
where we have defined $^\ddagger =\,^{\sharp T}$ analogously to $^\dagger =\, ^{*T}$ for Hermitian conjugation.
As we shall see momentarily, this naturally extends to conjugation of polarizations (or any particle representation) with complex momenta too,
or any other higher-spin particle representation.
In general, as explained in some more detail immediately below,
an amplitude constructed from these complex-momentum spinor-helicity representations will have the same transformation properties,
which naturally (and perhaps unsurprisingly) connects to the well-known analytic property that e.g. for some analytically-continued complex invariant $s$, 
a scattering amplitude should obey $\mathcal{A}(s^*) = \mathcal{A}(s)^*$,
whence the Optical Theorem:
this is nothing but the self-adjoint relation~\eqref{eqn:csselfadj} with respect to $s$.
For lack of a name, we shall refer to this $\sharp$ involution as ``$C^\sharp$-conjugation''.\,\footnote{%
To our knowledge, the first mention of this holomorphy-preserving analytic property in the context of amplitudes is found in Ref.~\cite{Landau:1959fi},
and thus one might perhaps refer to this as ``Landau'' conjugation. 
This is, however, less notationally convenient for specifying the arguments under which holomorphy is preserved.}
As context requires, we specify the arguments (such as $\zeta$) with respect to which holomorphy is preserved, 
in superscript in parentheses: for example $C^{\sharp(\zeta)}$.
We shall see in Sec.~\ref{sec:holohqet} that the same sense of conjugation is vital also to the analytic properties of HQET constructed with complex velocities.

The complex shift~\eqref{eqn:zshift} modifies the light cone projection in Eq.~\eqref{eqn:lcproj}, such that 
\begin{equation}
	\pc_\mu(z) = \tpc_\mu + \frac{\pc[2]}{2 k \cdot \pc} k_\mu + z \zeta_\mu\,.
\end{equation}
In order to preserve the spinor-helicity construction of the polarizations, it is therefore natural to choose the reference momentum $k =\zeta$, 
with the immediate result that at the simple pole the lightcone projection becomes
\begin{equation}
	\label{eqn:zlcproj}
	\hpc_\mu = \tpc_\mu +  \frac{\hmc[2]}{2 \zeta \cdot \pc} \zeta_\mu\,,
\end{equation}
and the polarizations have representation at the simple pole
\begin{equation}
	\label{eqn:epsphat}
	\epsilon^\pm_\mu(\hpc) = \pm \frac{\big\langle \zeta^{\mp} \big| \sigma_\mu \big| \tpc[\mp] \big\rangle}{\sqrt{2} \big\langle \zeta^{\mp} \big| \tpc[\pm] \big\rangle}\,, 
	\qquad \epsilon^{0}_\mu(\hpc) = i\Big( \frac{\tpc_\mu}{\hmc} - \frac{\hmc}{2 \zeta \cdot \pc} \zeta_\mu \Big)\,.
\end{equation}
The importance of Eqs.~\eqref{eqn:zlcproj} and~\eqref{eqn:epsphat} is that one may choose
 the reference momentum for the \emph{unshifted} ($z=0$) momenta to be the complex momentum $\zeta$ satisfying Eqs.~\eqref{eqn:zetanull}.
That is $\pc_\mu = \tpc_\mu + \pc[2]/(2\zeta \cdot \pc) \zeta_\mu$, with $\tpc$ the same complex momentum as in Eq.~\eqref{eqn:zlcproj}.
Under this choice, the complex shift to the pole at $z=\hat{z}$ manifests purely as 
\begin{equation}
	\sqrt{\pc[2]} \to \hmc\,,
\end{equation}
in the lightcone projection of the momentum and in $\epsilon^0$, while $\epsilon^\pm$ remain unchanged.
Under $C^{\sharp(\zeta,\hmc)}$-conjugation, per Eq.~\eqref{eqn:csspinor} observe further that $(\epsilon^\lambda)^\sharp = \epsilon^{\bar\lambda}$.
Thus we expect the analytic properties and conjugation relations of the amplitudes to be defined by this conjugation,
that is, an amplitude and its conjugate should be related by $C^{\sharp(\zeta,\hmc)}$-conjugation.
This follows also from the general idea that spin-conjugate amplitudes involving the same resonance should both feature the same (i.e. unconjugated) simple pole.
More generally, because the full amplitude $\mathcal{A}(0)$ is determined by the simple pole per the recursion relation~\eqref{eqn:recrel}, 
and because the choice of reference momentum $\zeta$ should be unphysical,
one then expects the full amplitude should be holomorphic (whether explicitly or implicitly) with respect to $\hmc$.

This sense of conjugation extends also to the action of discrete symmetries. 
For instance, complexity of the momenta means that action of $C$-parity on a spinor must be generalized 
to $C|\zeta^\pm\rangle = -i \sigma^2 |\zeta^\pm\rangle^{\sharp} = |\zeta^{\mp}\rangle$.  
Similarly, we shall see later in Sec.~\ref{sec:holohqet} that $C^\sharp$-conjugation is intimately related 
to analytic properties of HQET matrix elements under $T$ (or $PT$) symmetry, when HQET is constructed with complex velocities.
By contrast, space-like parity transformations generalize differently when momenta become complex.
For instance, under the choice, $|\zeta^+\rangle = \Big(\sqrt{\zeta^0 + \zeta^3}, (\zeta^1 - i \zeta^2)/\sqrt{\zeta^0 + \zeta^3}\Big)^T$, 
the parity transformation $P|\zeta^\pm \rangle \equiv \sigma^0 |P\zeta^\pm \rangle =  \pm  \sqrt{(\zeta^1 \mp i \zeta^2)/(\zeta^1 \pm i \zeta^2)} |\zeta^\mp\rangle$.
The prefactor is a pure phase when $\zeta$ is real---in spherical polar coordinates the phase is simply the azimuthal twist around the $e_3$ axis---but 
for complex $\zeta$ it has non-unit modulus. 
For the polarizations~\eqref{eqn:epsphat}, up to the usual parity action on the free Lorentz index, 
this parity transformation maps $\epsilon^{\pm}$ to $\epsilon^{\mp}$ with an additional prefactor $-(\tpc_1 \pm i\tpc_2)/(\tpc_1 \mp i\tpc_2)$.
Thus we expect matrix elements involving currents of definite parity and opposite polarizations to similarly be related by this prefactor:
We will revisit this behavior later in Sec.~\ref{sec:BD1*} below.

\section{Holomorphic HQET}
\label{sec:holohqet}
\subsection{Heavy quark power expansion}
\label{sec:hqpe}
Analytic continuation of the $b\to c$ matrix element under the complex shift~\eqref{eqn:zshift} to the simple pole 
requires us to understand the analytic continuation of the form factor representation of hadronic matrix elements,
and in particular, how to construct a HQ expansion.
In the standard construction of HQET---what we will hereafter call ``real HQET''---a very convenient (though not required) prescription 
for the HQ velocity (defined below) is simply the hadron velocity $v = p/m$.
This choice ensures that no residual momenta enter into the hadron mass expansion.
This is essential to obtaining the standard form of the Schwinger-Dyson relations among the HQ matrix elements,
as well as to the implementation of short-distance mass schemes for the hadron mass parameters, 
that becomes important beyond leading order in the HQ expansion.

Under the complex shift~\eqref{eqn:zshift} to the simple pole, 
the HQ velocities in this prescription become complex, 
\begin{equation}
	\label{eqn:vbvcdef}
	\vb = \hat{p}/m\,,\quad \text{and} \quad \vc = \hpc/\hmc\,,
\end{equation}
(recalling $\hat{p} = p(\hat{z})$; similarly we place a `` $\hat{}$ '' on complex velocities and associated quantities evaluated at the simple pole).
The analytic continuation of the matching of QCD onto HQET is thus realized by
matching QCD matrix elements whose states have complex momenta onto HQET matrix elements whose states have complex HQ velocities.
That is, one must contemplate the construction of HQET generalized for a complex HQ velocity, $\vb$.

To this end, we begin with the QCD Lagrangian $\mLqcd = \Qbar(i\Dslash - m_Q) Q$, 
where $D^\mu$ is a gauge covariant derivative of QCD.
For a complex velocity $\vb$ satisfying $\vb[2] =1$,~\footnote{%
Not $\vb \cdot \vb[*] = 1$; just as the $\Hcp$ dispersion relation involves $\hpc[2]$ not $\hpc \cdot \hpc[*]$.}
it remains the case that
the projectors $\Piv{\pm} \equiv (1 \pm \vslash)/2$ are idempotent and orthogonal,
and furthermore $\Piv{+} + \Piv{-} \equiv 1$ (we follow the notation of Ref.~\cite{Bernlochner:2022ywh} throughout).
It is not the case, however, that the projectors of the conjugate velocity $\Piv[{\vb[*]}]{\pm}$ are idempotent or orthogonal with respect to $\Piv{\pm}$.
As a result, if one were to naively generalize the real HQET construction by taking the velocity to be complex
and constructing an effective field theory (EFT) in terms of the HQ fields $Q^{\vb}_{\pm}(x) = e^{+im_Q \vb \cdot x} \Piv{\pm} Q(x)$
and their Dirac conjugates $\Qbar^{\vb}_{\pm}(x) = e^{-im_Q \vb[*] \cdot x} \Qbar(x)\Piv[{\vb[*]}]{\pm}$,
then one would obtain an EFT with different algebraic properties to real HQET:
it would have a different structure for the resulting HQ power expansion and any associated Isgur-Wise functions.

Instead, we employ the field redefinition 
\begin{align}
	\label{eqn:QQbartrans}
	Q(x) & = e^{-im_Q \vb \cdot x} \big[\Qpv(x) +  \Qmv(x)\big]\,, \nn \\
	\Qbar(x) & = e^{+im_Q \vb \cdot x} \big[ \Qtpv(x) + \Qtmv(x)  \big]\,,
\end{align}
in which we emphasize $m_Q$ remains real and 
\begin{align}
	Q^{\vb}_{\pm}(x) & = e^{+im_Q \vb \cdot x} \Piv{\pm} Q(x)\,, \nn \\* 
	\Qtilde^{\vb}_{\pm}(x) & = e^{-im_Q \vb \cdot x} \Qbar(x)\Piv{\pm}\,. \label{eqn:Qvdef}
\end{align}
Because $\vb$ is complex, the HQ field $\Qtilde^{\vb}_{\pm}$ is \emph{not} the Dirac conjugate of $Q^{\vb}_{\pm}$, unlike in real HQET.
It is, however, a well-defined conjugate with respect to a different sense of conjugation---the $C^\sharp$-conjugation of Sec.~\ref{sec:sphlgc}---as 
discussed further below Eq.~\eqref{eqn:defconjv}.
As we shall proceed to show, based on identities that follow from the idempotency and orthogonality of $\Piv{\pm}$, 
it is the redefinition~\eqref{eqn:QQbartrans} that preserves the standard reorganization of the QCD Lagrangian 
into a complex generalization of HQET with the same algebraic properties as real HQET, and an HQ power expansion with the same structure as the latter.

Under the redefinition~\eqref{eqn:QQbartrans}, the QCD Lagrangian becomes
\begin{align}
	\mLqcd   	& = \big[ \Qtpv(x) + \Qtmv(x)  \big]e^{+im_Q \vb \cdot x} (i\Dslash - m_Q)e^{-im_Q \vb \cdot x} \big[\Qpv(x) +  \Qmv(x)\big]\nn\\*
			& \equiv \mLqcd(\vb), \label{eqn:QCDredef}
\end{align}
where in the last line we indicate the HQ velocity used to parametrize the Lagrangian by explicitly appending it as an argument, 
even though $\vb$ is unphysical in full QCD.
Except as required otherwise, we will write as usual $\Piv{\pm} = \Pi_\pm$ (and $\Piv[\vc]{\pm} = \Pi^\prime_\pm$).
Since in Eq.~\eqref{eqn:QQbartrans} we project both $Q$ and $\Qbar$ with respect to $\vb$ and not $\vb[*]$,
the transformation imposes a holomorphy on the resulting theory with respect to $\vb$,
such that it no longer appears to be explicitly Hermitian.
In particular, conjugating Eq.~\eqref{eqn:QCDredef} one finds instead
\begin{equation}
	\label{eqn:LqcdHC}
	\mLqcd(\vb)^*= \mLqcd(\vb[*]) \,.
\end{equation}
Because of the reparametrization invariance of $\mLqcd$, it must be that $\mLqcd(\vb[*]) = \mLqcd(\vb)$ 
and so the field redefinition~\eqref{eqn:QQbartrans} implicitly does preserve the hermiticity of the QCD Lagrangian.
We will see momentarily, however, that the HQET Lagrangian derived from this field redefinition is not Hermitian order by order
(but is instead self-adjoint under $C^{\sharp(\vb)}$-conjugation).

The standard reorganization of the QCD Lagrangian, based on identities that follow from the idempotency and orthogonality of $\Pi_{\pm}$, 
allows one to rewrite
\begin{equation}
	\mLqcd(v) = \Qtpv i \vcD \Qpv + \Qtpv i \Dslash_\perp \Qmv + \Qtmv i \Dslash_\perp \Qpv - \Qtmv (i \vcD + 2m_Q) \Qmv\,,
\end{equation}
just as in real HQET.~\footnote{We emphasize that this reorganization would not be possible if attempted in terms of $Q^{\vb}_{\pm}$ and the Dirac conjugate $\Qbar^{\vb}_{\pm}$.}
Here the transverse derivative $D^\mu_\perp = D^\mu - (\vcD)\vb[\mu]$;
as in Eq.~\eqref{eqn:QQbartrans}, it is defined with respect to $\vb$ and not $\vb[*]$. 
The equations of motion for $\Qmv$ and $\Qtmv$ remain in the same form as for the real velocity case
\begin{align}
	(i \vcD + 2m_Q) \Qmv & = i \Dslash_\perp \Qpv \,,\nn\\
	\Qtmv (-i  \vb \cdot \overleftarrow{D} + 2m_Q) & = -\Qtpv i \overleftarrow{\Dslash}_\perp\,, \label{eqn:eqmn}
\end{align}
although they are no longer related by Hermitian conjugation because $\vb$ is complex.

In the regime $m_Q \gg \lqcd \sim D$, we may proceed to integrate out $\Qmv$ yielding the effective theory
\begin{equation}
	\label{eqn:eftHQET}
	\mL_{\text{HQET}}(v) = \Qtpv i \vcD \Qpv +  \Qtpv i\Dslash_\perp \frac{1}{i \vcD + 2m_Q}  i\Dslash_\perp \Qpv\,,
\end{equation}
Because of its holomorphy in the HQ velocity, we refer to this effective theory as ``holomorphic HQET''.
This theory may be expanded order-by-order in $1/m_Q$.
Writing this power expansion as $\mL_{\text{HQET}}(\vb)  = \sum_{n = 0}\mL_n(\vb)/(2m_Q)^n$, then e.g. to first order
\begin{subequations}
\begin{align}
	\mL_0(\vb) & = \Qtpv i \vcD \Qpv \,,\\*
	\mL_1(\vb) & = -\Qtpv \Dslash_{\perp} \Dslash_{\perp} \Qpv  =  -\Qtpv \bigg[D^2 + \frac{g}{2} \sigma_{\ab} G^{\ab} \bigg]\Qpv \,,
\end{align}
\end{subequations}	
in which the field strength $ig\, G^{\ab} = [D^\alpha, D^\beta]$, $\sigma^{\ab} = \frac{i}{2}[\g^\alpha, \g^\beta]$, 
and one makes use of the equation of motion, $i \vcD\, \Qpv  = 0$ in the free effective theory.

As mentioned above, the complex HQ velocity results in $\mL_n$ no longer being Hermitian,
and in particular the free effective field theory defined by $\mL_0$ is no longer a Hermitian theory:
the imaginary part of the velocity incorporates some knowledge of the finite width of the hadrons in the QCD matrix element
and produces an evanescent-like wave factor.
However, just as in Eq.~\eqref{eqn:LqcdHC},
\begin{equation}
	\label{eqn:conjvvs}
	\mL_n(\vb)^* = \mL_n(\vb[*])\,,
\end{equation}
or equivalently, $\mL_n(\vb) = \mL_n(\vb[*])^*$. 
That is, with reference to the generalized conjugation introduced in Sec.~\ref{sec:sphlgc}, 
it is again natural to introduce the $C^\sharp$-conjugation operator, 
such that for any $f$ that is holomorphic function of $\vb$
one defines $f^\sharp(\vb) = [f(\vb[*])]^*$, that remains holomorphic in $\vb$.
In this language, Eq.~\eqref{eqn:conjvvs} immediately implies
\begin{equation}
	\label{eqn:defconjv}
	\mL_n^\sharp(\vb) = \mL_n(\vb)\,,
\end{equation}
under $C^{\sharp(\vb)}$-conjugation.
Note further with reference to the notation of Eq.~\eqref{eqn:csspinor}, $\Qtilde^{\vb}_{\pm} = [Q^{\vb}_{\pm}]^\ddagger \g^0$: 
an analog of Dirac conjugation in which complex conjugation is replaced  
by $C^{\sharp(\vb)}$-conjugation.
We shall refer to this as ``$C^{\sharp}$-Dirac'' conjugation, and generally denote by $\gentilde{X}$ the $C^{\sharp}$-Dirac conjugate of $X$; 
whenever $\gentilde{X} = X$ we say it is ``$C^\sharp$-Dirac-self-adjoint''.
Note the equations of motion~\eqref{eqn:eqmn} are related by $C^\sharp$-Dirac conjugation.
We shall explore in Sec.~\ref{sec:anprop} the features of a $C^\sharp$-self-adjoint theory, 
in particular conjugation relations that arise for the kernels of $PT$-symmetric operators within hadronic matrix elements.

Similarly to Eq.~\eqref{eqn:eftHQET}, the QCD source term $\Jbar Q + \Qbar J$ becomes $\Jtilde^{\vb} \mJ_{\text{HQET}}(\vb) \Qpv + \Qtpv \mJtilde_{\text{HQET}}(\vb) J^{\vb}$,
in which the current corrections
\begin{align}
	\mJ_{\text{HQET}}(\vb) & = 1 + \frac{1}{i \vcD + 2m_Q} i \Dslash_\perp\,,\nn \\*
	\mJtilde_{\text{HQET}}(\vb) & = 1 -  i \overleftarrow{\Dslash}_\perp \frac{1}{-i \vb \cdot \overleftarrow{D} + 2m_Q}\,.
\end{align}
Just as in real HQET, these may be expanded order-by-order in $1/m_Q$.
Writing this expansion as
\begin{equation}
	\mJ_{\text{HQET}}(\vb) = 1 + \Pi_-\sum_{n = 1}\mJ_n(\vb)/(2m_Q)^n\,,
\end{equation} 
then to first order $\mJ_1(\vb) = i \Dslash$, $\mJ_2(\vb) = -\Dslash\Dslash$ and so on.
The $C^\sharp$-Dirac conjugate $\mJtilde_n(\vb) \equiv \g^0 \overleftarrow{\mJ}^\ddagger_n(\vb)\g^0$.

\subsection{Matching and interaction basis}
\label{sec:transmats}
We are interested here in $\Hb \to \Hcp$ transitions generated by a weak current of general form $\cbar\,\Gamma\,b$,
in which $\Hb$ and $\Hcp$ are on-shell states with complex momenta $\hat{p}= m \vb$ and $\hpc = \hmc \vc$.
As usual, we choose the operator basis for the weak currents to be
\begin{equation}
	\label{eqn:Gdef}
	J_S = \cbar \, b\,, \quad  J_P = \cbar \, \g^5\, b\,, \quad
	J_V = \cbar \,\g^\mu\, b\,, \quad  J_A = \cbar \, \g^\mu\g^5\, b\,, \quad
	J_T = \cbar \, \sigma^{\mu\nu}\, b\,,
\end{equation}
where $\sigma^{\mu\nu} \equiv (i/2)[\g^\mu,\g^\nu]$.
The pseudotensor matrix element is determined by the identity $\sigma^{\mu\nu} \g^5 \equiv \pm(i/2)\epsilon^{\mu \nu \rho \sigma}
\sigma_{\rho \sigma}$, in which the sign is determined by a convention choice:
For $\Bbar \to \Dx$, the usual sign convention  is $\sigma^{\mu\nu} \g^5 \equiv -(i/2)\epsilon^{\mu \nu \rho \sigma}
\sigma_{\rho \sigma}$, which implies $\text{Tr}[\g^\mu\g^\nu\g^\rho\g^\sigma\g^5] = +4i \epsilon^{\mu\nu\rho\sigma}$.
The opposite sign convention is typically used for $\Bbar \to D^{**}$ or $\Lambda_b \to \Lambda_c^{(*)}$. 
 
The QCD transition matrix element with complex-momentum on-shell states matches onto the path integral computed in holomorphic HQET via
\begin{multline}
	\frac{\langle \Hcp(\hpc) | \cbar \,\Gamma\, b | \Hb(\hat{p}) \rangle}{\sqrt{\hmc\, m}}
	 = \big\langle \Hcp[\vc] \big|\frac{1}{\mathcal{Z}}\int \mathcal{D} \ctvp \mathcal{D} \cvp \mathcal{D} \btv \mathcal{D} \bv \label{eqn:QCDmatch} \\*
	\quad \times 
	\exp\bigg\{i \!\int\! d^4x \big[\mathcal{L}'_{\text{HQET}}(\vc) + \mathcal{L}_{\text{HQET}}(\vb) \big](x) \bigg\} 
	\ctvp \mJtilde_{\text{HQET}}^{\prime}(\vc)\, \Gamma\, \mJ_{\text{HQET}}(\vb) \bv\big| \Hb^{\vb} \big\rangle\,. 
\end{multline}
In the specific context of a $b \to c$ transition matrix element, 
we write the HQ fields as $\bv$ and $\cvp$ rather than $\Qpv$ and $Q'^{\vc}_+$.
We otherwise use primes to denote charmed HQ parameters.
The HQET partition function $\mathcal{Z}$ and action are no longer Hermitian, but instead are $C^{\sharp(\vb,\vc)}$-self-adjoint.
Here $|H^{\vb}\rangle$ are the eigenstates of $\mL_0(\vb)$, normalized as $\langle H^{\vc}\!(k') | H^{\vb}(k) \rangle = 2\vb[0] \delta_{\vb \vc}(2\pi)^3 \delta^3(k-k')$.
To first order in the HQ power expansion, Eq.~\eqref{eqn:QCDmatch}  becomes
\begin{align}
	\frac{\langle \Hcp(\hpc) | \cbar \,\Gamma\, b | \Hb(\hat{p}) \rangle}{\sqrt{\hmc\, m}}  
	& \simeq \big\langle \Hcp[\vc] \big| \ctvp \, \Gamma \, \bv \big| \Hb^{\vb} \big\rangle  \label{eqn:QCDmatchexp}\\*
	& + \frac{1}{2m_c} \big\langle \Hcp[\vc] \big|  \big(\ctvp\mJtilde^{\prime}_1(\vc)+ \mL'_1(\vc) \circ \ctvp \big) \Gamma \, \bv \big| \Hb^{\vb} \big\rangle \nn \\*
	& + \frac{1}{2m_b} \big\langle \Hcp[\vc] \big| \ctvp  \, \Gamma  \big(\mJ_1(\vb)\bv + \bv \circ \mL_1(\vb) \big)  \big| \Hb^{\vb} \big\rangle~, \nn
\end{align}
in which the $\circ$ denotes a time-ordered operator product, as defined in Ref.~\cite{Bernlochner:2022ywh}.

\subsection{Analytic properties from $PT$-symmetry} 
\label{sec:anprop}
In Hermitian theories---i.e. theories with a real Hamiltonian---the 
kernels (or density matrices) of $PT$-symmetric and antisymmetric operators are constrained by conjugation relations.\footnote{%
It is often said in the Literature that $T$-symmetry implies form factors of transition matrix elements, and by extension Isgur-Wise functions, are real. 
This holds in a $0+1$ dimensional theory; in general $PT$-symmetry is required, as discussed in Appendix~\ref{sec:PTconjrel}.}
We rederive in Appendix~\ref{sec:hermTconj} in detail how these relations arise, 
and how in turn they control the analytic properties of the form factors and Isgur-Wise functions of the matrix elements mediated by such operators.
Explicit examples are provided in App.~\ref{sec:PTex} and the specific application of these results to (real) HQET is considered in App.~\ref{sec:hqetPTex}.

The central ideas and results are as follows: in a Hermitian theory, 
for a $\Hb(v) \to \Hcp(v')$ transition mediated by an HQ operator $\mathcal{O}$, 
the representation of the operator kernel in any given HQET matrix element can be written in the form
\begin{equation}
	\mathcal{O}(v',v) = \sum_i \mathcal{T}_i(v',v) W_i(w)\,,
\end{equation}
in which the (real) recoil parameter $w = v \cdot v'$, $\mathcal{T}_i$ are a basis of tensors of the velocities
and/or Dirac matrices according to the spin or representations of $\Hb$, $\Hcp$, and $\mathcal{O}$, and $W_i$ are Isgur-Wise functions.
The $PT$ transformation rule~\eqref{eqn:realPTtransop} requires that $[\mathcal{T}_i(v',v)]_{PT} = \genbar{\mathcal{T}}_i(v',v)$.
Imposition of this rule ensures the absence of ambiguities arising from equations of motion, 
that can change the apparent transformation properties of the kernel.
The $PT$-symmetry relation~\eqref{eqn:realPTsymop}, realized as Eq.~\eqref{eqn:realhqetPTsymop} for HQET, 
further implies that $PT$-(anti)symmetry of $\mathcal{O}$ is equivalent to the kernel being self-adjoint $\genbar{\mathcal{O}}(v',v) = \pm\mathcal{O}(v',v)$.
Thus if, in addition, the $\mathcal{T}_i$ are chosen such that $\mathcal{T}_i(v',v) = \pm\genbar{\mathcal{T}}_i(v',v)$, 
it immediately follows that $W_i$ are real functions of $w$.

We are interested here, however, in a $C^\sharp$-self-adjoint theory---holomorphic HQET---whose 
states have complex momenta according to the shift in Eq.~\eqref{eqn:zshift}.
We show in Appendix~\ref{sec:cshpTconj} that the above results generalize 
to the $PT$ transformation rule $[\mathcal{T}_i(\vc,\vb)]_{PT} = \gentilde{\mathcal{T}}_i(\vc,\vb)$,
and the $PT$-symmetry relation~\eqref{eqn:PTsymop} is realized as Eq.~\eqref{eqn:hqetPTsymop} for HQET, 
such that the representation of the HQ operator kernel becomes $C^{\sharp(\vc,\vb)}$-Dirac-self-adjoint. That is,
\begin{equation}
 	\gentilde{\mathcal{O}}(\vc,\vb) = \pm\mathcal{O}(\vc,\vb)\,,
\end{equation}
for a $PT$-(anti)symmetric operator,
and provided the basis of tensors are chosen such that $\mathcal{T}_i(\vc,\vb) = \pm\gentilde{\mathcal{T}}_i(\vc,\vb)$,
then the Isgur-Wise functions become $C^{\sharp(\wh)}$-self-adjoint holomorphic functions, 
\begin{equation}
	W_i^\sharp(\wh) = W_i(\wh)\,,
\end{equation}
with now 
\begin{equation}
	\label{eqn:whdef}
	\wh= \vb \cdot \vc \in \mathbb{C}\,.
\end{equation}
Put in other words, the Isgur-Wise functions are \emph{holomorphic functions in $\wh$ with real coefficients}: 
they are simply analytic continuations of real analytic functions into the $\wh$ complex plane. 
This powerful result means that complexity of the Isgur-Wise functions 
at the simple pole is fully captured by the complexity of the recoil parameter,
and they feature no unknown imaginary parameters. 
We emphasize that one could not obtain this result by 
simply taking the velocity to be complex within the real HQET construction, and generating an EFT using the $\Piv[{\vb[*]}]{\pm}$ and $\Piv[{\vb}]{\pm}$ projectors.
As mentioned above in Sec.~\ref{sec:sphlgc}, 
this result matches the expected analytic behavior for a scattering amplitude. 
Moreover this result greatly simplifies the parametrization of the Isgur-Wise functions, as only real coefficients are required.

\subsection{Radiative corrections}
In real HQET, perturbative $\mathcal{O}(\aS)$ (and higher-order) corrections 
are computed via matching of QCD radiative corrections onto HQET local operators at a matching scale $\mu$. 
That is, the $\mathcal{O}(\aS)$ correction to the $\cbar\, \Gamma\, b$ current matches as
\begin{equation}
	\delta(\cbar\, \Gamma\, b) \to \cbvp \Big[\haS \sum_j C_{\Gamma_j} \Gamma_i\Big] \bv[v]\,,
\end{equation}
where $\haS = \aS/\pi$,\,\footnote{%
This hatted notation for $\aS/\pi$ is distinct from the hatted notation representing complex-shifted quantities evaluated at the simple pole.}
$\Gamma_1 = \Gamma$, and $\Gamma_{j \ge 2}$ are a basis of 
descendant operators generated by all combinations of replacements $\gamma^\mu \to v^{(\prime)\mu}$ in $\Gamma$.
For the standard case of real HQ velocities and real quark momenta, 
the functions $C_{\Gamma_i} = C_{\Gamma_i}(w,m_c/m_b)$ are real analytic functions of the real recoil parameter $w = v \ccdot v'$, up to branch cuts,
and they further depend on the mass ratio $m_c/m_b$. 
Their $\mu$-(in)dependence is sensitive to the renormalization scheme. 
At $\mathcal{O}(\aS)$ the following operators are generated (in the notation of Ref.~\cite{Manohar:2000dt})
\begin{align}
\label{eqn:ascurrent}
	\delta(\cbar\, b)				&\to  \cbvp  \big[\haS\, C_S\big] \bv[v]\,, \nn\\*
	\delta(\cbar \g^5 b)			&\to  \cbvp \big[ \haS\, C_P \big] \g^5  \bv[v]\,,\nn\\
	\delta(\cbar \g^\mu b)		&\to  \cbvp \big[\haS\, C_{V_1} \g^\mu + \haS\, C_{V_2}\, v^\mu + \haS\, C_{V_3}\, v'^\mu \big] \bv[v] \,, \nn\\
	\delta(\cbar \g^\mu\g^5 b)		&\to  \cbvp \big[ \haS\, C_{A_1} \g^\mu + \haS\, C_{A_2}\, v^\mu + \haS\, C_{A_3}\, v'^\mu \big] \g^5 \bv[v]\, \nn \\
	\delta(\cbar \sigma^{\mu\nu} b)	&\to \cbvp \big[\haS\, C_{T_1} \sigma^{\mu\nu} 
		+ \haS\, C_{T_2}\, i(v^\mu\g^\nu - v^\nu\g^\mu) + \haS\, C_{T_3}\, i(v'^\mu\g^\nu - v'^\nu\g^\mu)  \nn \\*
  	& \qquad\quad + \haS\, C_{T_4}(v'^\mu v^\nu - v'^\nu v^\mu)\big] \bv[v]\,.
\end{align}
The $C_{\Gamma_j}$ were computed in Ref.~\cite{Neubert:1992qq}; explicit, closed-form expressions are given in Ref.~\cite{Bernlochner:2017jka}.

Under the complex momentum shift~\eqref{eqn:zshift} and the construction of holomorphic HQET in Sec.~\ref{sec:hqpe}, however,
the on-shell quark momenta $p_{b(c)} = m_{b(c)}\vb[(\prime)]$ become complex, 
though the quark masses $m_{c,b}$ remain real.
That is, we must consider matching of QCD radiative corrections involving complex external quark momenta $p_{b(c)} = m_{b(c)} \vb[(\prime)] + k^{(\prime)}$
onto holomorphic HQET corrections involving complex HQ velocities and HQ external momenta $k^{(\prime)}$.
(Usually one takes $k^{(\prime)} \to 0$  in order to match on-shell QCD quark amplitudes onto HQET amplitudes with zero momenta, 
except where a finite $k^{(\prime)}$ is required to regulate infrared divergences, such as in $2$-point functions.)
The general arguments regarding the analytic properties of matrix elements in Sec.~\ref{sec:anprop} imply that under such a matching 
(with the associated replacements $v^{(\prime)} \to \vb[(\prime)]$ and $\cbvp \to \ctvp$)
one should expect the radiative functions $C_{\Gamma_j}(\wh)$ to satisfy $C^\sharp_{\Gamma_j}(\wh) =  C_{\Gamma_j}(\wh)$---i.e. 
they become holomorphic functions of $\wh \in \mathbb{C}$ with real coefficients---and as such,
must be the analytic continuation of the standard radiative functions in Eq.~\eqref{eqn:ascurrent} into the complex $\wh$ plane.

Explicit verification of this result at $\mathcal{O}(\aS)$ requires verifying that the standard identities and master loop integrals hold for complex quark momenta and complex HQ velocities. 
We show this in Appendix~\ref{sec:mloop}.

\subsection{Hadron mass and width expansion}
\label{sec:hadmassexp}
The construction of the hadron mass expansion in real HQET derives from 
the vacuum transition matrix element for a HQ hadron $H$ with momentum $p$ and heavy quark $Q$.
This matrix element has the form $\langle 0| \genbar{X}_H(x) Q(x)| H(p) \rangle$, 
with $\genbar{X}_H$ the annihilator of the light degrees of freedom. 
Momentum conservation and integration by parts require $\langle0|i D_\mu\,[ \genbar{X}_H(x) Q(x)]|H(p)\rangle = p_\mu \langle0|\genbar{X}_H(x) Q(x)|H(p)\rangle$.
Taking the derivative of the corresponding HQ matrix element involving $\genbar{X}_H(x) \Qpv(x)$ 
and applying Eq.~\eqref{eqn:Qvdef} plus the complex shift~\eqref{eqn:zshift}, then
\begin{equation}
	\label{eqn:vactrans}
	i\partial_\mu \langle 0| \genbar{X}_H(x) \Qpv(x)| H(\hat{p}) \rangle \ceq (\hat{p}_\mu - m_Q\vb_\mu)\langle 0| \genbar{X}_H(x) \Qpv(x)| H(\hat{p}) \rangle\,.
\end{equation}
Here we use `$\ceq$' to denote a relation that follows from overall momentum conservation under composition with an external operator or current.

As above, we choose the HQ velocity to be the complex hadron velocity $\vb = \hat{p}/\hat{m}_H$,
where $\hat{m}_H = \sqrt{m_H^2 + i m_H \Gamma_H}$ for a resonance as per Eq.~\eqref{eqn:simplepole}, and otherwise is just $m_H$.
(For clarity, here and in the following we add a contextual ``$H$'' subscript to label the hadron masses and other parameters.)
Thus we write the right-side prefactor of Eq.~\eqref{eqn:vactrans} as $(\hat{m}_H - m_Q)\vb_\mu$.
We further define an HQ expansion of the splitting
\begin{equation}
	\label{eqn:massexp}
	\hat{m}_H - m_Q = \sum_{n=0} \frac{\Delta \hat{m}_{n+1}^{\sellP}}{(2m_Q)^n}\,,
\end{equation}
where $\sellP$ denotes the quantum numbers of the light degrees of freedom in $H$ (including the spin-parity $s_\ell^{P_\ell}$) 
that uniquely define the HQ (super)multiplet to which $H$ belongs.

Contracting Eq.~\eqref{eqn:vactrans} with $\vb[\mu]$, applying the $\Qpv$ equation of motion, 
and matching at leading order, one finds
\begin{equation}
	\label{eqn:dhm1}
	\Delta \hat{m}^{\sellP}_1 = \frac{\langle 0| \genbar{X}_H(x) i \vb \ccdot \overleftarrow{D} \Qpv(x)| H(p) \rangle}{\langle 0| \genbar{X}_H(x) \Qpv(x)| H(p) \rangle} \equiv \LamB_{\sellP} - i \GamB_{\sellP}/2\,.
\end{equation}
As in real HQET, the real part, $\LamB_{\sellP}$, can be thought of as the energy of the light degrees of freedom at leading order in HQET.
The imaginary part, $\GamB_{\sellP}$, can be understood in an EFT context
as encoding the total strong decay width of the brown muck at leading order in HQET and in the chiral limit, 
while the heavy quark itself remains stable.
Observe that for an excited state HQ hadron $H$ decaying to $H_f$ in a lower HQ supermultiplet $\sellP[(H_f)]$ plus a light hadronic system $X$,
then the phase space in the chiral limit, in which $m_X$ can be neglected, 
is $|p_X| \simeq \LamB_{\sellP}  - \LamB_{\sellP[(H_f)]}$, at leading order in HQET.
As an $S$-wave decay is always available in the chiral limit for some angular momentum and hadron multiplicity configuration of $X$, 
then one further expects \emph{for an excited state}
\begin{equation}
	\label{eqn:glexp}
	\GamB_{\sellP} \sim \LamB_{\sellP}  - \LamB_{\sellP[(H_f)]}\,,
\end{equation}
and thus both $\LamB_{\sellP}$ and $\GamB_{\sellP}$ enter at the same (i.e. first) order in HQET, commensurate with general EFT expectations.
We will see in a fit below to data that Eq.~\eqref{eqn:glexp} is borne out in the $D_{1(2)}^{(*)}$ and $B_{1(2)}^{(*)}$ system.
By contrast, for a state in a ground-state HQ supermultiplet, one would expect $\GamB=0$ as any strong decay, even if kinematically available,
must vanish at all orders in HQET.~\footnote{%
An example is the $D^* \to D\pi$ strong decay, that occurs very close to threshold and is comparable in rate to the radiative $D\gamma$ mode. 
This decay can be thought of as arising solely from a first-order correction within chiral perturbation theory, 
while the $D^*$ partial width in HQET is zero to all orders in the HQ expansion in Eq.~\eqref{eqn:massexp}, and hence $\GamB = 0$.}
Thus Eq.~\eqref{eqn:massexp} can be thought of as a ``mass and width expansion'' of the hadron.
For the regime in which
$\Gamma_H \ll m_H$, then $\hat{m}_H \simeq m_H - i\Gamma_H/2$ 
and so $\Gamma_H \simeq \GamB_{\sellP}$ up to higher-order corrections in $1/m_{Q}$ (and $1/m_H$).

Just as for the hadron masses, describing the measurable splittings of the widths of various hadrons 
belonging to the same HQ supermultiplet with this expansion 
requires the inclusion of (at least) next-to-leading order corrections.
Matching of the QCD matrix element that involves the HQET Hamiltonian, i.e. the matrix element $\langle H \big| \Qtpv i \vcD \Qpv \big| H \rangle$, onto HQET 
allows derivation of such higher-order corrections (see e.g. Appendix A of Ref.~\cite{Bernlochner:2022ywh}).
In particular, matching at first order yields higher-order mass and width expansion terms
\begin{subequations}
\label{eqn:nlohadmatch}
\begin{align}
	\text{Re}\, \hat{m}_H & = m_Q + \LamB_{\sellP} - \frac{\lam{1}^{\sellP} + \bar{n}_H \lam{2}^{\sellP}}{2m_Q} + \ldots\,,\\
	\text{Im}\, \hat{m}_H & = -\frac{1}{2}\Big[ \GamB_{\sellP} - \frac{\kap{1}^{\sellP} + \bar{n}_H \kap{2}^{\sellP}}{2m_Q} + \ldots\Big]\,,
\end{align}
\end{subequations}
where $\bar{n}_H$ is the (signed) degrees of freedom of the partner hadron to $H$ in the HQ doublet.
The hadron mass parameters
\begin{subequations}
\label{eqn:masswidthexpnlo}
\begin{align}
	 \lam1^{\sellP} - i \kap1^{\sellP}/2 & = -\frac{\langle H^{\vb} | \Qtpv D^2 \Qpv | H^{\vb} \rangle}{\langle H^{\vb} | \Qtpv \Qpv | H^{\vb} \rangle} \,,\\
	 \bar{n}_H(\lam2^{\sellP} - i \kap2^{\sellP}/2) & = -\frac{\langle H^{\vb} |  \Qtpv (g \sigma_{\ab} G^{\ab}/2) \Qpv | H^{\vb} \rangle}{\langle H^{\vb} | \Qtpv \Qpv | H^{\vb} \rangle} \,.
\end{align}
\end{subequations}
Here $\kappa^{\sellP}_{1,2}$ are the imaginary parts of the kinetic energy and chromomagnetic corrections that arise because $\mL_1(\vb)$ is not Hermitian.
The two matrix elements in Eq.~\eqref{eqn:masswidthexpnlo}, however, both involve $PT$-symmetric operators.
The $C^{\sharp(\vb)}$-self-adjoint property of $\mL_1(\vb)$ then
ensures that the operator kernels must be represented by $C^{\sharp}$-self-adjoint functions or parameters
(see App.~\ref{sec:PTconjrel}).
That is, these HQET matrix elements must be holomorphic with respect to $ \lam{1,2}^H - i \kap{1,2}^H/2$, 
and thus so is any HQET matrix element related to these parameters by Schwinger-Dyson relations (see Sec.~\ref{sec:schdy} below).
 
Let us examine the compatibility of the second-order expansion~\eqref{eqn:nlohadmatch} with data. 
In particular, we consider the measured masses and widths of the charmed and beauty $s_\ell^P = \frac{3}{2}^+$ orbitally-excited HQ supermultiplet,
furnished by the $D_1(2420)$, $D_2^*(2460)$ and the $B_1(5721)$, $B_2^*(5747)$ states, respectively.
We consider the neutral modes only, which have more precisely measured widths in the beauty system,
and use the charged-pion modes to compute the kinematics 
(switching to an isospin-average introduces only small corrections in the following).
The PDG-averaged hadron widths are shown in Table~\ref{tab:masswidth32}~\cite{ParticleDataGroup:2024cfk}~\footnote{%
\label{ft:gav}%
In using the PDG-averaged width data, caution is warranted as various measurements use different lineshape models and width definitions, 
leading to an unaccounted systematic uncertainties. We take the PDG averages at face value, while keeping in mind the uncertainties are likely underestimated.}
as well as the dominant decay modes.
These states are sufficiently narrow that we may apply the approximation 
$\text{Re}[\hat{m}_H] \simeq m_H$ and $\text{Im}[\hat{m}_H] \simeq \Gamma_H/2$ on the left side of Eqs.~\eqref{eqn:nlohadmatch}.

\begin{table}[tb]
\renewcommand*{\arraystretch}{1.5}
\newcolumntype{X}{ >{\centering\arraybackslash $} c <{$}}
\newcolumntype{C}{ >{\centering\arraybackslash $} m{2.5cm} <{$}}
\newcolumntype{D}{ >{\centering\arraybackslash $} m{3cm} <{$}}
\newcolumntype{H}{>{\setbox0=\hbox\bgroup$}c<{$\egroup}@{}}
\begin{tabular}{XCCDXCH}
\hline
 \text{Hadron} & \text{Mass [MeV]} & \text{Width [MeV]} & \text{Decay modes} & \mathcal{B}_0\, \text{\footnotesize{(HQ limit)}} & |\vec{p}^*_\pi|\,\text{[MeV]} & \mathcal{P}_\pi(H_J) \\
 \hline\hline
 D_1^0 & 2422.1(0.6) & 31.3(1.9) & D^{*} \pi & 1 & 355 &0.040\\
 D_2^{*0} & 2461.1(0.8) & 47.3(0.8) & D^{*} \pi,\, D \pi & \frac{3}{5},\, \frac{2}{5} & 390,\, 507& 0.133 \\
 B_1^0 & 5726.1(1.3) & 27.5(3.4) & B^{*} \pi & 1 & 364 &0.046 \\
 B_2^{*0} & 5739.5(0.7) & 24.2(1.7) & B^{*} \pi,\, B \pi & \frac{3}{5},\, \frac{2}{5} & 377,\,420&0.070\\
\hline
\end{tabular}
\caption{Measured masses and widths for the neutral orbitally-excited charmed 
and beauty $s_\ell^{P_\ell} = \frac{3}{2}^+$ HQ supermultiplet mesons~\cite{ParticleDataGroup:2024cfk}. 
(The reported charmed measurements combine results from neutral and charged modes.)
Also shown are the permitted two-body decay modes and their respective HQ-limit branching fractions, denoted by $\mathcal{B}_0$, 
as well as the corresponding (charged) pion momentum in the parent rest frame.}
\label{tab:masswidth32}
 \end{table}

To compare the measured widths with the expansion in Eq.~\eqref{eqn:nlohadmatch}, 
one must take into account that the physical decays of the $\frac{3}{2}^+$ states (predominantly) 
proceed through an $L=2$ partial wave to the $\frac{1}{2}^-$ ground-state supermultiplets---the $D^{(*)}$ and $B^{(*)}$---plus a pion whose mass is non-negligible.
Their decay widths therefore feature an overall $|\vec{p}^*_\pi|^{5}$ phase space suppression factor, 
where $|\vec{p}^*_\pi|$ is the pion momentum in the parent rest frame.
To properly characterize the hadron width expansion as a power expansion in the small parameter $\sim \lqcd/(2m_Q)$ 
one must include an overall $(|\vec{p}^*_\pi|/\Lambda_{\chi \text{SB}})^{5}$ factor.
We take the chiral symmetry breaking scale $\Lambda_{\chi \text{SB}} = 4\pi f_\pi/\sqrt{3} \simeq 0.67$\,GeV.
As can been seen in Table~\ref{tab:masswidth32}, 
the large size of the $D^*$--$D$ mass splitting leads to an enhancement
of $|\vec{p}^*_\pi|$ in the $D_2^* \to D \pi$ mode.
This introduces an additional significant source of HQ symmetry breaking between the various hadrons, that must be included.
With reference to the allowed decay modes and respective HQ limit branching fractions in Table~\ref{tab:masswidth32}, 
we therefore define a branching-fraction-weighted phase space suppression factor
\begin{equation}
	\label{eqn:PSsupp}
	\mathcal{P}_\pi(H_J) = \sum_{H_f = H, H^*} \mathcal{B}_0[H_J  \to H_f \pi]\bigg[\frac{|\vec{p}^*_\pi|_{H_J  \to H_f \pi}}{\Lambda_{\chi \text{SB}}}\bigg]^5\,,
\end{equation}
for $H = B$ or $D$ and $J=1$ or $2$, and $\mathcal{B}_0$ denotes the branching fraction in the HQ limit.
The resulting values for the phase space suppression factors are 
$\mathcal{P}_\pi(D_1) = 0.040$,  $\mathcal{P}_\pi(D_2^*) = 0.133$, $\mathcal{P}_\pi(B_1) = 0.046$, and $\mathcal{P}_\pi(B_2^*) = 0.070$.
Just as for $|\vec{p}^*_\pi|$, the value of $\mathcal{P}_\pi(D_2^*)$ is notably larger and comports with the correspondingly larger measured $D_2^*$ width.  

Incorporating all these effects, Eqs.~\eqref{eqn:nlohadmatch} become explicitly
\begin{subequations}
\label{eqn:nloBCmatch}
\begin{align}
	m_{D_1} & \simeq m_c + \LamB' - \frac{\lam{1}' + 5 \lam{2}'}{2m_c}\,, 	& \Gamma_{D_1} & \simeq \mathcal{P}_\pi(D_1) \bigg[\GamB' - \frac{\kap{1}' +  5 \kap{2}'}{2m_c}\bigg]  \,,\\
	m_{D_2^*} & \simeq m_c + \LamB' - \frac{\lam{1}' - 3 \lam{2}'}{2m_c}\,, & \Gamma_{D_2^*} & \simeq \mathcal{P}_\pi(D_2^*) \bigg[ \GamB' - \frac{\kap{1}' - 3 \kap{2}'}{2m_c}\bigg]  \,,\\
	m_{B_1} & \simeq m_b + \LamB' - \frac{\lam{1}' + 5 \lam{2}'}{2m_b}\,, 	& \Gamma_{B_1} & \simeq \mathcal{P}_\pi(B_1) \bigg[\GamB' - \frac{\kap{1}' +  5 \kap{2}'}{2m_b}\bigg]  \,,\\*
	m_{B_2^*} & \simeq m_b + \LamB' - \frac{\lam{1}' - 3 \lam{2}'}{2m_b}\,, & \Gamma_{B_2^*} & \simeq \mathcal{P}_\pi(B_2^*) \bigg[ \GamB' - \frac{\kap{1}' - 3 \kap{2}'}{2m_b}\bigg]  \,.
\end{align}
\end{subequations}
Here we have applied the conventional primed notation for the mass parameters of hadrons belonging to the $\frac{3}{2}^+$ supermultiplet. 
For numerical evaluation, we apply the $1S$ mass scheme~\cite{Hoang:1998ng, Hoang:1998hm, Hoang:1999ye}, 
in which $\mbS$ is defined as half of the perturbatively computed $\Upsilon(1S)$ mass,
such that the pole mass 
\begin{equation}
	\label{eqn:mb1S}
	m_b(\mbS) \simeq \mbS(1 +2\aS^2/9 )\,,
\end{equation}
and we choose HQET to QCD matching scale $\mu_{bc} = \sqrt{m_b m_c} \simeq 2.5$\,GeV, such that $\aS \simeq 0.27$.
The charm quark mass is then fixed via the mass splitting $\dmbc \equiv m_b - m_c$, 
which is extracted along with $\mbS$ from precision fits to inclusive $b \to c$ spectra~\cite{Bauer:2002sh,Bauer:2004ve,Ligeti:2014kia}.
(The $1S$ scheme admits a systematic approach to the cancellation of renormalon ambiguities, but we do not discuss this in the present work.)
Using the numerical inputs
\begin{equation}
	\mbS = (4.71\pm 0.05)\,\GeV\,,\qquad \dmbc = (3.40\pm0.02)\,\GeV\,,
\end{equation}
one obtains from a fit to the width data in Table~\ref{tab:masswidth32} (the fit has $4-3 = 1$ degrees of freedom),
\begin{equation}
	\label{eqn:gfit}
	\GamB' = 0.36 \pm 0.03\,\text{GeV}\,,\quad \kap{1}' = -0.45 \pm 0.11\,\text{GeV}^2\,, \quad \kap{2}' = -0.15 \pm 0.02\,\text{GeV}^2\,.
\end{equation}
The HQ hadron mass expansion fit to the mass data in Table~\ref{tab:masswidth32} yields $\LamB' = 0.90\pm0.05$~GeV. 
Since for the ground states $\LamB = 0.57 \pm 0.05$~GeV in the $1S$ mass scheme (see e.g. Ref.~\cite{Bernlochner:2017jka}), 
then one observes $\LamB' - \LamB \simeq 0.33$~GeV in good agreement with the fit value for $\GamB'$ in Eq.~\eqref{eqn:gfit},
and thus the EFT expectations in Eq.~\eqref{eqn:glexp}.

\begin{figure}[t]
	\includegraphics[width = 8cm]{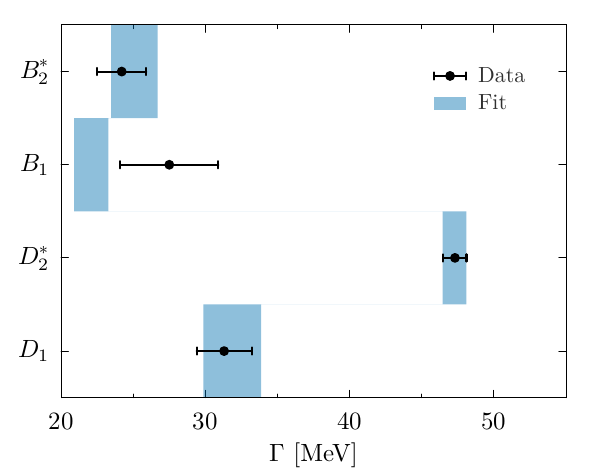}
	\caption{Fit $1\sigma$ CLs for the $3$-parameter fit of the second-order width expansion~\eqref{eqn:nloBCmatch} (blue shaded) 
	to the four measured widths for the $s_\ell^{\pi_\ell} = 3/2^+$ orbitally-excited $D_{1(2)}^{(*)}$ and $B_{1(2)}^{(*)}$ states (black points; see Table~\ref{tab:masswidth32}).
	Including fit correlations, the fit agrees with the data at the $1.7\sigma$ level ($p =10\%$).}
	\label{fig:BD32widthfit} 
\end{figure}

The recovered meson widths from the fit are
\begin{align}
	\Gamma_{D_1} & = 31.9(2.0)\,\text{MeV}\,, & \Gamma_{D_2^*} & = 47.3(0.8)\,\text{MeV}\,, \nn \\
	\Gamma_{B_1} & = 22.1(1.2)\,\text{MeV}\,, & \Gamma_{B_2^*} & = 25.1(1.6)\,\text{MeV}\,,
\end{align}
which are shown in Fig.~\ref{fig:BD32widthfit} compared to the measured widths in Table~\ref{tab:masswidth32}:
including correlations they agree with data at the $1.7\sigma$ level ($p = 10\%$).
Thus, once HQ symmetry breaking from phase space is included, 
the second-order expansion~\eqref{eqn:nlohadmatch} exhibits acceptable compatibility with data (keeping in mind also footnote~\ref{ft:gav}).
Note, however, that when working only at first order in HQET 
and in the regime $\Gamma_H \ll m_H$,
to incorporate this additional breaking it should be sufficient to simply take $\GamB_{\sellP} \simeq \Gamma_H$ for each hadron $H$ in Eq.~\eqref{eqn:dhm1}.

\subsection{Schwinger-Dyson relations}
\label{sec:schdy}
Similar to real HQET, the hadron mass and width parameters play a key role in the Schwinger-Dyson relations (also referred to as modified Ward identities) 
between various $\Hb \to \Hcp$ HQET matrix elements entering at different orders in the HQ power expansion.
These relations arise because, analogously to Eq.~\eqref{eqn:vactrans}, 
the derivative of the QCD matrix element containing the HQET current operator $J_{\Gamma+}(z) = \ctvp(z) \Gamma \bv(z)$ 
can be related to the matrix element itself via overall momentum conservation with an external operator or current 
(see App.~B of Ref.~\cite{Bernlochner:2022ywh}) such that
\begin{align}
	i\partial^z_\mu \langle \Hcp |  J_{\Gamma+}(z) | \Hb \rangle 
		& \ceq \big[(\hat{p} - m_b\vb)_\mu - (\hpc - m_c \vc)_\mu\big]  \langle \Hcp |  J_{\Gamma+}(z) | \Hb \rangle\,,\label{eqn:genSDrel}\\
		& = \Big[\big(\LamB - i \GamB + \ldots\big)\vb_\mu - \big(\LamB' - i \GamB' + \ldots\big) \vc_\mu  \Big] \langle\Hcp |  J_{\Gamma+}(z) | \Hb \rangle\,. \nn
\end{align}
In the second line we have kept only the leading-order terms of the mass and width expansions in Eq.~\eqref{eqn:massexp} for both $\Hb$ and $\Hcp$,
denoted by unprimed and primed parameters, respectively.
Note $\Hb$ and $\Hcp$ typically belong to different HQ supermultiplets when considering decays to excited states,
and if $\Hb$ or $\Hcp$ is a ground state then $\GamB = 0$ or $\GamB' = 0$, respectively.

Matching the left and right side of Eq.~\eqref{eqn:genSDrel} onto HQET via (the analogue of) 
Eq.~\eqref{eqn:QCDmatchexp} at first order in the HQ expansion
leads to the leading-order Schwinger-Dyson relation between HQET matrix elements
\begin{multline}
	\label{eqn:LOSDrel}
	\big\langle \Hcp[\vc] \big|  \ctvp(z) i\overleftarrow{D}^z_\mu \, \Gamma \, \bv(z)  \big| \Hb^{\vb} \big\rangle 
	+ \big\langle  \Hcp[\vc] \big|  \ctvp(z) \, \Gamma i\overrightarrow{D}^z_\mu  \bv(z) \big| \Hb^{\vb} \big\rangle \\
	\ceq \Big[\big(\LamB - i \GamB\big) \vb_\mu - \big(\LamB' - i \GamB'\big)\vc_\mu  \Big]  \big\langle \Hc^{v'} \big| J_{\Gamma+}(z) \big| \Hb^v \big\rangle\,.
\end{multline}
Higher-order Schwinger-Dyson relations can be similarly derived~\cite{Falk:1992wt, Bernlochner:2022ywh}.

Not only do these relations constrain higher-order HQET matrix elements in terms of lower order ones,
but they also impose constraints on the analytic properties of the hadron mass and width parameters.
In particular, as the $i D_\mu$ operator is $PT$-symmetric, the representation of its operator kernel must be $C^{\sharp}$-self-adjoint
(see App.~\ref{sec:PTconjrel}).
The Schwinger-Dyson relation, however, imposes that this representation is nothing but $\LamB^{(\prime)} - i \GamB^{(\prime)}$, 
and thus it imposes a correspondence between $C^{\sharp(\vb,\vc)}$-conjugation and $C^{\sharp(\hat{m}, \hmc)}$-conjugation.
This is to be expected: the complex structure of $\vb$ and $\vc$ derives from the choice of the complex shift momentum $\zeta$ in Eq.~\eqref{eqn:zshift}
and the simple pole at $\hmc$ (as $\hat{m} = m$ remains real),
but only $\hmc$ should be physical in the full resonance amplitude, 
which itself should be holomorphic with respect to $\hmc$
based on the general considerations in Sec.~\ref{sec:sphlgc}.

\section{Example: $\Bbar \to \Dv(1^-) \to D\pi$}
\label{sec:BD1*}

To illustrate the construction of a resonant amplitude with the above on-shell recursion methods and holomorphic HQET 
(which we refer to hereafter as ``OSR+HQET''), 
we consider the example of $\Bbar \to \Dv \to D\pi$, 
in which the resonance is an orbitally-excited $D^*$ with spin-parity $J^P = 1^-$.\footnote{%
Although it is common in the literature to write the broad $D'_1(2430)$ as the $D^*_1$, 
the formal hadron naming convention~\cite{Workman:2022ynf} reserves a ``$^*$'' superscript for natural spin-parity states.
We follow this convention.}
The observed $\Dv(2600)$ is such a state, but we will consider hypothetical $\Dv$'s over a range of narrower widths 
in order to better explore the transition from the narrow to broad limits.

Although the main motivation for examining resonant processes are to describe the $\Bbar \to D^{**}$ semileptonic decays,
we consider in this work a $\Dv(1^-)$ as a first example
because expressions for the $0^- \to 1^-$ form factors and amplitudes are much more broadly known and familiar
from extensive studies of the $\Bbar \to D^*$ system in the literature.
Thus it may be easier to understand the deformations introduced into these expressions by the on-shell construction.
Derivation of the $\Bbar \to D^{**}$ amplitudes with the methods of this work are reserved for a future study. 
Also for the sake of simplicity, we consider only SM interactions in the following example.

\subsection{Recursion relation}
With reference to the master recursion relation~\eqref{eqn:recrel}, in the $\Bbar \to (\Dv \to D\pi)l\nu$ system 
the recursion relation becomes (assuming constructibility)
\begin{equation}
	\label{eqn:BDvconrecrel}
	\mathcal{A}_{s_l s_\nu}[\Bbar \to (\Dv \to D\pi)l\nu] = -\frac{1}{(\mDpi^2 - \hmDv[2])} \sum _{\lambda'=\pm,0} \mathcal{A}_{\lambda's_l s_\nu}^{\Bbar \to \Dv l \nu}(\hat{z}) \times  \mathcal{A}_{\lambda'}^{\Dv \to D\pi}(\hat{z})\,,
\end{equation}
in which $s_{l\,,\nu}$ are the spin quantum numbers of the external lepton and neutrino.
In order to preserve the notational simplicity of the preceding discussion and results, 
we write here the $\Dv$ mass and width with the primed notation, $\mDv$ and $\GDv$, respectively,
noting also $\hmDv[2] = \mDv[2] + i \mDv \GDv$ remains the location of the simple pole.
Similarly the $\Dv$ momentum, which is the momentum of the $D\pi$ system, is written as $\pDv$, 
although we write the invariant mass $\pDv[2] = (p_D + p_\pi)^2 = \mDpi^2$ to distinguish from the $\mDv$ pole mass.
In Eq.~\eqref{eqn:BDvconrecrel} we have contracted the on-shell complex $\Dv$ polarizations, $\epsilon^{\lambda'}_\mu(\hpc)$, 
into the semileptonic $\Bbar \to \Dv l \nu$ and $\Dv \to D\pi$ hadronic decay amplitudes on the left and right.
This produces a sum over Lorentz-invariant amplitudes that must be computed on the simple pole, 
under the momentum shift $\hat{z} \zeta$ per Eq.~\eqref{eqn:zshift}.

One may as usual decompose the $\Bbar \to \Dv l \nu$ amplitude into hadronic and leptonic factors,
by applying completeness relations to the propagator of the charged current mediator, 
which is assumed to be heavy and narrow.
In the SM, this results in a sum over spin $\lambda = \pm,0$ of an exchanged (far off-shell) $W$ plus a longitudinal term
\begin{equation}
	\label{eqn:amplSMbcln}
	\mathcal{A}_{\lambda' s_l s_\nu}^{\Bbar \to \Dv l \nu}(\hat{z}) \simeq \frac{2\sqrt{2}G_FV_{cb}}{\sqrt{w^2-1}}\sqrt{\frac{\mB}{\hmDv}} \sqrt{q^2-m_l^2}
	\sum_{\lambda = \pm,0,\ell} A_{\lambda' \lambda}^{\Bbar \to \Dv W}(\hat{z}) \times A_{\lambda s_l s_\nu}^{W \to l \nu}\,,
\end{equation}
in which we have pulled out a convenient overall prefactor.
The spin $\lambda = \ell$ corresponds to the longitudinal $W$ exchange,
corresponding to the $-q_\mu q_\nu/q^2$ term in decomposition of the numerator of the $W$ propagator under the polarization completeness relation
\begin{equation}
	-g_{\mu\nu} + q_\mu q_\nu/m_W^2 \simeq \sum_{\lambda = \pm,0} {\epsilon^*_\mu}_\lambda(q) {\epsilon_\nu}_\lambda(q) - q_\mu q_\nu/q^2\,,
\end{equation}
similar to Eq.~\eqref{eqn:decampl}.
Here we have kept only the leading electroweak order term in $\mathcal{O}(q^2/m_W^2)$.

\subsection{Kinematic configuration}

In the narrow-width limit, in which all on-shell momenta remain real and unshifted, 
it is convenient to compute the (Lorentz-invariant) $\Bbar \to \Dv W$, $W \to l\nu$, and $\Dv \to D\pi$ amplitudes in the $\Bbar$, dilepton, and $\Dv$ rest frames, 
respectively: when using standard helicity angles to represent the phase space in each rest frame, 
these amplitudes reduce to (linear combinations of) Wigner-$d$ functions (times form factors).
Because here, however, the complex shift momentum $\zeta$ is defined via its vanishing inner products with the $\Bbar$ and $\pi$ momenta, 
as in Eq.~\eqref{eqn:zetanull},
it is natural to compute the $\Bbar \to \Dv W$ and $\Dv \to D\pi$ amplitudes in the same frame:
We choose the $\Bbar$ rest frame just as in  Fig.~\ref{fig:BRefFrame}.

\begin{figure}[tb]
   \begin{subfigure}[b]{0.3\textwidth}
    	\centering
	\includegraphics[width = 3.5cm]{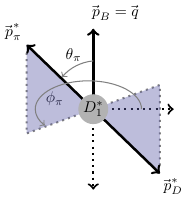}\hfil
	\caption{$\Dv$ rest frame.\label{fig:Dvframe}}
    \end{subfigure}\hfil
    \begin{subfigure}[b]{0.3\textwidth}
    	\centering
   	\includegraphics[width = 3.5cm]{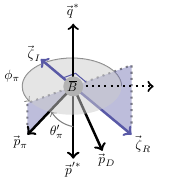}\hfil
	\caption{$\Bbar$ rest frame.\label{fig:Bbarframe}}
   \end{subfigure}\hfil
    \begin{subfigure}[b]{0.3\textwidth}
    	\centering
   	\includegraphics[width = 3.5cm]{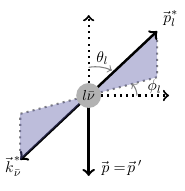}
	\caption{Dilepton rest frame.\label{fig:dilframe}}
    \end{subfigure}
    \caption{Kinematic configuration and phase space coordinates for the $\Bbar \to (\Dv \to D\pi)(W \to l \nu)$ cascade. 
    (\subref{fig:Dvframe}): The standard helicity angles parametrizing $\Dv \to D\pi$ phase space in the $\Dv$ rest frame.
    Superscript ``$*$''s on spatial momenta indicate daughter momenta in the rest frame of the parent.
    (\subref{fig:Bbarframe}): Configuration of the $\Bbar \to (\Dv \to D\pi)W$ system in the $\Bbar$ rest frame, including the complex shift momentum $\zeta = \zeta_R + i \zeta_I$. 
    Note the real part $\zeta_R$ is chosen to lie in the plane defined by $\vec{p}_\pi$ and $\vec{p}_D$, while the imaginary part $\zeta_I$ lies perpendicular to it.
   (\subref{fig:dilframe}): The standard helicity angles parametrizing $W \to l\nu$ phase space in the $l\nu$ dilepton rest frame.}
    \label{fig:refframes}
\end{figure}

Without loss of generality we align the recoil and $\Dv$ momentum, $q$ and $\pDv$, with the positive and negative $z$ axis, respectively.
The kinematic configuration of the $\Bbar \to (\Dv \to D\pi)W$ cascade in this frame is shown in Fig.~\ref{fig:refframes},
with explicit momenta
\begin{subequations}
\begin{align}
	p_B^\mu 		& = \big( \mB\,,~\vec{0} \big)\,, \\
	\pDv [\mu]		& = \big( \mDpi\cosh\yDv\,,~0\,,~0\,,~-\mDpi\sinh\yDv\big)\,, \\
	q^\mu 		& = \big( \big[\mDpi^2\sinh^2\yDv +q^2\big]^{1/2}\,,~0\,,~0\,,~\mDpi\sinh\yDv\big)\,, \\
	p_{\pi}^\mu	& = \big( \big[ |\vec{p}_{\pi}|^2+m_{\pi}^2\big]^{1/2}\,,~|\vec{p}_{\pi}|\sech\etpi\cos\phpi\,,~|\vec{p}_{\pi}|\sech\etpi\sin\phpi\,,~-|\vec{p}_{\pi}|\tanh\etpi \big)\,, \label{eqn:expppi}
\end{align}
\end{subequations}
and $p_D = \pDv - p_\pi$. 
Here $\phpi$ is the azimuthal helicity angle of the pion momentum in the $\Dv$ rest frame, oriented positively with respect to $\vec{q}$, as defined in Fig.~\ref{fig:Dvframe}.
Further, $\yDv$ is the rapidity of the $\Dv$---i.e. the $D\pi$ system---in the $\Bbar$ rest frame and $\etpi$ is the pseudorapidity of the $\pi$, 
defined with respect to the $\Dv$ momentum in the $\Bbar$ rest frame.
The pseudorapidity is related to the polar angle $\theta_{\pi}'$, defined in Fig.~\ref{fig:Bbarframe}, as usual via $\cos\theta_{\pi}' = \tanh\etpi$.
As we shall see below, $|\vec{p}_{\pi}|$ and $\eta_\pi$ are the natural kinematic coordinates for the $\Bbar \to (\Dv \to D\pi)W$ amplitudes, 
rather than the polar helicity angle $\theta_\pi$ and  rest frame momentum $|\vec{p}_\pi^*|$ defined in Fig.~\ref{fig:Dvframe}.

The momentum shift in the $\Bbar$ rest frame $\zeta^\mu = \zeta_R^\mu + i \zeta_I^\mu$ is chosen to be
\begin{subequations}
\begin{align}
    \zeta_R^\mu & = \big(0\,,~-|\vec{p}_{\pi}|\tanh\etpi \cos\phpi\,,~-|\vec{p}_{\pi}|\tanh\etpi\sin\phpi\,,~-|\vec{p}_{\pi}|\sech\etpi\big)\,,\\*
    \zeta_I^\mu & = \big(0\,,~|\vec{p}_{\pi}|\sin\phpi\,,~-|\vec{p}_{\pi}|\cos\phpi\,,~0\big)\,,
\end{align}
\end{subequations}
which is manifestly a solution of Eq.~\eqref{eqn:zetanull}, 
having the same structure as discussed in Sec.~\ref{sec:recrel}.
As discussed in Sec.~\ref{sec:sphlgc} we also choose $\zeta$ to be the null reference momentum that defines the $\Dv$ polarization basis,
as in Eq.~\eqref{eqn:epsphat}.
With this choice and in these kinematic coordinates, 
the $P$-parity transformation prefactor relating the complex-momentum $\epsilon^\pm$ polarizations of the $\Dv$, discussed at the end of Sec.~\ref{sec:sphlgc},
becomes
\begin{equation}
	\label{eqn:Pexppref}
	-(\tpc_1 \pm i\tpc_2)/(\tpc_1 \mp i\tpc_2) = e^{\pm2(i\phpi - \etpi)}\,,
\end{equation}	
and thus one expects this to arise as a relative prefactor between $\lambda' = \pm$ amplitudes.
Finally, the helicity angles of the $W\to l\nu$ system are defined as usual in Fig.~\ref{fig:dilframe}, 
noting the azimuthal helicity angle $\phi_l$ has the same positive orientation as $\phpi$. 
Thus one expects only the phase difference $\phpi - \phi_l$ to be physical.

Following the choices of Sec.~\ref{sec:holohqet}, the heavy quark velocities are chosen to be the hadron velocities
\begin{equation}
	v = p_B/\mB\,,\quad v'= \pDv/\mDpi\,,
\end{equation}
with the recoil parameter 
\begin{equation}
	w = v \cdot v' = \frac{\mB^2 + \mDpi^2 -q^2}{2\mB\mDpi} = \cosh\yDv\,, 
\end{equation}	
i.e. as usual the recoil is equivalent to the boost of the $\Dv$ in the $\Bbar$ frame.
Under the complex shift to the simple pole, per Eqs.~\eqref{eqn:vbvcdef} and~\eqref{eqn:whdef} we also use the complex recoil parameter
\begin{equation}
	\wh = \vb \cdot \vc = \frac{\mB^2 + \hmDv[2] - q^2}{2\mB\hmDv}
\end{equation}
because $q = p_B - \pDv = \hat{p}_B - \hpDv$.

\subsection{Differential rate and phase space}
Just as in the standard narrow-width limit calculation,
the $4$-body phase space is naturally decomposed into products of $2$-body phase space,
such that the fully-differential rate
\begin{equation}
	\label{eqn:diffrate}
	d\Gamma = \sum_{s_l s_\nu} \frac{1}{16\mB(2\pi)^{7}} \Big| \mathcal{A}_{s_l s_\nu}[\Bbar \to (\Dv \to D\pi)l\nu] \Big|^2 \frac{|\vec{p}^{\prime*}|}{m_B}\frac{|\vec{p}_\pi^*|}{\mDpi}\frac{|\vec{p}_{l}^*|}{\sqrt{q^2}} \,d\mDpi^2dq^2 d\Omega^*_\pi d\Omega^*_l\,,
\end{equation}
in which $|\vec{p}^{\prime*}|$, $|\vec{p}_\pi^*|$, and $|\vec{p}_{l}^*|$ are the spatial momenta 
of the $\Dv$, $\pi$, and lepton in the $\Bbar$, $\Dv$, and $l\nu$ rest frames, respectively (see Fig.~\ref{fig:refframes}).
Further, $d\Omega^*_{\pi,l}$ are the integration measures over the helicity angles defined in the $\Dv$ and $l\nu$ rest frames per Figs~\ref{fig:Dvframe} and~\ref{fig:dilframe}, 
so that $d\Omega_{\pi,l}^* = d\cos\theta_{\pi,l}\,d\phi_{\pi,l}$.
We emphasize that, as shown in Fig.~\ref{fig:refframes}, $|\vec{p}_\pi^*|$ is not the same as the $\Bbar$ frame pion momentum $|\vec{p}_\pi|$ in Eq.~\eqref{eqn:expppi}.
Note also the measure in Eq.~\eqref{eqn:diffrate} includes the $D\pi$ invariant mass $\mDpi$, that encodes the $\Dv$ lineshape dependence.

Although as mentioned above (and we shall see below)
the $\Bbar \to (\Dv\to D\pi)W$ amplitude is naturally expressed in terms of $|\vec{p}_\pi|$ and $\eta_\pi$ in the $\Bbar$ frame
rather than $|\vec{p}^*_\pi|$ and $\theta_\pi$, 
it remains natural to express the phase space measure itself in terms of the latter.
Thus to compute the differential rate, one must express $|\vec{p}_\pi|$ and $\eta_\pi$ as functions of $|\vec{p}^*_\pi|$ and $\theta_\pi$, 
and the other kinematic coordinates,
\begin{subequations}
\label{eqn:etpippi}
\begin{align}
	|\vec{p}_\pi|^2 & = \big[E_\pi^*\sinh\yDv -  |\vec{p}^*_\pi|\cosh\yDv\cos\theta_\pi\big]^2 + |\vec{p}^*_\pi|^2\sin^2\theta_\pi\,,\\
	\etpi & = \tanh^{-1}\bigg[\frac{E_\pi^*\sinh\yDv -  |\vec{p}^*_\pi|\cosh\yDv\cos\theta_\pi}{|\vec{p}_\pi|}\bigg]\,,
\end{align}
\end{subequations}
with $E_\pi^* = \sqrt{m_\pi^2 +  |\vec{p}^*_\pi|^2}$.

\subsection{Form factors and ratios}
Under the shift to the simple pole, the $\Bbar \to \Dv$ matrix elements involve on-shell complex-momentum states,
and as discussed  in Secs~\ref{sec:sphlgc} and~\ref{sec:schdy} they should be (implicitly) holomorphic with respect to the simple pole at $\hmDv$.
Thus the usual spin and parity selection rules and associated form factor representation for on-shell transitions still apply:
In the SM, the complex-shifted $\Bbar \to \Dv$ matrix elements are then represented by four independent form factors.
In order to match these matrix elements onto holomorphic HQET, it is convenient to choose the usual HQS basis for the form factors,
that are holomorphic functions of the complex recoil $\wh$.
We denote these form factors as $\hat{h}_X = \hat{h}_{X}(\wh)$, $X = V$, $A_{1,2,3}$, defined via
\begin{subequations}
\label{eqn:defBDvSMff}
\begin{align}	
	\frac{\big\langle {\Dv}_\lambda(\hpDv) |\, \cbar \g_\mu b\, | \Bbar(\hat{p}_B) \big\rangle}{\sqrt{m_{B} \hmDv}}
		 & = i \Hhv\, \varepsilon_\mu^{\nu\alpha\beta}\, 
			{\heps_\nu^{\,\sharp}{}}_\lambda\,\vc_\alpha\, \vb_\beta \,,\label{eqn:DsV}\\
	\frac{\big\langle {\Dv}_\lambda(\hpDv) |\, \cbar \g_\mu \g^5 b\, | \Bbar(\hat{p}_B) \big\rangle}{\sqrt{m_{B} \hmDv}}
		 & =	\Hha1\,(\wh+1){\heps_\mu^{\,\sharp}{}}_\lambda
  		- \Hha2\,({\heps_\mu^{\,\sharp}{}}_\lambda \cdot \vb)\vb_\mu 
 		- \Hha3\, ({\heps_\mu^{\,\sharp}{}}_\lambda \cdot \vb)\vc_\mu \,,\label{eqn:DsA}
\end{align}
\end{subequations}
where the polarizations $\heps_\lambda = \epsilon_\lambda(\hpDv)$, $\lambda = \pm$, $0$ as defined in Eq.~\eqref{eqn:epsphat}.

As in the narrow-width real on-shell limit, the on-shell amplitudes can be simply expressed in terms of three (complex) form-factor ratios
\begin{subequations}
\label{eqn:defBdvR}
\begin{align}
	\Rg1(\wh) & = \frac{\Hhv}{\Hha1}\,,\\
	\Rg2(\wh) &= \frac{\Hha3+ \hrDv\,\Hha2}{\Hha1}\,,\\
         \Rg0(\wh) & = \frac{(\wh+1)\Hha1 - (\wh -\hrDv)\Hha3 - (1 - \wh\hrDv)\Hha2}{(1+\hrDv)\, \Hha1}\,,
\end{align}
\end{subequations}
where $\hrDv = \hmDv/\mB$. However, the additional pieces of the amplitude
that enter proportional to the off-shell factor $\hmDv[2] - \mDpi^2$, that is, 
the pieces of the amplitude that vanish in the narrow-width limit,
are conveniently expressed in terms of the \emph{auxiliary} form-factor ratios
\begin{subequations}
\label{eqn:defBdvRaux}
\begin{align}
	\Rg{\pm\pm} 	& = \mp \Rg1\Big[1-\rDpi\,e^{\yDv}\Big]- \Rg2\,, & \Rg{\pm\mp} 	& = \pm \Rg1\Big[1-\rDpi\,e^{-\yDv}\Big] + \Rg2\,,\\
	\Rg{\pm 0} 	& =  \pm \sqrt{2}\Rg2 (w\rDpi - 1) -\frac{\sqrt{2}q^2\Rg1}{\mB^2}\,, & \Rg{0 \pm} 	& = \mp 2\hrDv\Big[(\wh+1) + \wh\Rg2\Big]\,,\\
	\Rg{0 0}		& = 2\sqrt{2}\, \hrDv\,(w\rDpi - 1)\Big[(\wh+1) + \wh\Rg2\Big]\,,
\end{align}
\end{subequations}
noting the deliberate distinctions between $\wh$ and $w$, and $\rDpi = \mDpi/\mB$ and $\hrDv= \hmDv/\mB$.

\subsection{Representation of $\Bbar \to \Dv$ HQET matrix elements}
\label{sec:hqetBDv}
The $\Bbar$ and the $\Bbar^*$ belong to an HQ-spin symmetry doublet, denoted as usual by $\Bxbar$,
formed by the tensor product of a heavy quark with light muck of definite spin-parity $s_\ell^{P_\ell} = \frac{1}{2}^-$.
Together with a partner pseudoscalar $D_0(0^-)$, 
the $\Dv(1^-)$ resonance similarly belongs to an $s_\ell^{P_\ell} = \frac{1}{2}^-$  HQ-spin symmetry doublet,
which we denote as $\Djx$.
Because both doublets remain on-shell under the complex shift~\eqref{eqn:zshift},
they retain the same spacetime on-shell representation within the trace formalism as in real HQET~\cite{Falk:1990yz, Bjorken:1990rr, Falk:1991nq},
but suitably generalized to incorporate a complex velocity and associated $C^{\sharp(\vb)}$-Dirac conjugation for holomorphic HQET, i.e.
\begin{equation}
	|H^{\vb}\rangle \mapsto H(\vb) = \Pi_+\big[V^{\vb} \slashed{\epsilon}(\vb) - P^{\vb}\g^5 \big]\,, 
	\qquad \langle H^{\vb} |  \mapsto \Htilde(\vb)  =  \big[V^{\vb} \slashed{\epsilon}^\sharp(\vb) + P^{\vb} \g^5 \big] \Pi_+ \,,
\end{equation}
(recalling as above the $C^\sharp$-Dirac conjugate $\gentilde{X}(\vb)  \equiv  \g^0 X^{\sharp}(\vb) \g^0$).
Thus one may write the corresponding leading order HQET matrix element in Eq.~\eqref{eqn:QCDmatchexp} in the usual form
\begin{equation}
	\label{eqn:LOtrace}
	\big\langle \Djx[\vc] \big| \ctvp \, \Gamma \, \bv \big| \Bbar^{\vb} \big\rangle = -\xi(\wh)\Tr\big[ \Htilde(\vc) \Gamma H(\vb) \big]\,,
\end{equation}
in which $\xi(\wh)$ is the leading Isgur-Wise function. 
As shown in Sec.~\ref{sec:anprop} and App.~\ref{sec:PTconjrel}, $\xi(\wh)$ is a holomorphic function of $\wh$ with real coefficients.

Although both $\Bxbar$ and $\Djx$ doublets have the same definite spin-parity and the same HQET trace representation,
they nonetheless have different principle quantum numbers as the latter is a radially excited state. 
They thus must be treated as belonging to different HQET representations, 
with two different sets of hadron mass and width parameters.
Further the HQET matrix elements need not satisfy normalization constraints in the equal mass and zero recoil limit.
As a result, it need not be the case that $\xi(1) = 1$ 
(as would be required for ground state to ground state $\Bxbar \to \Dx$ transitions).

To keep the subleading Isgur-Wise functions dimensionless, 
at first order in the HQ power expansion we normalize the trace representation 
of the HQET matrix elements with respect to the geometric mean of the real hadron mass parameters $\LamB$ and $\LamB'$ for the $\Bxbar$ and $\Djx$ doublets, respectively.
As seen for the examples considered in App.~\ref{sec:PTconjrel},
choosing a real normalization manifestly ensures the subleading Isgur-Wise functions are holomorphic functions of $\wh$ with real coefficients.\footnote{%
As discussed in Sec.~\ref{sec:schdy}, because the overall theory must be $C^{\sharp(\hmc)}$-self adjoint, i.e. with respect to the simple pole itself,
HQET operator kernels must be $C^\sharp$-self adjoint with respect to all terms in the mass and width expansion for both hadrons.
Noting the complex structure of $\wh$ itself also explicitly and solely derives from $\hmc$, 
one could have instead chosen to normalize by e.g. $\LamB' - i\GamB'$, without disrupting the analytic structure of the Isgur-Wise functions.}
In particular, with reference to Eq.~\eqref{eqn:QCDmatchexp},
the trace representation of the first-order current correction HQET matrix elements is
\begin{align}
	\big\langle \Djx[\vc] \big| \ctvp \Gamma i \overrightarrow{D}_\mu \bv \big| \Bxbar[\vb] \big\rangle &  = -\sqLLp\, \Tr[\Htilde(\vc) \Gamma H(\vb) \Xi^{(b)}_\mu(\vc,\vb)]\,, \nn\\
	\big\langle \Djx[\vc] \big| \ctvp (-i \overleftarrow{D}_\mu) \Gamma  \bv \big| \Bxbar[\vb] \big\rangle &  = -\sqLLp\, \Tr[\Htilde(\vc) \Gamma H(\vb) \gentilde{\Xi}^{(c)}_\mu(\vb,\vc)]\,, \label{eqn:NLOcurrent}
\end{align}
in which the kernel $\Xi^{(Q)}_\mu(\vc,\vb) = \xi^{(Q)}_+(\wh) (\vb+\vc)_\mu + \xi^{(Q)}_-(\wh)(\vb-\vc)_\mu - \eta^{(Q)}(\wh) \g_\mu$, 
and $Q = c$, $b$. 
Because $\Bxbar$ and $\Djx$ formally belong to different HQET representations, 
there are correspondingly two different sets of (subleading) Isgur-Wise functions representing operators acting on $\bv$ versus $\cvp$.

The HQET matrix elements for first-order Lagrangian corrections are similarly represented as
\begin{subequations}
\label{eqn:NLOchromo}
\begin{align}
	\big\langle \Djx[\vc] \big| \ctvp \Gamma \bv \circ \big[\btv D^2 \bv \big]\big| \Bxbar[\vb] \big\rangle & = \sqLLp\, \Tr[\Htilde(\vc) \Gamma H(\vb) X^{(b)}_0(\vc,\vb)]\,,\nn\\*
	\big\langle \Djx[\vc] \big| \ctvp \Gamma  \bv \circ \big[\btv \frac{g}{2}\sigma^{\ab} G_{\ab} \bv \big] \big| \Bxbar[\vb] \big\rangle & = \sqLLp\, \Tr[ \Htilde(\vc)  \Gamma \Pi_+ \sigma^{\ab} H(\vb) X^{(b)}_{\ab}(\vc,\vb)]\,,
\intertext{and for the conjugate matrix elements}	
	\big\langle \Djx[\vc] \big| \big[\ctvp D^2 \cvp\big] \circ \ctvp \Gamma  \bv \big| \Bxbar[\vb] \big\rangle & = \sqLLp\, \Tr[\Htilde(\vc) \Gamma H(\vb) \gentilde{X}^{(c)}_0(\vb,\vc)]\,,\nn\\*
	\big\langle \Djx[\vc] \big| \big[\ctvp \frac{g}{2}\sigma^{\ab} G_{\ab} \cvp\big] \circ \ctvp \Gamma  \bv \big| \Bxbar[\vb] \big\rangle & =  \sqLLp\, \Tr[\Htilde(\vc) \sigma^{\ab} \Pi'_+\Gamma H(\vb) \gentilde{X}^{(c)}_{\ab}(\vb,\vc)]\,,\!
\end{align}
\end{subequations}
with the kernel $X^{(Q)}_0(\vc,\vb) = 2\chi^{(Q)}_1(\wh)$ and $X^{(Q)}_{\ab}(\vc,\vb) = -i\chi^{(Q)}_2(\wh) (\vb-\vc)_{[\alpha} \g_{\beta]}  + 2\chi^{(Q)}_3(\wh) \sigma_{\ab}$.
As shown in App.~\ref{sec:hqetPTex} the latter is not a well-defined operator kernel, 
and one must apply the identity~\eqref{eqn:sigmaid} in order to show the Isgur-Wise functions $\chi^{(Q)}_{2,3}$ are holomorphic functions of $\wh$ with real coefficients.

\subsection{Matching $\Bbar \to \Dv$ onto holomorphic HQET}
\label{sec:matchBDv}
At first order in the HQ power expansion, HQ symmetry permits the QCD-HQET matching to be expressed in the form
\begin{align}
	\frac{\langle \Djx | \cbar \,\Gamma\, b | \Bxbar \rangle}{ \sqrt{\smash[b]{\hat{m}_{\Djx}}\smash[b]{m_{B_{\phantom{J}}^{(*)}}}} } 
	& = -\xi(\wh) \bigg\{ \Tr\big[ \Htilde(\vc) \Gamma H(\vb) \big] \nn \\* 
	& + \ec\, \Tr\big[ \Htilde^{(c)}(\vc,\vb) \Gamma H(\vb) \big]  + \eb\, \Tr\big[ \Htilde(\vc) \Gamma H^{(b)}(\vb,\vc) \big] \bigg\}\,, \label{eqn:HQSmatch}
\end{align}
in which 
\begin{subequations}
\label{eqn:HQSrepn}
\begin{align}
	H^{(b)}(\vb,\vc) & = \Pi_+\Big\{ P^{\vb} \hL1b (-\g^5) + V^{\vb} \big(\hL2b \slashed{\epsilon} + \hL3b \epsilon \ccdot \vc\big) \Big\}  \nn \\*
		& \qquad \qquad \qquad  + \Pi_-\Big\{ P^{\vb} \hL4b (-\g^5) + V^{v} \big(\hL5b \slashed{\epsilon} + \hL6b \epsilon \ccdot \vc\big) \Big\}\,,\\
	\Htilde^{(c)}(\vc,\vb) & = \Big\{ P^{\vc} \hL1c \g^5 + V^{\vc} \big(\hL2c \slashed{\epsilon}^{\prime\sharp} + \hL3c \epsilon^{\prime\sharp} \ccdot \vb\big) \Big\}\Pi'_+  \nn \\*
		& \qquad \qquad \qquad  + \Big\{ P^{\vc} \hL4c \g^5 + V^{\vc} \big(\hL5c \slashed{\epsilon}^{\prime\sharp} + \hL6c \epsilon^{\prime\sharp} \ccdot \vb\big) \Big\}\Pi'_-\,.
\end{align}
\end{subequations}
The leading order Isgur-Wise function $\xi(\wh)$ has been factored out in Eq.~\eqref{eqn:HQSmatch}, and the HQ expansion parameters are defined to be the real ratios
\begin{equation}
	\eQ = \sqLLp/(2m_Q)\,.
\end{equation}
This choice matches the normalization choice for trace representation of the HQET matrix elements in Eqs~\eqref{eqn:NLOcurrent} and~\eqref{eqn:NLOchromo}.
In a standard notational convention, functions that are normalized by the leading Isgur-Wise function are typically denoted with a ``$\hat{\phantom{x}}$'':
that is $\hat{X}(w) = X(w)/\xi(w)$. 
To distinguish this from the (unfortunate) collision with the notation for complex shifted quantities, 
we have altered this convention to $\dhat{X}(\wh) = X(\wh)/\xi(\wh)$---one 
``$\hat{\phantom{x}}$'' for the complex shift and one ``$\hat{\phantom{x}}$'' for the normalization with respect to $\xi$---whence 
the notation for the HQS functions $\hL{i}{Q}(\wh)$.

The HQS functions $\hL{i}{Q}$ can be determined by matching Eqs~\eqref{eqn:NLOcurrent} and~\eqref{eqn:NLOchromo} onto Eqs~\eqref{eqn:HQSmatch},
along with application of the Schwinger-Dyson relation~\eqref{eqn:LOSDrel}.
As usual, because $\chi_{1}^{(c,b)}$ arise from the same HQET trace as the leading order term~\eqref{eqn:LOtrace}, 
HQ symmetry guarantees that these Isgur-Wise functions always occur in the combination $\xi + 2\ec \chi_1^{(c)} + 2\eb \chi_1^{(b)}$,
and thus $\chi_1^{(c,b)}$ may be reabsorbed into $\xi$ via the redefinition $\xi + 2\ec\chi_{1}^{(c)} + 2\eb\chi_{1}^{(b)} \to \xi$,
up to induced second-order power corrections.
After this redefinition, one finds
\begin{subequations}
\begin{align}
	\hL{1}{Q} & = 
		 -4(\wh-1)\hchi{2}^{(Q)} + 12\hchi{3}^{(Q)}\,,\\*
	\hL{2}{Q} & = 
		- 4\hchi{3}^{(Q)}\,,\\*
	\hL{3}{Q} & = 4\hchi{2}^{(Q)}\,,\\
	\hL{4}{c} & = \frac{\LamB - \wh(\LamB' - i \GamB')}{\sqLLp\,(\wh-1)} + 2\heta^{(c)}\,, & \hL{4}{b} & = \frac{(\LamB' - i \GamB') - \wh\LamB}{\sqLLp\,(\wh-1)} + 2\heta^{(b)} \,,\\*
	\hL{5}{c} & = \frac{\LamB - \wh(\LamB' - i \GamB')}{\sqLLp\,(\wh-1)} \,, & \hL{5}{b} & = \frac{(\LamB' - i \GamB') - \wh\LamB}{\sqLLp\,(\wh-1)}  \,,\\*
	\hL{6}{c} & = 2\frac{\LamB - \wh(\LamB' - i \GamB')}{\sqLLp\,(\wh^2-1)} - \frac{2\heta^{(c)}}{\wh+1}\,, & \hL{6}{b} & = 2\frac{(\LamB' - i \GamB') - \wh\LamB}{\sqLLp\,(\wh^2-1)} - \frac{2\heta^{(b)}}{\wh+1}\,,
\end{align}
\end{subequations}
where we have set $\GamB = 0$ as the $\Bxbar$ doublet is a ground state HQET representation.
Only $\hL{1}{b}$ and $\hL{4}{b}$ enter into the $\Bbar \to \Djx$ form factors---we do not consider $\Bbar^{*}$ decays---and
HQ symmetry ensures that $\eb\hL{1}{b}$ also always occurs in the same linear combination with the leading order term.
Thus one may further reabsorb $\chi_{2,3}^{(b)}$ into $\xi$, up to induced second-order power corrections, via the further redefinition
\begin{equation}
	\xi + \eb L_1^{(b)} \to \xi\,,
\end{equation}
such that in total we have redefined $\xi + 2\ec\chi_{1}^{(c)} + 2\eb\big[\chi_{1}^{(b)} - 2(\wh-1)\chi_{2}^{(b)} + 6\chi_{3}^{(b)}\big] \to \xi$.

The resulting SM $\Bbar \to \Dv$ form factors are, including also the perturbative corrections~\eqref{eqn:ascurrent},
\begin{subequations}
\label{eqn:ffshqetnlo}
\begin{align}
	\hHv   & = 1+ \haS \Cv1 + \ec \big(\hL2c-\hL5c\big) - \eb\hL4b\,,\\
	\hHa1 & = 1+ \haS \Ca1 + \ec \hL2c  - \frac{\wh-1}{\wh+1}\big[\ec\hL5c + \eb \hL4b \big]\,,\\
	\hHa2 & = \haS \Ca2 + \ec \big(\hL3c+\hL6c\big) \,,\\
	\hHa3 & = 1+ \haS\big(\Ca1+\Ca3\big) +\ec \big(\hL2c-\hL3c-\hL5c+\hL6c\big)  -  \eb \hL4b \,.
\end{align}
\end{subequations}
Analogous results for the form factors for beyond SM currents can be similarly derived.

\subsection{Parametrization of Isgur-Wise functions}
\label{sec:paramIW}
In real HQET it is standard to consider parametrizations of the leading Isgur-Wise function $\xi(w)$, expanding in the small parameter $(w-1)$, 
or under the conformal transformation $q^2 \mapsto \mathpzc{z}(q^2,q_0^2)$~\cite{Boyd:1995sq,Boyd:1997kz} that maps the $q^2$ interval 
above the crossed $bc$ pair-production threshold onto the unit circle in the complex $\mathpzc{z}$ plane, 
with center at $q^2 = q_0^2$,
while keeping the physical recoil regime on the real axis.\footnote{Analytic continuation in $\mathpzc{z}(q^2,q_0^2)$ accesses the subthreshold $bc$ pole structure in $q^2$.
As noted below Eq.~\eqref{eqn:recrel}, this is different to the complex shift~\eqref{eqn:zshift}, which leaves $q^2$ unchanged, and instead accesses the pole structure of the charmed resonance system.}

Under the general complex shift~\eqref{eqn:zshift}, the complex recoil
\begin{equation}
	w(z) = \frac{p_B(z) \cdot \pDv(z)}{\mB \sqrt{\pDv[2](z)}}  = \frac{ \pDv \cdot p_B + z \zeta \cdot \pDv}{\mB \sqrt{\mDpi^2 + 2 z \zeta \cdot \pDv}} \sim \sqrt{z}
\end{equation}
in the large-$z$ limit. 
Constructibility of the recursion relation~\eqref{eqn:recrel} leads one to expect that the 
complex-shifted leading Isgur-Wise function $\xi(w(z))$ should be parametrized as an expansion in powers of $w(z)^{-1}$,
starting with $w(z)^{-2} \sim 1/z$.
The (higher-order) pole introduced by inverse powers of $w(z)$ at $z_w = -\pDv \cdot p_B/(\zeta \cdot \pDv)$ lies on the real line above $\zbr$, 
because from Eq.~\eqref{eqn:zbr} $z_w/\zbr = 2\mDpi \mB w/(\mDpi^2 - (m_D + m_{\pi})^2) >1$ 
(and in fact is $\gg1$ over the physically allowed range of $\mDpi$ in the $\Bbar \to \Dv \to D\pi$ system),
and thus we assume its (multiple-$z$-derivative) contributions to the recursion relation may be neglected whenever the branch cut discontinuity integral may be, as discussed in Sec.~\ref{sec:recrel}.
Similarly, the branch point introduced by odd powers of $w(z)$ at $z_{w,\text{br}}=-\mDpi^2/ (2\zeta \cdot \pDv)$ lies above $\zbr$.
This additional branch point may convert the keyhole contour into a so-called dogbone contour enclosing $\zbr$ and $z_{w,\text{br}}$,
with the effect of simply restricting the branch cut integration range to lie between them:
We assume this restricted discontinuity integral may be neglected whenever the full one may be.

Per Sec.~\ref{sec:anprop}, as the Isgur-Wise functions must be holomorphic functions of $\wh$ with real coefficients, 
one is then led to deduce the Laurent series parametrization
\begin{equation}
	\xi(\wh) = \sum_{k\ge2} \frac{a_k}{\wh^{k}}\,,\qquad a_k \in \mathbb{R}\,,
 \label{eq:IWparam}
\end{equation}
which is holomorphic everywhere except at $\wh = 0$.
For the case of ground-state $\Bx \to \Dx$ transitions, the $\xi(1) = 1$ normalization constraint simply implies $\sum a_k = 1$.
When restricted to back the real axis, the scaling of $\xi(w)$ with inverse powers of $w$ is commensurate 
with e.g. the well-known and/or expected behavior of the leading Isgur-Wise function for $\Bx \to \Dx$ transitions:
when Taylor expanded in powers of $(w-1)$ it has a negative slope and positive curvature for $w >1$.
It also aligns with the expectations outlined in Ref.~\cite{Neubert:1991td} 
(see also Ref.~\cite{Veseli:1995fr}), 
based on the understanding that the hadronic matrix elements encode a wavefunction overlap of the two heavy hadrons 
and should therefore vanish as $w$ increases.
This is seen also in quark model-based calculations (see e.g. Refs~\cite{Korner:1987kd,Isgur:1988gb,Scora:1995ty}).

One may parametrize subleading Isgur-Wise functions as similar Laurent series.
For the sake of the toy numerical examples we consider below for the $\Bbar \to \Dv$ system, 
we consider only the heavy quark limit for the sake of simplicity and similarly take just 
\begin{equation}
	\label{eqn:IWparamHQL}
	\xi(\wh) \sim 1/\wh^2\,.
\end{equation}

\subsection{On-shell subamplitudes}
\label{sec:ossubamp}
It remains to now compute the complex-shifted on-shell subamplitudes, 
$A_{\lambda' \lambda}^{\Bbar \to \Dv W}(\hat{z})$, $A_{\lambda'}^{\Dv \to D\pi}(\hat{z})$, and $A_{\lambda s_l s_\nu}^{W\to l\nu}$.
The $\Dv \to D\pi$ decay proceeds through $P$-wave, mediated by the operator $\Dv{}^\mu (D\partial_\mu \pi - \pi \partial_\mu D)$.
The corresponding amplitude under the complex shift  $A_{\lambda'}^{\Dv \to D\pi}(\hat{z}) = g_\pi \, \heps_{\lambda'} \ccdot (p_D - \hat{p}_\pi) = 2 g_\pi \, \heps_{\lambda'} \ccdot p_D$,
because of on-shell transversity $\heps \cdot \hpDv = 0$ and momentum conservation.
Here $g_\pi$ is in principle a form factor, dependent on $\pDv[2] = \mDpi^2$ and $(p_D-p_\pi)^2$, 
the latter of which is also fixed up to $\mDpi$ dependence as the $D$ and $\pi$ are on-shell.
Under the shift to the simple pole, then $g_\pi = g_\pi(\hmDv[2])$, which we treat as an overall constant: 
in the narrow-width limit $g_\pi^2 \simeq 6\pi \mDv[2]\Gamma[\Dv \to D\pi]/|\vec{p}^{*}_\pi|^3$.

Using the representation of the polarizations in Eq.~\eqref{eqn:epsphat},
one finds
\begin{subequations}
\label{eqn:DvDPiampl}
\begin{align}
	A_{\pm}^{\Dv \to D\pi}(\hat{z}) & = \pm \sqrt{2}g_{\pi} e^{\mp(\etpi-i\phpi)}\Big[\sqrt{m_{\pi}^2+|\vec{p}_\pi|^2}\pm|\vec{p}_\pi|\Big]\,,\\
	A_{0}^{\Dv \to D\pi}(\hat{z}) & = -g_{\pi}\frac{\hmDv[2]-m_{D}^2 +m_{\pi}^2}{\hmDv} = -2g_\pi\hat{E}_\pi^*\,, 
\end{align}
\end{subequations}
in which we emphasize $|\vec{p}_\pi|$ is the pion momentum in the $\Bbar$ frame,
and $\hat{E}_\pi^*$ is energy of the complex-shifted pion momentum in the $\Dv$ rest frame.
Note the prefactors $e^{\mp(\etpi-i\phpi)}$ comply generate the relative prefactor for the $\lambda' = \pm$ amplitudes expected from Eq.~\eqref{eqn:Pexppref}.

Being careful to recall that under the generalized $C^{\sharp(\hmDv)}$-conjugation 
$(\epsilon^{\lambda'})^\sharp = \epsilon^{\bar\lambda'}$ (see Sec.~\ref{sec:sphlgc}),
so that $[A^{\Dv \to D\pi}_{\lambda'}]^\sharp = A^{\Dv \to D\pi}_{\bar{\lambda}'}$,
observe that the spin sum of the $C^{\sharp(\hmDv)}$ square modulus 
\begin{align}
	\sum_{\lambda'=\pm,0}\big|A_{\lambda'}^{\Dv \to D\pi}(\hat{z})\big|^2 & \equiv \sum_{\lambda'=\pm,0} A_{\lambda'}A_{\lambda'}^\sharp  =  A_+ A_- + A_- A_+ + A_0 A_0\nn\\
	& = 4g_\pi^2(\hat{E}_\pi^* - m_\pi^2) = 4g_\pi^2|\vec{p}_\pi^*(\hat{z})|^2\,,
\end{align}
where $|\vec{p}_\pi^*(\hat{z})|$ is the complex-shifted spatial momentum of the pion in the $\Dv$ rest frame.
That is, we obtain merely the complex shift of the usual spin sum square modulus for a $P$-wave decay with real momenta, 
$\sum_{\lambda'} \big|A_{\lambda'}^{\Dv \to D\pi}(0)\big|^2 = 4g_\pi^2|\vec{p}_\pi^*|^2$.
Put a different way, for amplitudes involving complex on-shell momenta the $C^{\sharp(\hmDv)}$-conjugation 
preserves the kinematic dependence expected from angular momentum selection rules.

Using the form factor definitions in Eq.~\eqref{eqn:defBDvSMff} combined with the form factor ratios~\eqref{eqn:defBdvR} 
and auxiliary ratios~\eqref{eqn:defBdvRaux}, one finds for the $\Bbar \to \Dv W$ amplitudes
\begin{subequations}
\label{eqn:BDvWamplSM}
\begin{align}
A^{\Bbar \to \Dv W}_{\pm\pm}(\hat{z}) & = \Hha1\bigg\{\pm\frac{1}{4} e^{\yDv} \mDpi \Rg{1} \sqrt{w^2-1}-\frac{1}{4} e^{\yDv} \hmDv (\wh + 1)\nn \\*
	& \qquad + \frac{(\hmDv[2] - \mDpi^2) \Rg{\pm\pm}}{8 \mDpi}\bigg\}\,,\\
A^{\Bbar \to \Dv W}_{\pm0}(\hat{z}) & = \Hha1\bigg\{e^{\pm(\etpi - i \phpi)}\bigg[\pm\frac{\mB \mDpi \Rg{2} \big[w^2-1\big]}{2 \sqrt{2} \sqrt{q^2}} \pm \frac{\hmDv (\mDpi-\mB w) (\wh + 1)}{2 \sqrt{2} \sqrt{q^2}}\nn \\*
	& \qquad + \frac{\mB (\hmDv[2] - \mDpi^2) \Rg{\pm0}}{8 \mDpi \sqrt{q^2}}\bigg]\bigg\}\,,\\	
A^{\Bbar \to \Dv W}_{\pm\mp}(\hat{z}) & = \Hha1\bigg\{e^{\pm2 (\etpi - i \phpi )}\bigg[\pm\frac{1}{4} e^{-\yDv} \mDpi \Rg{1} \sqrt{w^2-1}+\frac{1}{4} e^{-\yDv} \hmDv (\wh + 1)\nn \\*
	& \qquad + \frac{(\hmDv[2] - \mDpi^2) \Rg{\pm\mp}}{8 \mDpi}\bigg]\bigg\}\,,\\	
A^{\Bbar \to \Dv W}_{0\pm}(\hat{z}) & = \Hha1\bigg\{e^{\mp(\etpi-i \phpi) }\bigg[\frac{\hmDv \Rg{1} \sqrt{w^2-1}}{2 \sqrt{2}}\mp\frac{\mDpi (\wh + 1)}{2 \sqrt{2}}\nn \\*
	& \qquad + \frac{\mB (\hmDv[2] - \mDpi^2) \Rg{0\pm}}{8 \sqrt{2} \mDpi \hmDv}\bigg]\bigg\}\,,\\	
A^{\Bbar \to \Dv W}_{00}(\hat{z}) & = \Hha1\bigg\{\frac{\mB \mDpi^2 \Rg{2} w \big[w^2-1\big]}{2 \hmDv \sqrt{q^2}}+\frac{\mDpi w (\mDpi-\mB w) (\wh + 1)}{2 \sqrt{q^2}}\nn \\*
	& \qquad + (\hmDv[2] - \mDpi^2) \bigg[\frac{\mDpi \Rg{2} \big[w^2-1\big]}{4 \hmDv \sqrt{q^2}}+\frac{\mB^2 \Rg{00}}{8 \sqrt{2} \mDpi \hmDv \sqrt{q^2}}\bigg]\bigg\}\,.	
\end{align}
\end{subequations}	
Note that, with reference to Eq.~\eqref{eqn:amplSMbcln}, we have pulled out a prefactor 
$2\sqrt{2}G_FV_{cb}/\sqrt{w^2-1} \times \sqrt{\mB/\hmDv}$ in defining $A^{\Bbar \to \Dv W}_{\lambda'\lambda}(\hat{z})$.
Further note once again the deliberate distinctions between $\wh$ and $w$.
For the longitudinal $W$ amplitudes, one finds
\begin{subequations}
\label{eqn:BDvWLamplSM}
\begin{align}
A^{\Bbar \to \Dv W}_{\pm\ell}(\hat{z}) & = \pm\Hha1\bigg\{\frac{e^{\pm(\etpi - i \phpi)} \hmDv (\mB+\hmDv) \Rg{0} \sqrt{w^2-1}}{2 \sqrt{2} q^2}\bigg\}\,,\\
A^{\Bbar \to \Dv W}_{0\ell}(\hat{z}) & = \Hha1\bigg\{\frac{\hmDv (\mB+\hmDv) \Rg{0} \wh \sqrt{w^2-1}}{2 q^2}\bigg\}\,.
\end{align}
\end{subequations}

In Eqs.~\eqref{eqn:BDvWamplSM} we have kept explicit terms proportional to $\hmDv[2] - \mDpi^2$.
These terms vanish in the narrow-width real on-shell limit in which $\mDpi \to \hmDv \to m_{\Dv}$.
Applying this limit (which also implies $\wh \to w$ and $\hat{z} \to 0$) to Eqs.~\eqref{eqn:DvDPiampl}, \eqref{eqn:BDvWamplSM} and~\eqref{eqn:BDvWLamplSM} 
results in $\Dv \to D\pi$ and $\Bbar \to \Dv W$ amplitudes that are the same as the usual helicity basis amplitudes up to a transformation of the $\Dv$ spin basis.
In this case, the $\Dv$ polarization axis that defines the $\Dv$ spin basis can be thought of as having been rotated 
from the usual choice of (anti)alignment with the $\Dv$ momentum in the $\Bbar$ frame 
to $\vec\zeta = \vec\zeta_R + i \vec\zeta_I$ lying in complex $3+1$ dimensional space.
One may check that in this limit $\sum_{\lambda'=\pm,0} A^{\Bbar \to \Dv W}_{\lambda' \lambda}(\hat{z}) \times A^{\Dv \to D\pi}_{\lambda'}$ 
corresponds to the usual result obtained in the helicity basis.

When composed with the $1/(\hmDv[2] - \mDpi^2)$ residue in the full amplitude~\eqref{eqn:BDvconrecrel}, 
the $\hmDv[2] - \mDpi^2$ terms become regular in $\mDpi$, 
and thus fall off more slowly far above or below the resonance peak compared to what one would expect 
from a naive Breit-Wigner parametrization under the narrow-width approximation.
We shall see in the next section (Sec.~\ref{sec:toynumex}) that such terms produce lineshape tails that broadly resemble those seen in data.

Finally, the $W \to l\nu$ amplitudes, $A^{W \to l\nu}_{\lambda s_l s_\nu}$ are 
not affected by the complex shift and remain the same as in the standard calculation.
The charged lepton and neutrino spin quantum numbers take the values $s_l=\pm$ and $s_\nu=\pm$.\footnote{%
To incorporate subsequent $\tau$ decays, using the conventions of Ref.~\cite{Ligeti:2016npd} for massive spinors on internal lines, 
one would instead label the charged lepton spin by $s_l = 1$ and $2$, rather than $-$ and $+$, respectively.}
In the SM, with only left-handed neutrinos, all $s_\nu = +$ amplitudes vanish.
We use the spinor phase conventions of Ref.~\cite{Ligeti:2016npd}, 
which amounts to the inclusion of an additional phase factor $e^{i\phi_l}$ for $s_l = \up$ amplitudes.
The resulting $W \to l\nu$ amplitudes are
\begin{subequations}
\label{eqn:WlnuamplSM}
\begin{align}
	A^{W \to l\nu}_{+ \up -} & = -\sqrt{2} e^{-i\phi_l} \sin^2(\theta_l/2)\,, & A^{W \to l\nu}_{+ \dn -} & = \frac{m_l e^{-i\phi_l} }{\sqrt{2q^2}}\sin(\theta_l)\,,\\
	A^{W \to l\nu}_{0 \up -} & = \sin(\theta_l)\,, & A^{W \to l\nu}_{0 \dn -}& = -\frac{m_l}{\sqrt{q^2}}\cos(\theta_l)\,,\\
	A^{W \to l\nu}_{- \up -} & = \sqrt{2} e^{i\phi_l} \cos^2(\theta_l/2)\,, &A^{W \to l\nu}_{- \dn -} & = \frac{m_l e^{i\phi_l} }{\sqrt{2q^2}}\sin(\theta_l)\,.
\end{align}
\end{subequations}
Note that, with reference to Eq.~\eqref{eqn:amplSMbcln}, we have pulled out a prefactor 
$\sqrt{q^2 - m_l^2}$ in defining $A^{W \to \ell\nu}_{\lambda s_l s_\nu}$.
The amplitudes for the longitudinal $W$ are
\begin{equation}
\label{eqn:WLlnuamplSM}
	A^{W \to l\nu}_{\ell \dn -} = -m_l\,, \qquad A^{W \to l\nu}_{\ell \up -} = 0\,.
\end{equation}

From Eqs.~\eqref{eqn:DvDPiampl}, \eqref{eqn:BDvWamplSM} and~\eqref{eqn:BDvWLamplSM}, 
observe that the phase structure of the $\Bbar \to (\Dv \to D\pi)W$ amplitudes
\begin{equation}
	\label{eqn:BDvDPiamplSM}
	\sum_{\lambda' = \pm,0} A^{\Bbar \to \Dv W}_{\lambda'\lambda}(\hat{z}) A_{\lambda'}^{\Dv \to D\pi}(\hat{z})  \sim 
		\begin{cases} e^{\mp(\etpi-i\phpi)}\,,  & \lambda = \pm\,, \\ 1\,, & \lambda = 0,\ell\,.\end{cases}
\end{equation}
As can be seen from Eqs.~\eqref{eqn:WlnuamplSM} and~\eqref{eqn:WLlnuamplSM}, 
the $\phpi$ dependence in Eq.~\eqref{eqn:BDvDPiamplSM} ensures only the phase combination 
$\phpi - \phi_l$ appears in the full amplitudes, as expected.

With reference to Eqs.~\eqref{eqn:etpippi}, 
at the pair production threshold $\mDpi \to m_D + m_\pi$ the pion momentum $|\vec{p}^*_\pi| \to 0$ and so the pseudorapidity $\etpi \to\infty$.
The $\etpi$ dependence in Eq.~\eqref{eqn:BDvDPiamplSM} then leads to a singularity in the $\lambda = -$ amplitudes at production threshold,
while the $\lambda = +$ amplitudes vanish.
As discussed in Sec.~\ref{sec:recrel}, in the near-threshold regime $|\hat{z}|$ diverges (because the denominator $\zeta \cdot \pDv \sim  |\vec{p}^*_\pi|^2 \to 0$), 
which is commensurate with this divergent behavior of the amplitudes.
Moreover, the on-shell recursion relation~\eqref{eqn:recrel} itself should fail in the near-threshold regime 
because of a non-negligible contribution from the production threshold branch cut:
the threshold singularity in the amplitudes~\eqref{eqn:BDvDPiamplSM} 
suggests that in this regime the branch cut contribution must be similarly divergent 
in order to regulate the full amplitude.

From Eqs.~\eqref{eqn:etpippi}, in the soft limit $|\vec{p}^*_\pi| \to 0$, 
the exponential factor that arises in the square amplitude $e^{2\etpi} \simeq (m_\pi^{2} \sinh^2\yDv)/(|\vec{p}^*_\pi|^2\sin^2\theta_\pi)$,
and so the threshold singularity in the differential rate~\eqref{eqn:diffrate} is not regulated by its vanishing $|\vec{p}_\pi^*|$ phase space factor.
We shall observe this behavior further in the numerical examples of the next section (Sec.~\ref{sec:toynumex}).

\subsection{Toy numerical examples}
\label{sec:toynumex}

To illustrate the characteristic features of the $\Dv$ lineshape obtained in the OSR+HQET approach,
we consider some toy numerical examples for the $\Bbar \to \Dv (\to D\pi) l \nu$ system.
Ultimately we are interested (subsequent to this work) in applying this framework to $\Bbar \to D^{**}$ decays.
With respect to the narrow $s_\ell^{\pi_\ell} = 3/2^+$ $D^{**}$ states, 
angular momentum and parity selection rules forbid the decay $D_1(1^+) \to D\pi$ but $D_2^*(2^+) \to D\pi$ is permitted,
while the $D^*\pi$ final state can be accessed from both the $D_1$ and $D_2^*$.
Thus the $D\pi$ final state affords the opportunity to examine the lineshape of a single $D^{**}$ resonance---the $D_2^*$, via $\Bbar \to (D_2^* \to D\pi)l\nu$---with 
only a highly subdominant component from the $D_0^*$.
With this in mind, we compute the lineshape for a fictitious $\Dv$ state with mass and width
\begin{equation}
	\label{eqn:toymasswidth}
	\mDv  = 2460\,\MeV\,, \qquad \GDv = 40\,\MeV\,,
\end{equation}
chosen to be generically similar to the $D_2^*$,
with an eye towards comparing the computed lineshape to those measured for $\Bbar \to (D_2^* \to D\pi)l\nu$~\cite{Belle:2022yzd}. 
We emphasize from the beginning that the different spin of the $D_2^*$ obviously modifies both the structure of the amplitudes and the form factors
at the $\mathcal{O}(1)$ level and thus this comparison can be roughly qualitative at best.
Further, as this is a toy example, we consider only the heavy quark limit, in which the form factor ratios $R_{1,2,0}(\wh) \to 1$
and we take $\hHa1(\wh) \to \xi(\wh) = 1/\wh^2$ per Eq.~\eqref{eqn:IWparamHQL}.
Including higher-order terms would be expected to lead to corrections at the $\mathcal{O}(\lqcd/2m_c) \sim 20\%$ level or smaller.

\begin{figure}[t]
\centering
\begin{subfigure}{.49\linewidth}
	\centering
	\includegraphics[width = 8cm]{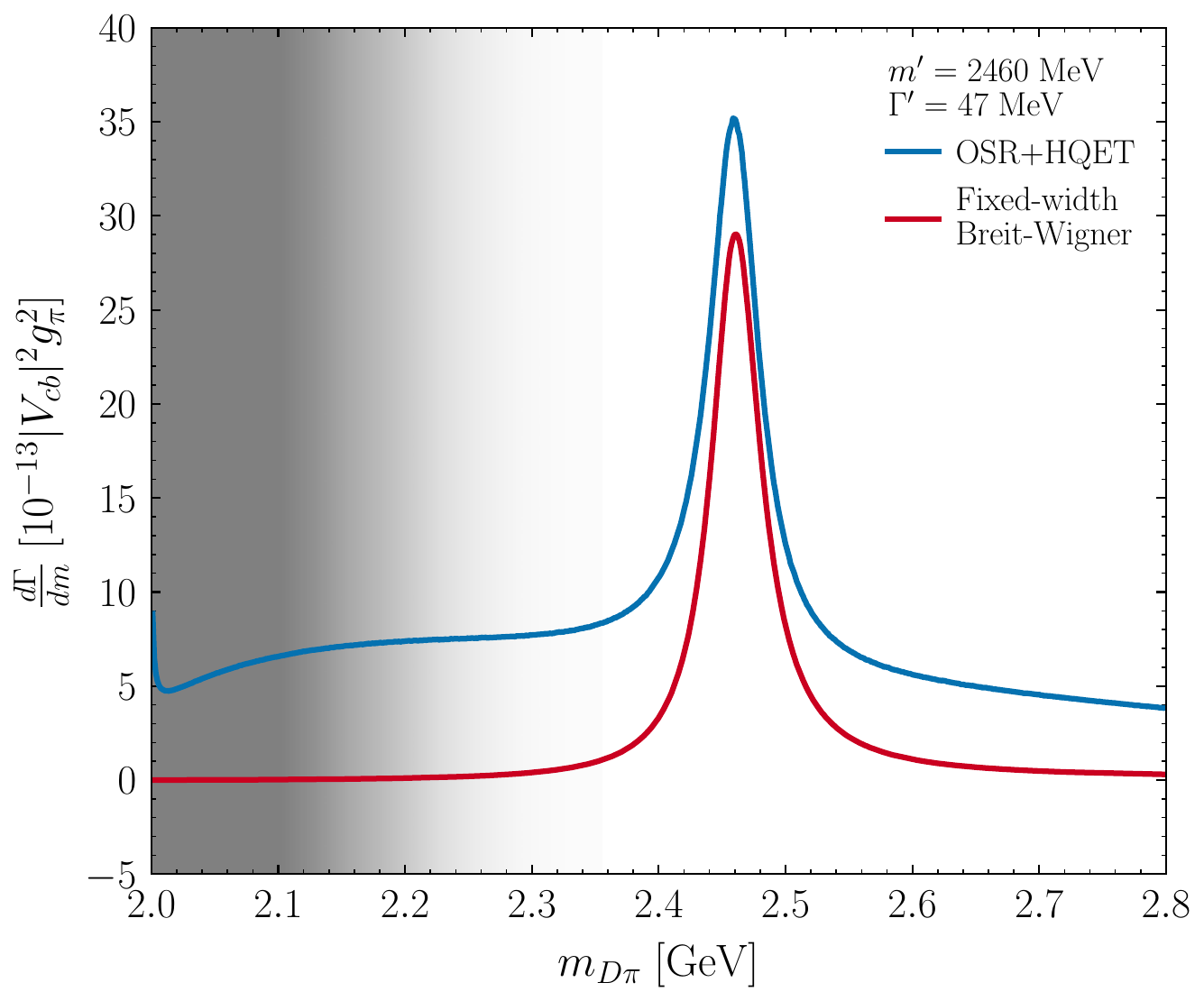}
	\caption{}\label{fig:singleRes1}
\end{subfigure}
\hfil
\begin{subfigure}{.49\linewidth}
	\centering
	\includegraphics[width = 8cm]{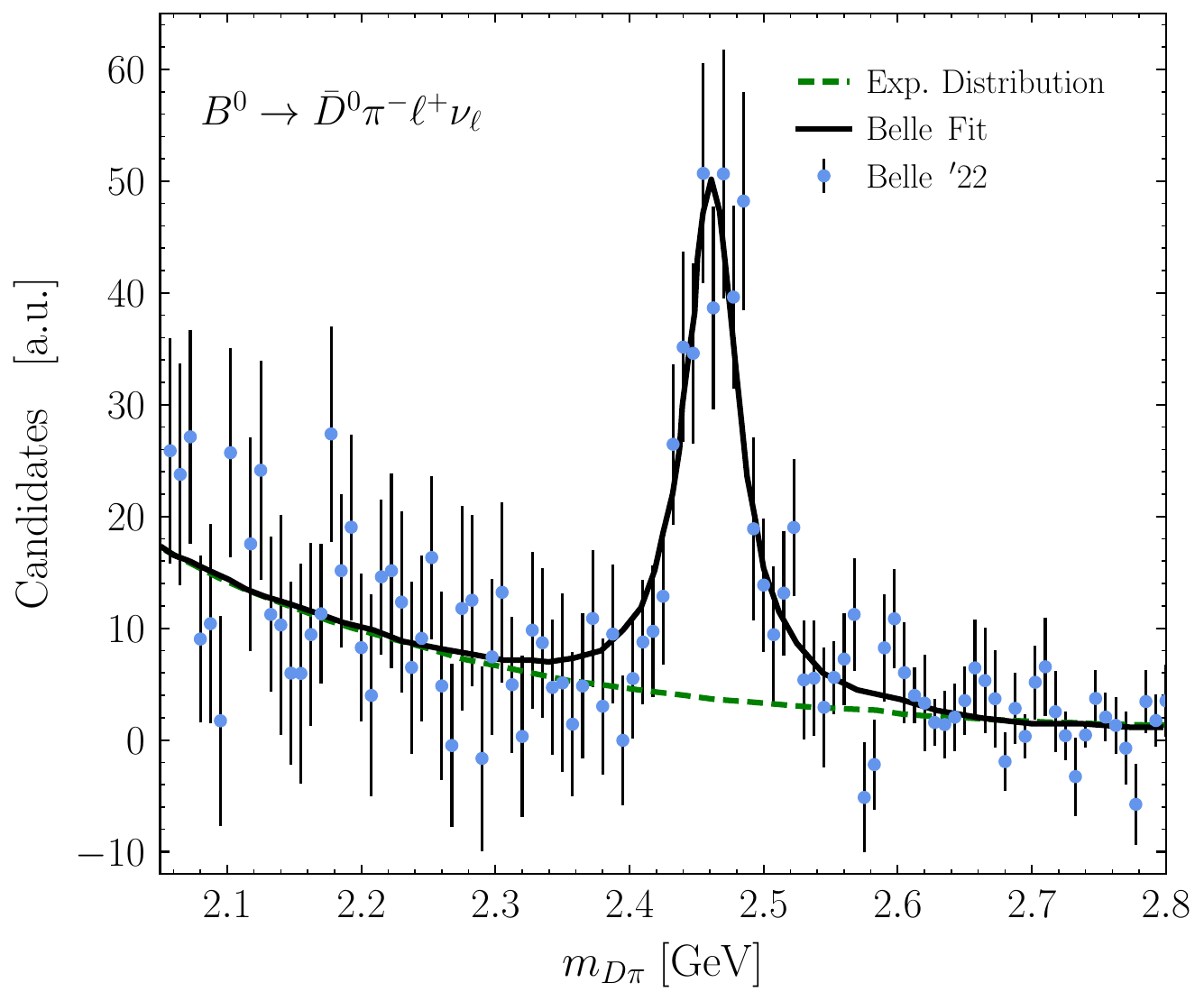}
	\caption{}\label{fig:singleRes2}
\end{subfigure}
\caption{(\subref{fig:singleRes1}): 
Toy numerical results for marginal differential distribution $d\Gamma[\Bbar \to \Dv (\to D\pi) l \nu]/d\mDpi$ derived in the OSR+HQET approach (blue curve)
for the mass and width values of Eq.~\eqref{eqn:toymasswidth}
compared to the same for a naive fixed-width Breit-Wigner parametrization (red curve).
The gray shaded region denotes the regime in which the on-shell recursion relation~\eqref{eqn:recrel} 
is expected to to receive additional non-negligible corrections from the branch cut discontinuity.
(\subref{fig:singleRes2}): 
The invariant mass distribution for $m_{\bar{D}^0\pi^-}$ in $B^0\to \bar{D}^0\pi^-\ell^+\nu_{\ell}$ as measured by Belle~\cite{Belle:2022yzd}. 
The main $D_2^*$ resonance is clearly visible. 
Also shown is the Belle analysis' parametric fit of the lineshape (black curve), 
that models the dominant $D_2^*$ (and subdominant $D_0^*$) resonant contribution as a Breit-Wigner convolved with a Gaussian 
plus an additional exponential distribution to account for the sizeable tails.}
\label{fig:singleRes}
\end{figure}

Fig.~\ref{fig:singleRes1} shows the marginal differential rate for $d\Gamma[\Bbar \to \Dv (\to D\pi) l \nu]/d\mDpi$ in the OSR+HQET approach (blue curve)
obtained from numerical integration of the fully-differential rate~\eqref{eqn:diffrate} using the {\it{Vegas}} (v.5.6) Monte Carlo integration algorithm~\cite{Lepage:1977sw}.
(Analytic integration is moderately complicated by the use of the $\Bbar$-frame pion pseudorapidity $\etpi$ 
and the $\Bbar$-frame pion momentum $|\vec{p}_\pi|$ in the amplitudes,
that have nontrivial forms~\eqref{eqn:etpippi} when expressed in the phase space measure coordinates of Eq.~\eqref{eqn:diffrate}.
We leave this for future work.)
For comparison we also show (red curve) the same distribution in the narrow-width approximation
using a fixed-width Breit-Wigner for the lineshape.
Though such a treatment would not represent a state-of-the-art approach versus, 
e.g. a dynamic-width Breit-Wigner parametrization with Blatt-Weisskopf factors,
it allows a straightforward characterization of the effects of the OSR+HQET approach on the tails and peak of the lineshape.
In particular, one notes that the OSR+HQET approach yields a slight downward shift in the location of the peak,
a broadening of the peak lineshape, and roughly constant sizeable tails over a broad $\mDpi$ range.

In the near threshold regime $\mDpi \to m_D + m_\pi \simeq 2\,\GeV$ one sees a turnover and a divergence
associated with the threshold singularity discussed in Sec.~\ref{sec:ossubamp}.
This singularity arises deep into the heuristic regime specified in Eq.~\eqref{eqn:heurthres}, 
in which the branch cut discontinuity integral is expected to begin to be non-negligible
and the on-shell recursion relation~\eqref{eqn:recrel} begins to receive additional such corrections.
In this toy example, we consider this region to roughly begin somewhere in the interval 
between $|\hat{z}|/\zbr \simeq 3$ (corresponding to $\mDpi \simeq 2.36\,\GeV$)
and $|\hat{z}|/\zbr \simeq 1$ (corresponding to $\mDpi \simeq 2.24\,\GeV$),
which is indicated by the gray shading.
We note also that for $\pc[2] \gg |\hmc[2]|$, i.e. well above the peak, the far tail drops relatively slowly, 
such that branch cut contributions in the far tail are expected to remain as similarly negligible compared to the $\hat{z}$ residue contribution
as they are near the resonance peak (see Sec.~\ref{sec:recrel}).

For a qualitative comparison, 
Fig.~\ref{fig:singleRes2} shows the $\mDpi$ differential distribution for $B^0\to \bar{D}^0\pi^-\ell^+\nu_{\ell}$ 
as measured by Belle~\cite{Belle:2022yzd} in the interval $2.05$--$2.8\, \GeV$. 
Also shown is a parametric fit to the data obtained by the Belle analysis (black curve).
This fit models the dominant $D_2^*$ (and subdominant $D_0^*$) resonant contribution as a Breit-Wigner convolved with a Gaussian,
and also includes an \emph{ad hoc} exponential distribution (gray dashed) to account for the sizeable tails observed after background subtraction.
The qualitative size and shape of these tails, however, is commensurate with and characteristic of the behavior of the tails derived in the OSR+HQET approach.
A full fit to this data using the OSR+HQET predictions for $\Bbar \to (D_{2,0}^* \to D\pi)l\nu$ is therefore well-motivated,
but left for future work: apart from deriving the OSR+HQET expressions, 
a self-consistent analysis will likely require a signal plus background fit to the underlying $B^0\to \bar{D}^0\pi^-\ell^+\nu_{\ell}$ data.

\begin{figure}[t]
\centering
\begin{subfigure}{.49\linewidth}
	\includegraphics[width = 8cm]{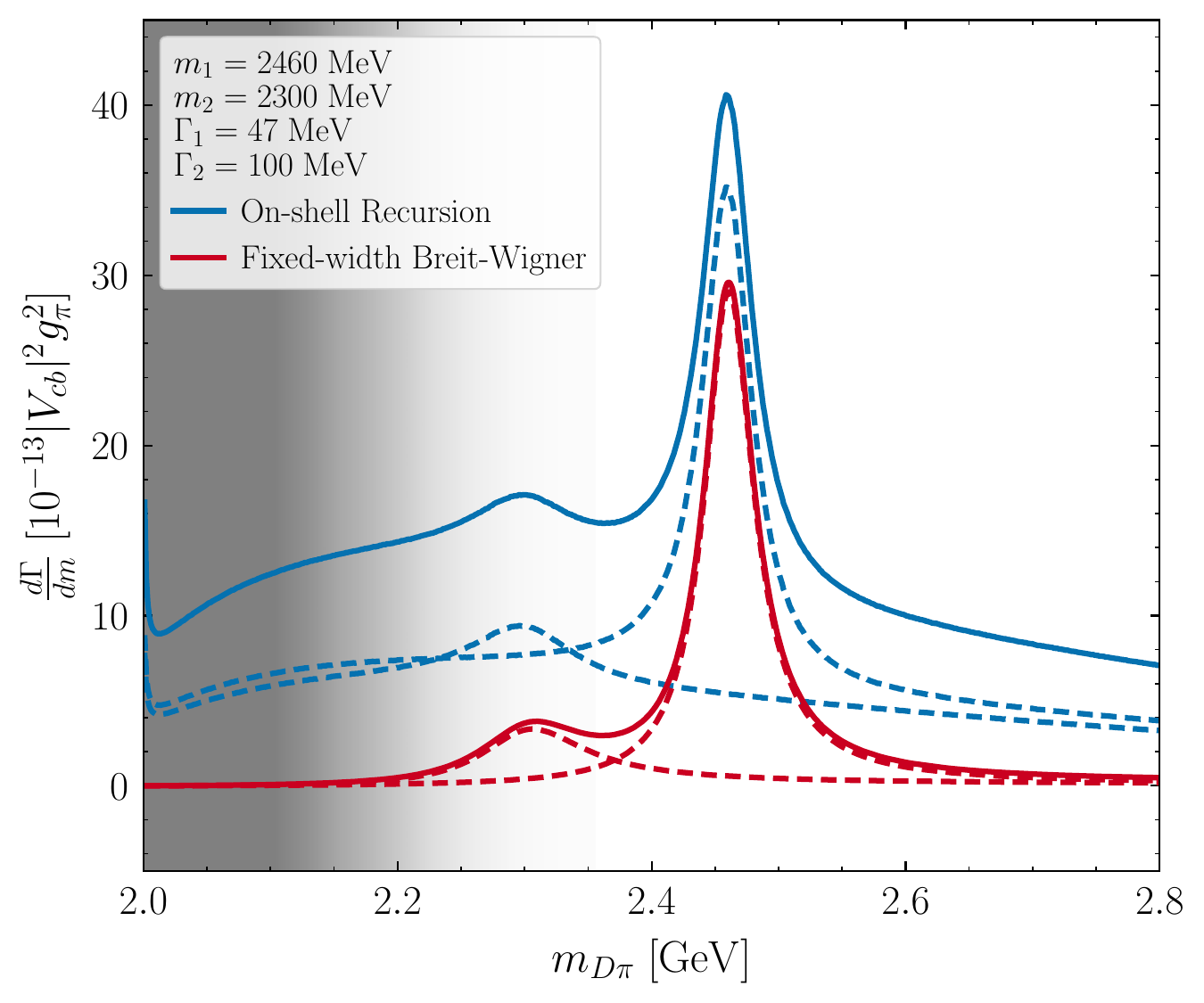}
	\caption{Linear scale}
	\label{fig:doubleRes1}
\end{subfigure}
\hfil
\begin{subfigure}{.49\linewidth}
	\includegraphics[width = 8cm]{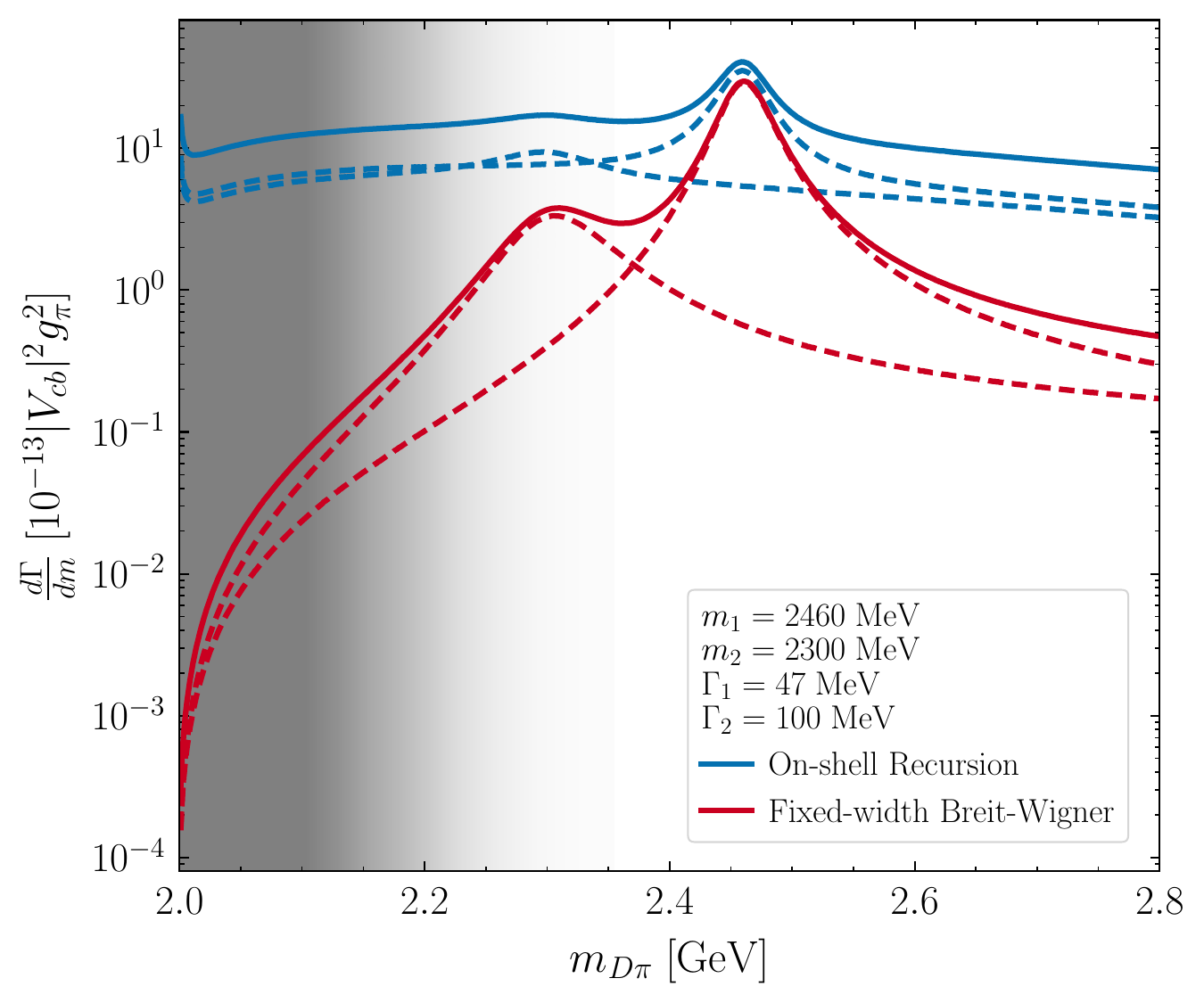}
	\caption{Log scale}
	\label{fig:doubleRes2}
\end{subfigure}
\caption{Toy numerical results for marginal differential distribution $d\Gamma[\Bbar \to \Dv (\to D\pi) l \nu]/d\mDpi$ derived in the OSR+HQET approach (blue curve)
on a linear (\subref{fig:doubleRes1}) and a logarithmic scale (\subref{fig:doubleRes2}),
for two resonances: one with mass and width values of Eq.~\eqref{eqn:toymasswidth} and one with mass and width $2300\,\MeV$ and $100\,\MeV$, respectively.
The single-resonance contributions (dashed curves) to the lineshape as well as their sum (solid curve) are shown,
and are compared to the same for a naive fixed-width Breit-Wigner parametrization (red curve).
The gray shaded region denotes the regime in which the on-shell recursion relation~\eqref{eqn:recrel} 
is expected to to receive additional non-negligible corrections from the branch cut discontinuity.}
\label{fig:doubleRes}
\end{figure}

Finally, so far we have examined the properties of an OSR+HQET lineshape for a relatively narrow resonance as in Eq.~\eqref{eqn:toymasswidth}.
One expects the difference between the OSR+HQET and fixed-width Breit-Wigner approaches to become more pronounced for broader resonances,
as the narrow-width approximation further breaks down.
To characterize this, we show in Fig.~\ref{fig:doubleRes} the lineshape for two $\Dv$ resonances:
one narrow with the same mass and width as in Eq.~\eqref{eqn:toymasswidth} and another broader state 
with mass and width $2300\,\MeV$ and $100\,\MeV$, respectively.
We set the effective $g_\pi$ couplings to be the same for both resonances.

Both the single-resonance contributions (dashed curves) to the lineshape as well as their sum (solid curve) are shown 
on a linear (Fig.~\ref{fig:doubleRes1}) and a logarithmic scale (Fig.~\ref{fig:doubleRes2})
for the OSR+HQET approach (blue curves).
These are again compared to a naive fixed-width Breit-Wigner (red curves).
One sees that the OSR+HQET resonance peak is notably broadened and slightly downward shifted,
as well as an order of magnitude larger.
The putative breakdown of the on-shell recursion relation~\eqref{eqn:recrel} is again indicated by a gray shaded region,
determined by whichever of the two resonances first enters the heuristic regime of Eq~\eqref{eqn:heurthres}:
in this example, this occurs first for the narrow resonance and thus the range is the same as in Fig.\eqref{fig:singleRes}.
Based on these examples, one might speculate that using an OSR+HQET model in a signal plus background fit analysis
may well lead to recovered $D_0^*$ or $D_1'$ widths that are somewhat narrower than the currently-reported averages
and/or alter the reported $\Bbar \to D_0^*$ or $D_1'$ relative branching ratios, 
with possible implications for the so-called spin-$\frac12$--$\frac32$ puzzle~\cite{Bigi:1997fj,LeYaouanc:2000cj,Uraltsev:2000qw,Uraltsev:2004ra}.
A full analysis using OSR+HQET $\Bbar \to D^{**}$ predictions and a consistent signal plus background fit is required to explore this further.

\section{Summary and outlook}
\label{sec:summout}
In this work we developed a new framework for the description of heavy quark hadronic resonances within HQET,
properly incorporating the effects of off-shell and longitudinal mode contributions.
This framework makes use of on-shell recursion techniques to express an amplitude involving a resonant, off-shell state
in terms of a product of on-shell subamplitudes.
This not only enables the use of a form-factor representation for the off-shell hadronic matrix elements,
but also allows for the use of HQET techniques to obtain a HQ expansion.
The price for this approach is the shift of the momenta of the external states into the complex plane,
which requires a generalization of the standard spinor-helicity construction to incorporate complex momenta, 
and a generalized sense of conjugation---$C^\sharp$-conjugation---that preserves holomorphy with respect to the resonance simple pole.

In order to match QCD matrix elements with complex momenta onto an HQET, 
we developed holomorphic HQET from first principles.
This effective field theory is able to consistently incorporate complex HQ velocities into a HQ expansion of QCD,
and it is $C^\sharp$-self-adjoint with respect to the HQ velocity.
A significant result of this work was to show that $PT$-(anti)symmetric operators of a $C^\sharp$-self-adjoint theory
have an operator kernel representation that are (anti)$C^\sharp$-self-adjoint.
As a corollary, the Isgur-Wise functions of holomorphic HQET 
are holomorphic functions of the complex HQ recoil $\wh$ with real coefficients,
and thus simply the analytic continuation of the Isgur-Wise functions of real HQET.
The same argument and result applies to the perturbative corrections, that we explicitly verified at one-loop order.
This powerful result comports with the expectation that HQET should ``look the same'' whether the hadron is on-shell or off-shell,
as the dynamics of the light degrees of freedom should be agnostic to this status.

We showed that holomorphic HQET generates the usual hadron mass expansion, 
but suitably generalized with imaginary parts to incorporate an expansion of the hadron width.
When including terms to second order in the HQ power expansion, 
and accounting for HQ symmetry breaking from phase space,
we showed that this expansion yields a three parameter fit in excellent agreement with the well-measured data 
for the widths of the four $s_\ell^{\pi_\ell} = 3/2^+$ orbitally-excited $D_1$, $D_2^*$, $B_1$, and $B_2^*$ states.
We further showed that holomorphic HQET exhibits Schwinger-Dyson relations 
in terms of the hadron mass and width expansion parameters,
with respect to which the representation of the HQ matrix elements is holomorphic.

As an explicit demonstration of the use of this on-shell recursion and holomorphic HQET framework,
we considered as a first example the $\Bbar \to (\Dv(1^-) \to D\pi)l\nu$ radially-excited decay.
Computation of the on-shell subamplitudes within this framework involves a $\Dv$ polarization axis in complex-space,
that is very different to the standard helicity basis.
We derived an appropriate and convenient set of kinematic coordinates 
that results in relatively compact expressions for the complex-shifted SM subamplitudes,
which we computed explicitly.
We further computed the holomorphic HQET expansion of the Standard Model (SM) form factors to first order,
showing how to incorporate the generalized trace representation of the HQ matrix elements and the Schwinger-Dyson relations,
and we found that constructibility of the on-shell recursion relation implies that the Isgur-Wise functions should be a Laurent series in $\wh$, 
with the leading-order Isgur-Wise function have a leading term $\sim 1/\wh^4$.
Finally, we examined a toy numerical study of the marginal differential rates,
differential rate as a function of the $\Dv$ lineshape invariant mass, 
and find that this framework generates HQ resonance lineshapes that are characteristic of those seen in data.

An understanding of the underlying correspondence between the approach in this work 
and the HHChPT-based approach of Ref.~\cite{Papucci:2024qbt} is yet to be developed,
but may reveal deeper insights into the properties of HQET and/or QCD.
The on-shell recursion relation itself must fail near the pair production threshold,
as the branch cut discontinuity integral becomes important. 
Though we have used a very conservative heuristic for when this failure begins, 
further study is needed to formalize the regime of validity of the on-shell recursion approach.
The pending application of this framework to the $\Bbar \to D^{**}$ system will provide an HQET-based 
and therefore more predictive (and for the case of BSM studies, more self-consistent) 
parametrization and description of the $D^{**}$ lineshapes.
These will be crucial for future precision studies of exclusive and inclusive semileptonic $b \to cl\nu$ decays,
in which the excited state decays contribute to backgrounds or signal, respectively.

\acknowledgements
DJR is deeply thankful to Marat Freytsis for several years of collaborative exploration of other approaches 
to constructing an HQET-based description of off-shell heavy quark resonances,
and for many patient discussions.
We thank Michele Papucci and Ryan Plestid for discussions concerning their alternative heavy hadron chiral perturbation theory-based approach to this problem.
We thank all of the abovementioned, Florian Bernlochner, Zoltan Ligeti, Markus Prim, and Vladimir V.~Gligorov for their comments on the manuscript. 
CAM and DJR are supported by the Office of High Energy Physics of the U.S. Department of Energy under contract DE-AC02-05CH11231. CAM is supported also by the NSF grant PHY-2210390.

\appendix
\makeatletter
	\def\p@subsection{\thesection.}
\makeatother

\section{$T$ and $PT$-symmetry relations}
\label{sec:PTconjrel}
\subsection{Hermitian theories}
\label{sec:hermTconj}
It is instructive to first derive the $T$ and $PT$-relations for the case of a Hermitian theory.
Starting with a $0+1$-dimensional theory with just a time coordinate, $t$, the usual spectral decomposition for a in-state is
\begin{equation}
	|\psi\rangle = \int_{\mathbb{R}} \frac{d p}{2\pi}\, \psi(p) | p \rangle  = \int_{\mathbb{R}} dt\, \psi(t) |t\rangle\,,
\end{equation}
in which the time and energy eigenstate bases obey $\langle t | p \rangle = e^{i p t}$, $\langle p | q \rangle = 2\pi \delta(p-q)$, and $\langle t | t' \rangle = \delta(t-t')$,
such that the wavefunction $\psi(t) = \langle t | \psi \rangle = \int_{\mathbb{R}} \frac{d p}{2\pi}\ \psi(p) e^{i pt}$.
The corresponding out-state is generated by
\begin{equation}
	\langle \psi | = |\psi \rangle^* = \int_{\mathbb{R}} \frac{d p}{2\pi}\, \psi^*(p) \langle p |\,.
\end{equation}
Here and hereafter the $^*$ superscript should be understood to denote complex conjugation 
contextually generalized to include any additional transpositions or unitary transformations required when e.g. $|\psi\rangle$ carries nontrivial spin
or has a nontrivial (Lorentz) representation,
such as Dirac conjugation in the case of a Dirac spinor.

We are interested here in the transitions $\psi \to \phi$ mediated by an operator $\mathcal{O}$, with spectral decomposition 
\begin{equation}
	\mathcal{O} = \iint_{\mathbb{R}} \frac{d p}{2\pi} \frac{d q}{2\pi}\, \mathcal{O}(p,q) |p \rangle \langle q |\,,	
\end{equation}
in which $\mathcal{O}(p,q)$ is the \emph{operator kernel} (or density matrix).
The corresponding matrix element becomes
\begin{equation}
	\label{eqn:transmatelem}
	\langle \phi | \mathcal{O} | \psi \rangle = \iint_{\mathbb{R}} \frac{d p}{2\pi} \frac{d q}{2\pi}\, \phi^*(q) \mathcal{O}(q,p) \psi(p) \,.
\end{equation}
For definite spins (and other quantum numbers) of the external states, the kernel has a particular \emph{representation}, 
that can always be expressed as a sum over spacetime representations times form factors
(see App.~\ref{sec:PTex} for examples).
When composed with a spacetime representation of external state wavefunctions, one then obtains a form-factor representation of the matrix element of interest.
The analytic properties of these form factors are constrained by the symmetries of $\mathcal{O}$, as we now show.

Before proceeding, however, it is important to observe that equations of motion for the wavefunctions, 
of the generic form $K_\psi(p) \psi(p) = m_\psi \psi(p)$, 
allows a spectral decomposition $\langle \phi | \mathcal{O} | \psi \rangle = \iint_{\mathbb{R}} \frac{d p}{2\pi} \frac{d q}{2\pi}\, \phi^*(q) K^*_\phi(q)/m_\phi\, \mathcal{O}(q,p)\, K_\psi(p)/m_\psi  \psi(p)$.
As explained further below, 
this may introduce ambiguities into the identification of a ``true representation'' of the operator kernel $\mathcal{O}(q,p)$ 
versus a representation of $K^*_\phi(q)\mathcal{O}(q,p)K_\psi(p)$,
which may have different or ``wrong''  symmetry transformation properties.

As is very well-known, if $\psi(t)$ is a solution to the equation of motion $i \partial_t \psi(t) = H \psi(t)$, and the Hamiltonian $H$ is real (in the sense $H^* = H$),
then it follows from conjugation of the equation of motion that $\psi^*(-t) = \int_{\mathbb{R}} \frac{dp}{2\pi} \psi^*(p) e^{i pt}$ is also a solution.
Hence the theory features a $T$-conjugate in-state\,\footnote{We denote action by a discrete symmetry via a subscript. 
An equivalent notation would be $|\psi_T \rangle = |T\psi \rangle$, and for operators $\mathcal{O}_T = T \mathcal{O}T^{-1}$.}
\begin{equation}
	\label{eqn:realTconjstate}
	|\psi_T \rangle = \int_{\mathbb{R}} \frac{d p}{2\pi}\, \psi^*(p) | p \rangle\,,
\end{equation}
in which the wavefunction is conjugated. 
It is straightforward to then show that for two states $|\psi \rangle$ and $|\phi\rangle$, $\langle \phi_T | \psi_T \rangle = \langle \psi | \phi \rangle^*$.
For an operator $\mathcal{O}$, one similarly defines a $T$-conjugated operator $\mathcal{O}_T$ in which the operator kernel $\mathcal{O}(p,q)$ is conjugated, so that
\begin{equation}
	\label{eqn:realTdefop}
	\mathcal{O}_T = \iint_{\mathbb{R}} \frac{d p}{2\pi} \frac{d q}{2\pi}\, \mathcal{O}^*(p,q) |p \rangle \langle q |\,.
\end{equation}
It is also straightforward to then show that $\langle \phi_T |\mathcal{O}_T |\psi_T \rangle = \langle \phi |\mathcal{O} | \psi \rangle^*$.

Eq.~\eqref{eqn:realTdefop} requires that application of $T$-conjugation to any true representation of an operator kernel must yield its conjugate 
(under the appropriately generalized sense of conjugation),
i.e.  
\begin{equation}
	\label{eqn:realTtransop}
	\mathcal{O}_T(p,q) = \mathcal{O}^*(p,q)\,. 
\end{equation}
By contrast, it may not be true that $\big[K^*_\phi(q)\mathcal{O}(p,q) K_\psi(p)\big]_T = [K^*_\phi(q)\mathcal{O}(p,q) K_\psi(p)]^*$.
Thus Eq.~\eqref{eqn:realTtransop} provides a nontrivial necessary condition for $\mathcal{O}(p,q)$ to be a true representation of an operator kernel,
to which $T$-symmetry relations may then be applied.
In particular, in 0+1 dimensions, from Eq.~\eqref{eqn:realTdefop} it follows that $\mathcal{O}$ is $T$-(anti)symmetric if and only if the kernel is (anti)self adjoint:
\begin{equation}
	\label{eqn:realTsymrel}
	0+1:\qquad \mathcal{O}_T = \pm \mathcal{O}  \quad \Leftrightarrow \quad \mathcal{O}^*(p,q) = \pm\mathcal{O}(p,q)\,.
\end{equation}
We will see, via examples and discussion in App.~\ref{sec:PTex} and~\ref{sec:hqetPTex}, that the analytic properties 
of the form factors and Isgur-Wise functions derive immediately from relations of the type in Eq.~\eqref{eqn:realTsymrel}.

Turning now to $3+1$ dimensions, the relations for the spatial and momentum eigenstate bases generalize to 
$\langle x | p \rangle = e^{i p \cdot x}$, $\langle p | q \rangle = (2\pi)^4 \delta^4(p-q)$, and $\langle x | x' \rangle = \delta^4(x-x')$
in which $p = (p^0, \vec{p})$ and $x = (t,\vec{x})$.
Reality of the Hamiltonian ensures that $\psi^*(-t,\vec{x})$ is a solution to the equation of motion, so that the in-state and its $T$-conjugate have spectra
\begin{equation}
	|\psi\rangle = \int_{\mathbb{R}^{3+1}} \frac{d^4 p}{(2\pi)^4}\, \psi(p) | p \rangle\,, \qquad |\psi_T\rangle = \int_{\mathbb{R}^{3+1}} \frac{d^4 p}{(2\pi)^4}\, [-\psi^*(\bar{p})] | p \rangle\,,
\end{equation}
in which $\bar{p} = (p^0, -\vec{p})$ is the spatial parity-reversed momentum,
and the overall sign arises from reversing the direction of spatial integration.
Similarly, the $T$-conjugate operator
\begin{equation}
	\mathcal{O}_T = \iint_{\mathbb{R}^{3+1}} \frac{d^4 p}{(2\pi)^4} \frac{d^4q}{(2\pi)^4}\, \mathcal{O}^*(\bar{p},\bar{q}) |p \rangle \langle q |\,.
\end{equation}
In $3+1$ dimensions the parity reversal of the momentum in $\mathcal{O}_T$ means that Eq.~\eqref{eqn:realTsymrel} no longer holds for $T$-symmetric operators.
Instead, one must consider the $PT$-conjugate state and operators, because the additional spatial parity reversal restores $\bar{p} \to p$, such that 
\begin{equation}
	\label{eqn:realPTstates}
	|\psi_{PT}\rangle = \int_{\mathbb{R}^{3+1}} \frac{d^4 p}{(2\pi)^4}\, \psi^*(p) | p \rangle\,, \qquad \mathcal{O}_{PT} = \iint_{\mathbb{R}^{3+1}} \frac{d^4 p}{(2\pi)^4} \frac{d^4q}{(2\pi)^4}\, \mathcal{O}^*(p,q) |p \rangle \langle q |\,.
\end{equation}
It follows that a true representation of an operator kernel must satisfy, analogously to Eq.~\eqref{eqn:realTtransop},
\begin{equation}
	\label{eqn:realPTtransop}
	\mathcal{O}_{PT}(p,q)  = \mathcal{O}^*(p,q)\,.
\end{equation}
Given a representation of the matrix element and a candidate representation of the operator kernel, 
Eq.~\eqref{eqn:realPTtransop} provides a necessary condition for the candidate representation to be a true one.
Finally, from Eq.~\eqref{eqn:realPTstates} is it clear that  $\mathcal{O}$ is $PT$-(anti)symmetric if and only if
the kernel is (anti)self-adjoint:
\begin{equation}
	\label{eqn:realPTsymop}
	3+1:\qquad \mathcal{O}_{PT} = \pm \mathcal{O} \quad \Leftrightarrow \quad \mathcal{O}^*(p,q) = \pm\mathcal{O}(p,q)\,.
\end{equation}
Thus, as we will see also in the following examples, it is $PT$-symmetry that determines the analytic properties of the form factors in $3+1$ dimensions.

\subsection{Examples}
\label{sec:PTex}

To illuminate the application of Eqs~\eqref{eqn:realPTtransop} and~\eqref{eqn:realPTsymop}, 
consider as an example the matrix element for a $J^P = 1/2^+ \to 1/2^+$ hadronic transition mediated by the vector current operator $\mathcal{O}^\mu = \bar{q} \g^\mu q$.
This operator is $PT$-symmetric, and the matrix element has a standard form-factor representation 
\begin{equation}
	\label{eqn:exvectorff}
	\langle H(p_2) | \mathcal{O}^\mu | H(p_1) \rangle = \bar{u}(p_2) \big[ f_1(q^2) \g^\mu + f_2(q^2) (p_1 + p_2)^\mu + f_3(q^2) q^\mu \big] u(p_1)\,,
\end{equation}
with $q = p_1 - p_2$.\footnote{The fact that there are three form factors follows from angular momentum and parity conservation, 
see e.g. Ref.~\cite{Papucci:2021pmj} for how such a counting proceeds.}
To see that the form factors $f_{1,2,3}$ are real via the $PT$-symmetry relation~\eqref{eqn:realPTsymop}, 
we must first determine a true representation of the operator kernel.
Taking the candidate representation of the operator kernel to simply be 
\begin{equation}
	\label{eqn:exrightrep}
	\mathcal{O}^\mu(p_2,p_1) = f_1(q^2) \g^\mu + f_2(q^2) (p_1 + p_2)^\mu +  f_3(q^2) q^\mu\,,
\end{equation}
and applying $PT$-conjugation directly,\footnote{Note 
momenta are $PT$ even, because $\partial_\mu$ is necessarily $PT$ odd and $PT$ is antilinear, so then $i \partial_\mu$ is $PT$ even.} 
one finds $\mathcal{O}_{PT}^\mu(p_2,p_1) = f^*_1(q^2) \g^\mu + f^*_2(q^2) (p_1 + p_2)^\mu +  f^*_3(q^2) q^\mu = \genbar{\mathcal{O}}^\mu(p_2,p_1)$,
the Dirac conjugate (i.e. $\g^0 \mathcal{O}^{\mu\dagger} \g^0$),
which is the appropriate generalization of the conjugation used in App.~\ref{sec:hermTconj} for Dirac spinors.
Thus the representation~\eqref{eqn:exrightrep} satisfies the necessary condition.~\eqref{eqn:realPTtransop}. 
Since $\mathcal{O}^\mu$ is $PT$-symmetric, the $PT$-symmetry relation~\eqref{eqn:realPTsymop} 
further requires that $\genbar{\mathcal{O}}^\mu(p_2,p_1) = \mathcal{O}^\mu(p_2,p_1)$
and thus, as each term in Eq.~\eqref{eqn:exrightrep} is Dirac self-adjoint up to its form factor, 
it immediately follows that the form factors must be real.

To see, by contrast, the apparent self-inconsistencies that can arise from equations of motion, 
observe that in Eq.~\eqref{eqn:exvectorff}, application of the spinor equation of motion $\slashed{p}_i u(p_i) = m_i u(p_i)$
allows one to pull out $\slashed{p}_{1,2}/m_{1,2}$ factors and thereby rewrite the matrix element representation into another well-known form
\begin{equation}
	\label{eqn:exvectorff2}
	\langle H(p_2) | \mathcal{O}^\mu | H(p_1) \rangle = \bar{u}(p_2) \big[ F_1(q^2) \g^\mu + i \sigma^{\mn} q_\nu F_2(q^2)  +   F_3(q^2)q^\mu\big] u(p_1)\,.
\end{equation}
The $F_{1,2,3}$ are real linear combinations of $f_{1,2,3}$ and thus still real.
If, however, one were to naively take the candidate kernel representation to be 
\begin{equation}
	\label{eqn:exwrongrep}
	\mathcal{O}^\mu(p_2,p_1) = F_1(q^2) \g^\mu + i \sigma^{\mn} q_\nu F_2(q^2)  +  F_3(q^2)q^\mu\,,
\end{equation}
and simply apply the $PT$-symmetry relation~\eqref{eqn:realPTsymop} to it, one would deduce that $F_2$ must be imaginary.
The apparent contradiction here is encapsulated by the failure of Eq.~\eqref{eqn:realPTtransop}: 
because $i \sigma^{\mn} q_\nu$ is $PT$-even but antisymmetric under Dirac conjugation, 
one has for the candidate representation~\eqref{eqn:exwrongrep} $\mathcal{O}_{PT}^\mu(p_2,p_1)\not= \genbar{\mathcal{O}}^\mu(p_2,p_1)$.
Thus Eq.~\eqref{eqn:exwrongrep} cannot be a true representation of the kernel, and one cannot apply Eq.~\eqref{eqn:realPTsymop}.

As one more example, consider the matrix element for a $J^P = 0^- \to 1^-$ hadronic transition, 
again mediated by $\mathcal{O}^{\mu}$.
In this case, the transition matrix element has a standard form-factor representation
\begin{equation}
	\langle H_\lambda(p_2) | \mathcal{O}^\mu | H (p_1) \rangle = i g(q^2) \varepsilon^{\mu\alpha\beta\gamma}  {\epsilon^*_{\lambda}}_\alpha {p_2}_\beta {p_1}_\gamma\,.
\end{equation}
It is key to note that the wavefunction of the $1^-$ state has representation $\psi_\lambda = i{\epsilon_\lambda^*}_\alpha$,
which satisfies the positive normalization condition $\psi_\lambda \cdot \psi_{\bar{\kappa}} = -\epsilon^*_\lambda \cdot \epsilon^*_{\bar{\kappa}} = +1$.
Thus one would deduce the candidate operator kernel $ \mathcal{O}^{\mu\alpha}(p_2, p_1) = g(q^2) \varepsilon^{\mu\alpha\beta\gamma} {p_2}_\beta {p_1}_\gamma$,
i.e. without the $i$ prefactor.
The Levi-Civita tensor is $PT$-even, so that $\mathcal{O}_{PT}^{\mu\alpha}(p_2, p_1)  =  \mathcal{O}^{*\mu}(p_2,p_1)$ satisfying Eq.~\eqref{eqn:realPTtransop},
and it follows from Eq.~\eqref{eqn:realPTsymop} that $g(q^2)$ must be real.

\subsection{Application to HQET}
\label{sec:hqetPTex}
Let us now consider a generic real HQET matrix element for a $\Hb \to \Hc$ transition, mediated by $\cbar \Gamma b$, 
with the insertion of charmed and/or beauty HQ operators (contact operators, operator products, or a mixture of both).
One appends velocity labels to the external HQ states and operators, 
such that e.g. $|\psi^v\rangle = \int_{\mathbb{R}^{3+1}} \frac{d ^4k}{(2\pi)^4}\, \psi(v, k) | k\rangle$.
The idempotency of the HQ projectors guarantees $\vslash \psi(v,k) = \psi(v,k)$. 
Combined with HQ symmetry---independence from the choice of $\Gamma$ in the current---this requires
an HQ operator kernel to depend on both $v$ and $v'$, even if it acts only on one side of the current.
The matrix element then has a spacetime representation in terms of a generalized trace, generically of the form
\begin{equation}
	\label{eqn:trrep}
	\langle \Hc^{v'} | \mathcal{O}_c^{v'} \Gamma \mathcal{O}_b^v | \Hb^{\vb} \rangle = \LamB^n \Tr\big[ \Hbar_c(v') \Gamma \Hb(v) \Xi_\mathcal{O}(v',v)\big]\,,
\end{equation}
where $\LamB^n$ is a real prefactor of appropriate mass dimension.
Here $\Xi_\mathcal{O}(v',v)$ is the representation of the HQ operator kernel, taking the generic form
\begin{equation}
	\Xi_\mathcal{O}(v',v) = \sum_i \mathcal{T}_i(v',v) W_i(w)\,,
\end{equation}	
where the functions $W_i(w)$ are Isgur-Wise functions 
and $\mathcal{T}_i$ are a basis of tensors of Dirac matrices and HQ velocities, as allowed by HQ symmetry (and transversity where applicable).
$\Hbar_c(v')$ and $\Hb(v)$ are spacetime representations of the wavefunctions of the in-~and out-states.
Note HQ flavor violation in HQET is captured fully by the HQ masses alone, 
so that the $b$ and $c$ labels in the operators in Eq.~\eqref{eqn:trrep} are included only as a notational convenience: 
the two HQ operators involved are parametrized purely by $v$ and $v'$.

As real HQET is a Hermitian theory,
we may apply the same $PT$ analysis in App.~\ref{sec:hermTconj} to HQ states and operators.
Thus, the $PT$-transformation property~\eqref{eqn:realPTtransop} requires that $[\mathcal{T}_i(v',v)]_{PT}= \genbar{\mathcal{T}}_{i}(v',v)$ 
under the appropriate generalization of conjugation in App.~\ref{sec:hermTconj} (usually Dirac or anti-Dirac conjugation), here denoted by a bar.
The $PT$-(anti)symmetry relation~\eqref{eqn:realPTsymop} becomes 
\begin{equation}
	\label{eqn:realhqetPTsymop}
	\mathcal{O}_{PT} = \pm \mathcal{O} \quad \Leftrightarrow \quad \overline{\Xi}_\mathcal{O}(v',v) = \pm\Xi_\mathcal{O}(v',v)\,,
\end{equation}
and provided one chooses the $\mathcal{T}_i$ such that $\mathcal{T}_i(v',v) = \pm\genbar{\mathcal{T}}_{i}(v',v)$, 
then $W(w) = W(w)$, i.e. the Isgur-Wise functions are real.

To take one example, consider the first-order current correction to the $B \to \Dx$ HQET matrix element, mediated by $\cbar \Gamma b$,
\begin{equation}
	\big\langle \Hc^{v'} \big| \cbvp \Gamma i \overrightarrow{D}_\mu \bv[v] \big| \Hb^{v} \big\rangle  = -\LamB \Tr[\Hbar_c(v') \Gamma H_b(v) \Xi_\mu(v',v)]\,,
\end{equation}
using the same notation as defined in Ref.~\cite{Bernlochner:2022ywh} (see also Sec.~\ref{sec:hqetBDv}).
The candidate kernel representation of the HQ operator $i \overrightarrow{D}_\mu$, which is $PT$ symmetric, is
\begin{equation}
	\Xi^\mu(v',v) =  \xi_+(w) (v+v')^\mu + \xi_-(w)(v-v')^\mu - \xi_3(w) \g^\mu\,.
\end{equation}
This is deduced by writing down the most general set of terms that are linearly independent under HQ equations of motion.
This representation satisfies the necessary $PT$ transformation~\eqref{eqn:realPTtransop} under Dirac conjugation. 
The $PT$ symmetry relation~\eqref{eqn:realPTsymop} then ensures $\overline{\Xi}^\mu(v,v') = \Xi^\mu(v,v')$, 
and it follows that the Isgur-Wise functions $\xi_{\pm,3}$ are all real.

As a different (and pathological) example, consider the second-order current correction in the same process,
\begin{equation}
	\label{eqn:exnnlocc}
	\big\langle \Hc^{v'} \big| \cbvp \Gamma  \frac{g}{2} G^{\ab} \bv[v] \big| \Hb^{v} \big\rangle  = \LamB^2\, \Tr[ \Hbar_c(v')  \Gamma H_b(v) \Phi^{\ab}(v,v')]\,,
\end{equation}
with candidate representation
\begin{equation}
	\Phi_{\ab}(v',v) = i\varphi_1(w) v_{[\alpha}v'_{\beta]}  - i\varphi_2(w) (v-v')_{[\alpha} \g_{\beta]}  + 2\varphi_3(w) \sigma_{\ab}\,.
\end{equation}
However, $[\Phi^{\ab}(v',v)]_{PT} \not= \overline{\Phi}^{\ab}(v',v)$ because of the $\varphi_3(w)$ term, 
so that $\Phi_{\ab}(v',v)$ cannot be a true representation of the operator kernel.
Instead, it follows from the (exterior) equation of motion $H(v)\vslash = -H(v)$ that
\begin{equation}
	\label{eqn:sigmaid}
	H_b(v)\big[\vslash[v]' \sigma_{\ab} \vslash[v] - \vslash[v] \sigma_{\ab} \vslash[v]'\big]\Hbar_c(v') = H_b(v)\big[4 i v'_{[\alpha}v_{\beta]}  - 2i (v-v')_{[\alpha} \g_{\beta]} + 2(w-1) \sigma_{\ab}\big]\Hbar_c(v')\,,
\end{equation}
which allows one to rewrite Eq.~\eqref{eqn:exnnlocc} with the candidate kernel representation 
\begin{equation}
	\Phi_{\ab}(v',v) = i\phi_1(w) v_{[\alpha}v'_{\beta]}  - i \phi_2(w) (v-v')_{[\alpha} \g_{\beta]}  + \phi_3(w)\big[ \vslash[v]' \sigma_{\ab} \vslash[v] - \vslash[v] \sigma_{\ab} \vslash[v]'\big]\,.
\end{equation}
Here $\phi_{1,2,3}$ are real linear combinations of $\varphi_{1,2,3}$.
The $\phi_3$ term is antisymmetric under Dirac conjugation, so that now $[\Phi^{\ab}(v',v)]_{PT} = \overline{\Phi}^{\ab}(v',v)$
as required by Eq.~\eqref{eqn:realPTtransop}.
As $G_{\ab}$ is $PT$ odd then it must be from Eq.~\eqref{eqn:realPTsymop} that $\Phi^{\ab}(v',v) = -\overline{\Phi}^{\ab}(v',v)$, 
from which it follows that $\phi_{1,2,3}$ are real, and thus so are $\varphi_{1,2,3}$.

\subsection{$C^\sharp$-self-adjoint theories}
\label{sec:cshpTconj}
Now we turn to a $C^\sharp$-self-adjoint theory, with $C^\sharp$-conjugation as defined in Sec.~\ref{sec:sphlgc}.
It's again useful to start with a $0+1$-dimensional theory with just a time coordinate, $t$.
We formally construct such a theory on a fixed horizontal contour $C_\eta = (-\infty + i\eta, \infty + i\eta)$ in complex energy space,
such that the spectra of all states and operators lie on $C_\eta$.
That is, the spectral decomposition for an in-state is defined via
\begin{equation}
	\label{eqn:statedef}
	|\psi\rangle = \int_{C_\eta} \frac{d p}{2\pi}\, \psi(p) | p \rangle  = \int_{\mathbb{R}} dt\, \psi(t) |t\rangle\,,
\end{equation}
for any $|\psi\rangle$.
This derives from the spectral decompositions of a Hermitian theory via a complex-momentum shift $p \to p + \hat{z} \zeta$ per Eq.~\eqref{eqn:zshift},
that uniformly shifts the spectra of all the states and operators belonging to the Hermitian theory by a constant imaginary term.
The time and energy eigenstate bases in Eq.~\eqref{eqn:statedef} still obey $\langle t | p \rangle = e^{i p t}$ and $\langle t | t' \rangle = \delta(t-t')$, 
and  $\langle p | q \rangle = 2\pi \delta(p-q)$ so long as both $p$ and $q \in C_\eta$.
By construction, the wavefunction $\psi(p)$ is holomorphic with respect to $p$.

An out-state must have its momentum spectrum on the same contour, such that
\begin{equation}
	\langle \psi | = 	\int_{C_\eta} \frac{d p}{2\pi}\, \psi^\sharp(p) \langle p |\,,
\end{equation} 
in which the $C^\sharp$-conjugation preserves holomorphy with respect to $p$:
$C^{\sharp(p)}$-conjugation in the notation of Sec.~\ref{sec:sphlgc}.
Thus we may also write $\langle \psi | = |\psi \rangle^\sharp$, 
in the sense that the holomorphy on the contour $C_\eta$ itself is preserved. 
We refer to this state-wise transformation as $C^{\sharp(C_\eta)}$-conjugation.
Just as in App.~\ref{sec:hermTconj}, here and hereafter the $\sharp$ superscript should be understood to denote $C^\sharp$-conjugation 
contextually generalized to include any additional transpositions or unitary transformations required when e.g. $|\psi\rangle$ carries nontrivial spin
or has a nontrivial (Lorentz) representation.

Similarly to the in-~and out-states, the kernel $\mathcal{O}(p,q)$ of an operator $\mathcal{O}$ is holomorphic with respect to $p$, $q \in C_\eta$, 
and its spectral decomposition
\begin{equation}
	\mathcal{O} = \iint_{C_\eta} \frac{d p}{2\pi} \frac{d q}{2\pi}\, \mathcal{O}(p,q) |p \rangle \langle q |\,.
\end{equation}
A theory being $C^\sharp$-self-adjoint corresponds to its Hamiltonian obeying $H^\sharp = H$, 
so that its kernel is $C^\sharp$-self-adjoint: $H^\sharp(p,q) = H(p,q)$.
Put in other words, the complex spectrum of the Hamiltonian is invariant under $C^{\sharp(C_\eta)}$-conjugation.

The $C^{\sharp(C_\eta)}$-conjugate of the equation of motion $i \partial_t \psi(t) = H \psi(t)$ then implies that if $\psi(t)$ is a solution, so is $\psi^\sharp(-t)$.
Thus in direct analogy to the Hermitian case in App.~\ref{sec:hermTconj}, one may define $T$-conjugate in-~and out-states
\begin{equation}
	|\psi_T\rangle = \int_{C_\eta} \frac{d p}{2\pi}\, \psi^\sharp(p) | p \rangle\,,\qquad \langle \psi_T | = \int_{C_\eta} \frac{d p}{2\pi}\, \psi(p) \langle p |\,,
\end{equation}
whose spectra lie on $C_\eta$.
Thus by construction, the antilinear $T$-transformation does not act on the imaginary parts of the complex momenta on $C_\eta$.
For any operator $\mathcal{O}$, its $T$-conjugate operator
\begin{equation}
	\mathcal{O}_T = \iint_{C_\eta} \frac{d p}{2\pi} \frac{d q}{2\pi}\, \mathcal{O}^\sharp(p,q) |p \rangle \langle q |\,.
\end{equation}
That is, one simply replaces the conjugation in App.~\ref{sec:hermTconj} everywhere with $C^{\sharp(C_\eta)}$-conjugation.
It immediately follows by the same arguments in App.~\ref{sec:hermTconj} leading to Eq.~\eqref{eqn:realTtransop} that 
a true representation of an operator kernel must satisfy $\mathcal{O}_T(p,q) =  \mathcal{O}^\sharp(p,q)$,
and further, $\mathcal{O}$ is $T$-(anti)symmetric if and only if the kernel is (anti)$C^\sharp$-self-adjoint:
\begin{equation}
	\label{eqn:Tsymrel}
	0+1:\qquad \mathcal{O}_T = \pm \mathcal{O} \quad \Leftrightarrow \quad \mathcal{O}^\sharp(p,q) = \pm\mathcal{O}(p,q)\,.
\end{equation}

Turning to consider a $3+1$ dimensional theory, we generalize the dimensionality of the imaginary momentum shift, 
such that the contours $C_{\eta_\mu} = (-\infty + i\eta_\mu, \infty + i\eta_\mu)$ and $C_\eta = C_{\eta^0} \times C_{\vec{\eta}}$.
The in-~and out-states are then defined by 
\begin{equation}
	\label{eqn:states31}
	|\psi\rangle = \int_{C_\eta} \frac{d^4 p}{(2\pi)^4}\, \psi(p) | p \rangle\,, \qquad \langle \psi | = \int_{C_\eta} \frac{d^4 p}{(2\pi)^4}\, \psi^\sharp(p) \langle p |\,,
\end{equation}
in which $\psi(p)$ and $\psi^\sharp(p)$ are both holomorphic with respect to $p \in C_\eta$,
so that $|\psi\rangle^\sharp = \langle \psi |$ under $C^{\sharp(C_\eta)}$-conjugation.
Under these definitions, applying $C^{\sharp(C_\eta)}$-conjugation to the equation of motion $i\partial_t \psi(x) = H(x) \psi(x)$, 
then the wavefunction of the conjugate solution
\begin{equation}
	\psi^\sharp(-t,\vec{x}) = \int_{C_\eta} \frac{d^4 p}{(2\pi)^4} \psi^\sharp(p) e^{i p^0 t + i \vec{p} \cdot \vec{x}} = \int_{C_{\eta^0} \times C^*_{\vec{\eta}}} \frac{d^4 p}{(2\pi)^4}  [-\psi^\sharp(\bar{p})] e^{i p \cdot x}\,,
\end{equation}
in which we have used the observation that $\vec{p} \to -\vec{p}$ is the same as complex conjugating the spatial integration contour, $C_{\vec{\eta}}$,
and reversing the direction of integration (recall $\bar{p} = (p^0,-\vec{p})$).
Hence the state $|\psi_T\rangle$ has a spectrum on the spatially-conjugate contour $C_{\eta^0} \times C^*_{\vec{\eta}}$,
as does the $T$-conjugate operator $\mathcal{O}_T$. 
This means that Eq.~\eqref{eqn:Tsymrel} no longer holds for $T$-symmetric operators in $3+1$ dimensions in $C^\sharp$-self-adjoint theories.

Instead, as in App.~\ref{sec:hermTconj}, for a $3+1$ dimensional theory we must consider the $PT$-conjugate state,
which has wavefunction $\psi^\sharp(-t,-\vec{x}) = \int_{C_\eta} \frac{d^4p}{(2\pi)^4} \psi^\sharp(p)e^{i p \cdot x}$,
noting the additional $P$-conjugation restores the spectrum to the $C_\eta$ contour.
That is, 
\begin{equation}
	\label{eqn:Ptstates31}
	|\psi_{PT}\rangle = \int_{C_\eta} \frac{d^4 p}{(2\pi)^4}\, \psi^\sharp(p) | p \rangle\,, \qquad \langle \psi_{PT} | = \int_{C_\eta} \frac{d^4 p}{(2\pi)^4}\, \psi (p) \langle p |\,,
\end{equation}
and similarly for the $PT$-conjugate of an operator
\begin{equation}
	\mathcal{O}_{PT} = \iint_{C_\eta} \frac{d^4 p}{(2\pi)^4} \frac{d^4 q}{(2\pi)^4}\, \mathcal{O}^\sharp(p,q) |p \rangle \langle q |\,,
\end{equation}
again with the understanding that the antilinear $PT$ operator does not act on the imaginary parts of the complex momenta on $C_\eta$.
As in App.~\ref{sec:hermTconj}, a true representation of the operator kernel must satisfy
\begin{equation}
	\label{eqn:PTtransop}
	 \mathcal{O}_{PT}(p,q) = \mathcal{O}^\sharp(p,q)\,.
\end{equation}
and further, $\mathcal{O}$ is $PT$-(anti)symmetric if and only if its kernel is (anti)$C^\sharp$-self adjoint:
\begin{equation}
	\label{eqn:PTsymop}
	3+1:\qquad \mathcal{O}_{PT} = \pm \mathcal{O} \quad \Leftrightarrow \quad \mathcal{O}^\sharp(p,q) = \pm \mathcal{O}(p,q)\,.
\end{equation}
This relation determines the analytic properties of the form factors in $3+1$ dimensions for a $C^\sharp$-self-adjoint theory.

As holomorphic HQET is such a $C^\sharp$-self-adjoint theory with respect to the HQ velocities, 
we may apply the same $PT$ analysis to holomorphic HQ states and operators.
With reference to the generic HQET matrix element in Eq.~\eqref{eqn:trrep}, but with complex velocities,
the $PT$-(anti)symmetry relation~\eqref{eqn:PTsymop} becomes
\begin{equation}
	\label{eqn:hqetPTsymop}
	\mathcal{O}_{PT} = \pm \mathcal{O} \quad \Leftrightarrow \quad \widetilde{\Xi}_\mathcal{O}(\vc,\vb) = \pm \Xi_\mathcal{O}(\vc,\vb)\,.
\end{equation}
Here the tilde denotes the appropriate generalization of the conjugation in Eq.~\eqref{eqn:PTsymop}  
(usually $C^\sharp$-Dirac or anti-$C^\sharp$-Dirac conjugation as defined in Sec.~\ref{sec:hqpe}).
Just as in App.~\ref{sec:hqetPTex},
provided one chooses the $\mathcal{T}_i$ tensors in $\Xi_\mathcal{O}(\vc,\vb)$ 
such that $\mathcal{T}_i(\vc,\vb) = \pm\widetilde{\mathcal{T}}_{i}(\vc,\vb)$, 
then this guarantees that the Isgur-Wise functions obey $W_i^\sharp(\wh) = W_i(\wh)$, under $C^{\sharp(\wh)}$-conjugation. 
They are therefore  holomorphic functions of $\wh = \vb \cdot \vc \in \mathbb{C}$ with real coefficients.

\section{$\mathcal{O}(\aS)$ matching and corrections}
\label{sec:mloop}
The perturbative functions $C_{\Gamma_j}$ are generated via the master matching relation (see e.g. Ref.~\cite{Manohar:2000dt})
\begin{align}
	C_{\Gamma_1} & = \frac{1}{2} \Big[ R_1^{(c)} - R_1^{h} \Big] + \frac{1}{2} \Big[ R_1^{(b)} - R_1^{h} \Big]  + V_1^{\Gamma,1} - V_{\text{eff}}^\Gamma\,, \nn \\
	C_{\Gamma_{j\ge 2}} & = V_1^{\Gamma,j}\,, \label{eqn:masteras}
\end{align}
in which $R_1^{(Q)}$ and $R_1^h$ are the 1-loop QCD and HQET field strength renormalizations of the on-shell quark and heavy quark field, respectively, 
while $V_1^{\Gamma,j}$ and $V_{\text{eff}}^{\Gamma}$ are generated by the 1-loop QCD and HQET vertex corrections 
(by HQ symmetry, there are no HQET vertex corrections involving local operators containing $\Gamma_{j\ge2}$).
They correspond to the truncated diagrams via
\begin{subequations}
\label{eqn:asdiag}
\begin{align}
	-i \Sigma_2^{(c,b)} & = \begin{aligned} \includegraphics[width = 0.3\linewidth]{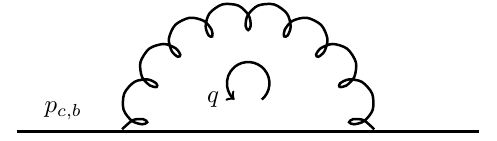} \end{aligned} & i \aS R_1^{h} & = \begin{aligned} \includegraphics[width = 0.3\linewidth]{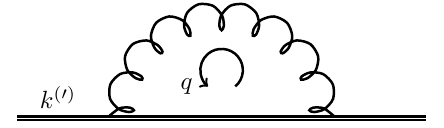} \end{aligned} \\
	-i \aS \Gamma_j V_1^{\Gamma,j} & = \begin{aligned} \includegraphics[width = 0.3\linewidth]{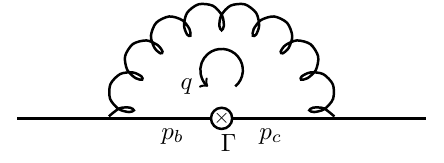} \end{aligned} & -i \aS \Gamma V_{\text{eff}}^\Gamma & = \begin{aligned} \includegraphics[width = 0.3\linewidth]{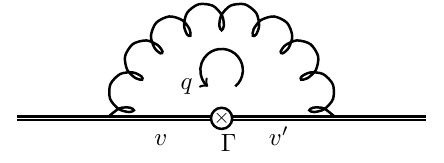}\end{aligned}
\end{align}
\end{subequations}
in which the double line denotes a HQ line, and $\aS R_1^{(Q)}$ is related to the $\mathcal{O}(\aS)$ quark 2-point function $\Sigma_2^{(Q)}(p) = A(p^2) m_Q + B(p^2) \slashed{p}$ via
$\aS R_1^{(Q)} = B(m_Q^2) + 2m_Q^2 [A'(m_Q^2) + B'(m_Q^2)]$.

As a first consideration, evaluation of the $2$-~and $3$-point 1-loop amplitudes~\eqref{eqn:asdiag}
relies on the Feynman parameter identities
\begin{align}
	\frac{1}{z_1^r z_2^s} & = \frac{\Gamma(r+s)}{\Gamma(r)\Gamma(s)} \int_0^1 \!dx_1 \int_0^1 \!dx_2\, \frac{\delta(1 - x_1 - x_2)}{(x_1 z_1 + x_2 z_2)^{r+s}}\,, \\
	\frac{1}{z_1^r z_2^s} & = \frac{2^s \Gamma(r+s)}{\Gamma(r)\Gamma(s)} \int_0^\infty \!d\lambda\, \frac{\lambda^{s-1}}{(z_1 + 2\lambda z_2)^{r+s}}\,.
\end{align}
Both relations hold in general for $z_1$, $z_2 \in \mathbb{C}$ and integers $r$, $s \ge 1$.
Application of these two identities to the complex-momentum loop amplitudes in Eqs.~\eqref{eqn:asdiag} leads to loop integrals of the generic form, 
\begin{equation}
	\label{eqn:unshftint}	
	\int_{\mathbb{R}^{n}} \frac{d^n q}{(2\pi)^n} \frac{X (\slashed{q} + \ldots)^{\alpha_1} Y (\slashed{q} + \ldots)^{\alpha_2} Z}{(q^2 + 2 q \cdot p + W^2)^\beta}\,, 
	\qquad \beta \ge \frac{3}{2} + \frac{\alpha_1 + \alpha_2}{2}\,,
\end{equation}
with numerator and denominator powers $\alpha_{1,2} (= 0$ or $1$) and $\beta$ positive integers
(where applicable, additional infrared regulators may be included for the gluon). 
Both $p$ and $W^2$ in the denominator can be complex.

In the standard approach, one completes the square of the denominator under the shift $q \to q - p$,
leading to standard master integrals of the form
\begin{equation}
	\label{eqn:masloop}
	\int_{\mathbb{R}^{n}} \frac{d^n q}{(2\pi)^n} \frac{(q^2)^\alpha}{(q^2 - M^2)^\beta}\,,
\end{equation}
which can be evaluated via dimensional regularization taking $n = 4 - \epsilon$.
When matching QCD matrix elements with complex momenta onto holomorphic HQET, because 
the momentum $p$ in the integral~\eqref{eqn:unshftint} is complex, one encounters an apparent difficulty:
the shift $q \to q - p$ moves the $n$-dimensional integration domain from $\mathbb{R}^{n}$ 
to a complex $n$-dimensional manifold analogous to the uniformly shifted contour $C_\eta$ defined in App.~\ref{sec:cshpTconj},
where here $\eta_\mu = \text{Im}(p_\mu)$.

The resolution of this difficulty lies in the observation that the degree $\beta$ of the denominator in Eq.~\eqref{eqn:unshftint} 
remains large enough to ensure that the integrand has no simple pole. 
One may therefore complete the square in the denominator via $ q \to q - p$, and then shift the integration contour back to the real domain $\mathbb{R}^n$.
The complex-momentum loop amplitudes can thus all be expressed in terms of the standard master loop integrals~\eqref{eqn:masloop}, 
whose evaluation under dimensional regularization holds even for complex $M^2$.
(Even in the regular matching of QCD onto real HQET, $M^2$ may be complex because of the Feynman prescription.)

As the Feynman parameter integrals evaluated subsequently to the master loop integrals~\eqref{eqn:masloop} 
are also agnostic to the complexity of the integrand,
the standard results follow for $R_1^{(Q)}$, $R_1^h$, $V_1^{\Gamma,j}$ 
and $V_{\text{eff}}^\Gamma$ in terms of $m_c/m_b$ and $w$,
with the exception that $w \to \wh$ is now analytically continued to the complex plane (recall $m_{c,b}$ remain real).
Thus one obtains from Eq.~\eqref{eqn:masteras} that the perturbative corrections are simply $C_{\Gamma_j}(\wh, m_c/m_b)$, 
the analytic continuation of the standard results for the perturbative functions in the standard matching of QCD onto real HQET with real momenta and velocities.

\addtocontents{toc}{\protect\vspace*{10pt}}
\bibliography{resHQET}

\end{document}